\documentclass[a4paper,fleqn,usenatbib]{mnras}

\usepackage[T1]{fontenc}
\usepackage{ae,aecompl}
\usepackage{natbib}
\usepackage{graphicx}
\usepackage{epstopdf}
\usepackage{hyperref}
\usepackage{amssymb}
\usepackage{amsmath}
\usepackage{threeparttable}
\usepackage{float}
\usepackage{mathtools}

\DeclarePairedDelimiter\abs{\lvert}{\rvert}%

\makeatletter
\let\oldabs\abs
\def\abs{\@ifstar{\oldabs}{\oldabs*}}

\title[Multiwavelength Analysis of the FSRQ 3C 279]{Multiwavelength Photometric and Spectropolarimetric Analysis 
of the FSRQ 3C 279}
\author[Pati\~no-\'Alvarez et al.]
{V. M. Pati\~no-\'Alvarez$^{1,2}$\thanks{E-mail:
victorp@inaoep.mx}, {S. Fernandes$^{3}$}, {V. Chavushyan$^{2}$}, {E. L\'opez-Rodr\'iguez$^{4}$},
\newauthor
{J. Le\'on-Tavares$^{5}$}, {E. M. Schlegel$^{3}$}, {L. Carrasco$^{2}$}, {J. Vald\'es$^{2}$}, {A. Carrami\~nana$^{2}$} \\
\\
$^{1}$Max-Planck-Institut f\"ur Radioastronomie, Auf dem H\"ugel 69, 53121 Bonn, Germany \\
$^{2}$Instituto Nacional de Astrof\'isica \'Optica y Electr\'onica (INAOE), Apartado Postal 51 y 216, 72000 Puebla, M\'exico\\
$^{3}$University of Texas at San Antonio, Department of Physics and Astronomy, One UTSA Circle, San Antonio Texas, 78249 TX, USA \\
$^{4}$SOFIA Science Center, NASA Ames Center, Mountain View, CA, USA \\
$^{5}$Centre for Remote Sensing and Earth Observation Processes (TAP). Flemish Institute for Technological Research (VITO), \\
Boeretang 282, 2400 Mol, Belgium.
} 

\date{Accepted XXX. Received YYY; in original form ZZZ}

\pubyear{2016}
\begin{document}
\label{firstpage}
\pagerange{\pageref{firstpage}--\pageref{lastpage}}
\maketitle


\begin{abstract}

In this paper, we present light curves for 3C 279 over a time period of six years; from 2008 to 2014. Our multiwavelength data comprise 1 mm to gamma-rays, 
with additional optical polarimetry. Based on the behaviour of the gamma-ray light curve with respect to other bands, we identified three different 
activity periods. One of the activity periods shows anomalous behaviour with no gamma-ray counterpart associated with optical and NIR flares. 
Another anomalous activity period shows a flare in gamma-rays, 1 mm and polarization degree, however, it does not have counterparts in the 
UV continuum, optical and NIR bands. We find a significant overall correlation of the UV continuum emission, the optical and NIR bands. 
This correlation suggests that the NIR to UV continuum is co-spatial. We also find a correlation between the UV continuum and the 1 mm data, 
which implies that the dominant process in producing the UV continuum is synchrotron emission. The gamma-ray spectral index shows statistically 
significant variability and an anti-correlation with the gamma-ray luminosity. We demonstrate that the dominant gamma-ray emission mechanism in 3C 279 changes over time. 
Alternatively, the location of the gamma-ray emission zone itself may change depending on the activity state of the central engine.

\end{abstract}



\section{Introduction}

Blazars, active galactic nuclei with a relativistic jet pointing almost directly into our line of sight \citep{Urry1995}, are the dominant population of gamma-ray sources in the sky, as seen by the Fermi Gamma-ray Telescope \citep{Nolan2012}. Blazars are highly variable at all frequencies, with non-thermal 
emission spanning over 10 decades in energy. According to the behaviour of their optical spectra, blazars are divided into BL Lac objects (BL Lacs) and 
flat spectrum radio quasars (FSRQs).

Some objects exhibit a high degree of linear polarization at optical wavelengths 
\citep[e.g.][and references therein]{Blinov2015}, however, this can also be a function of the activity state of the source 
\citep[e.g.][and references therein]{Itoh2015,Carnerero2015,Bhatta2015}.

3C 279, an FSRQ at $z=0.536$, was one of the first gamma-ray emitting quasars discovered with the Compton Gamma-Ray Observatory 
\citep{Hartman1992}. It has long been known that the emission of this source is highly variable in the radio \citep{Pauliny-Toth1966}, optical 
\citep{Oke67}, and gamma-rays \citep{Hartman1992}. For this reason, several intensive multiwavelength campaigns and theoretical studies
have led to important conclusions about the physical properties of 3C 279. \citet{Chatterjee2008} reported correlated changes between X-rays,
optical and radio wavebands for 3C 279 through an extensive multiwavelength campaign from 1996 to 2007. \citet{Collmar2010} studied the SED 
of 3C 279 in a high-state through multiwavelength observations. \citet{Larionov2008} found that the continuum spectra from 2006-2007 correspond
to a power-law with $\alpha = -1.6$. They also found co-rotation of the optical EVPA and VLBA core polarization {\it PA}, which showed that
the optical and radio emission were co-spatial within the uncertainties of their measurements. \citet{Bottcher2009} found a possible signature of a
decelerating jet through a plasmoid evolution simulation by fitting to {\it R}-band, {\it V}-band, and {\it I}-band light curves from January 2006.

\citet{Hayashida2012} report that, at radio wavelengths, the variability for this source appears to be much less rapid compared to the gamma-ray and optical bands; and the excess variance \citep[$F_{var}$,][]{Vaughan2003} in the radio regime is quite modest (for instance at 37, 15 and 5 GHz). \citep{Hayashida2015} reported rapid gamma-ray variability of 3C 279 during 2013-2014.

By analyzing multiwavelength observations and the SED of 3C 279 at three different epochs, \citet{Aleksic2011} found that the Very High Energy (VHE)
$\gamma$-ray emission detected in 2006 and 2007 by MAGIC challenge the standard one-zone model, based on relativistic electrons in a jet scattering
Broad Line Region (BLR) photons. Instead, they explored a two-zone model, where the VHE $\gamma$-ray emitting region is located just outside the BLR, 
while the standard optical-to-X-ray and the $\gamma$-ray emitting region is still inside the BLR, as well as a lepto-hadronic model, with both models fitting the
data reasonably well. \citet{Aleksic2014a} found a cutoff in the GeV range of the high energy SED of this source. This finding hints that the gamma-ray 
emission is coming from an inner region of the blazar, and therefore internally absorbed in the MAGIC energy range.

\citet{Janiak2012} modeled the time lag observed between optical {\it R}-band and gamma-rays in 3C 279, during the flare of February 2009 
(lasting $\sim$10 days). Their model showed that the flare was produced at a distance of a few parsecs from the central black hole; however, 
this is not the only possible model. Discerning between modeling such lags assuming single dissipative events \citep[as][]{Janiak2012}, 
or two-dissipative-zones scenarios is not an easy task. The data available and the cadence of the multiwavelength light curves do not yet allow 
distinguishing which mechanism is dominant in a given source \citep{Janiak2012}.

\citet{Lindfors2006} studied the synchrotron flaring behaviour of 3C 279, using data covering 10 years of monitoring from optical to radio frequencies.
The authors found that during high gamma-ray states, an early-stage shock component is normally present; while in low gamma-ray states, the time
since the onset of the last synchrotron outburst is significantly longer. They proposed that this supports the idea that gamma-ray flares are associated
with the early stages of shock components propagating in the jet.

Despite those studies, a general consensus about the location of the gamma-ray production zone in 3C 279 does not yet exist. 

To study the stratification of the emission regions for different wavelengths in 3C 279, we study multiwavelength light curves from 1 mm to gamma-rays, 
comprising a time-frame of six years, with an unprecedented cadence. We also wish to investigate the dominant emission mechanism for these 
wavelengths.

We arrange this paper as follows: Section 2 describes the observations used to carry out this work. Section 3 describes briefly the cross-correlation analysis between the different light curves. In Section 4 we address the variability of the different bands and the activity periods we identify. We present in Section 5 our results and discussion. We summarize our conclusions in Section 6.


\section[]{Observations}

\begin{figure*}
\includegraphics[width=0.9\textwidth]{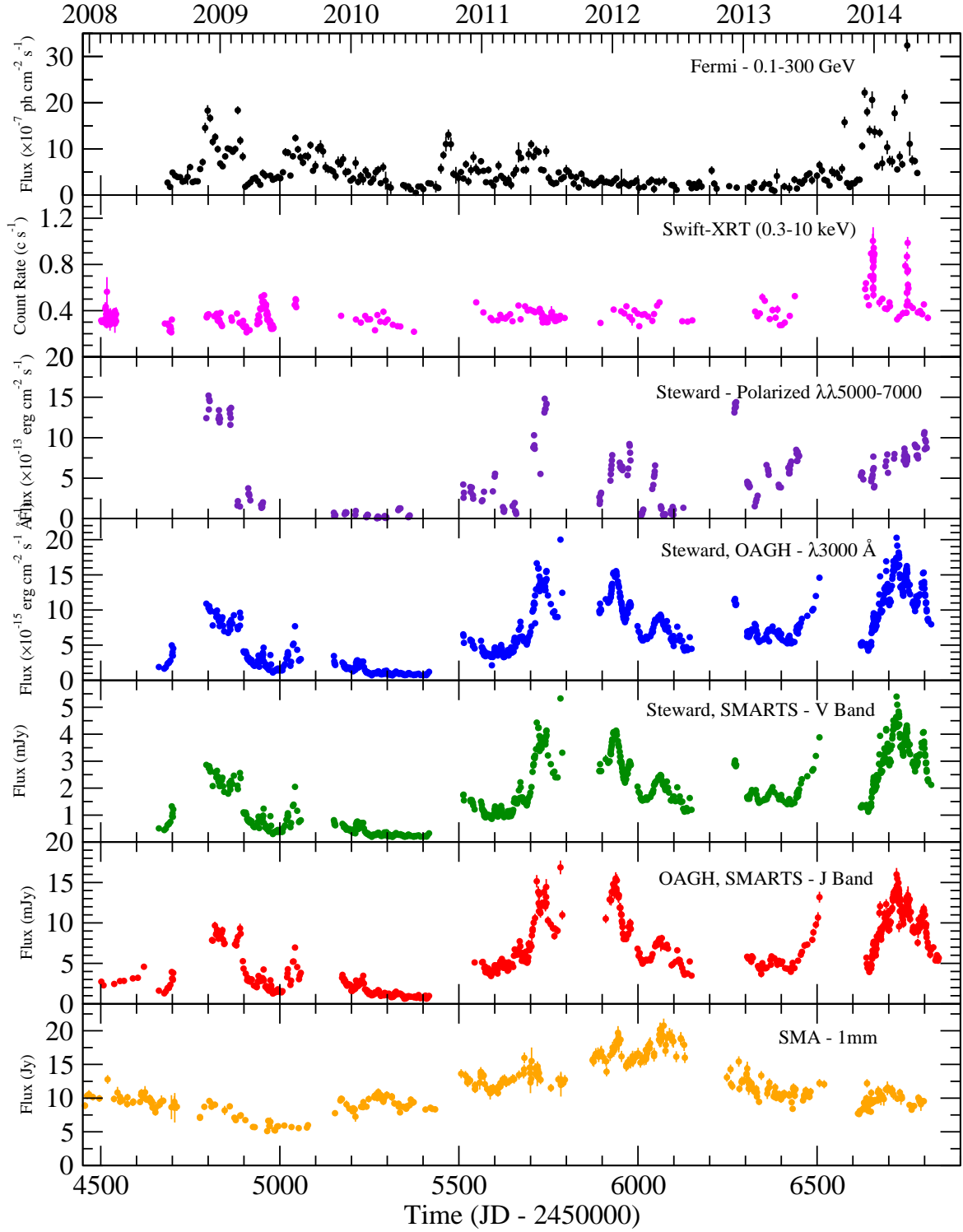}
\caption{Multiwavelength light curves of 3C 279. The band of observation and the origin of the data is labeled inside each panel. In the fourth panel, the 
points represent the observed spectra, as well as the points obtained using the relationship between the {\it V}-band and the 3000 \AA~continuum.}
\label{curves}
\end{figure*}

We obtained data from a variety of bands, including optical and near-infrared (NIR) photometry, optical spectra, millimeter, gamma-rays, X-rays, and spectropolarimetry. 
These datasets originate from a variety of sources. The light curves from the data are shown in Fig.~\ref{curves}. 

\subsection{Photometry}

To identify the emission mechanism of the source, we used optical ({\it V}-band) and Near-Infrared (NIR, {\it J}- {\it H}- and {\it K}-band) photometry.

We retrieved 263 {\it V}-band photometric points from The Steward Observatory Monitoring Program \citep{Smith2009} and 550 {\it V}-band photometric 
points from the SMARTS project \citep{Bonning2012}. The NIR photometric data points are from the Observatorio Astrof\'isico Guillermo Haro (OAGH) at Cananea, 
Sonora, Mexico using the Cananea Near-Infrared Camera \citep[CANICA,][]{Carraminana2009} and from the SMARTS project. We obtained the {\it J}-band photometric points from the OAGH (66 points) and the SMARTS project (545 points). The {\it H}-band photometric points were observed at the 
OAGH (70 points). The {\it K}-band photometric points were retrieved from the SMARTS project (487 points). To compare the optical and NIR amplitude 
variations with those observed at other wavelengths in linear scale, the optical and NIR photometry data have been converted to mJy. We converted the 
photometric data from magnitudes to fluxes using the absolute calibration of the photometry ($f_0$) from \citet{Carrasco1991}.

\subsection{Optical Spectra}

The optical spectra used in this work come from three different sources: 252 optical spectra were taken at the Steward Observatory as part of the Ground-based Observational Support of the Fermi Gamma-ray Space Telescope at the University of Arizona\footnote{http://james.as.arizona.edu/$\sim$psmith/
Fermi/} monitoring program. The optical spectra were taken with the SPOL CCD Imaging/Spectropolarimeter \citep{Smith2009}. The data in the archive at 
the University of Arizona include flux calibrated spectra for 3C 279, which have already been reduced; details on the observational setup and reduction 
process are presented in \cite{Smith2009}. The spectral coverage is from 4000 to 7500 \AA. The resolution ranges from 15-25 
\AA, depending on the slit width used for the observation. In this work, we only use spectra that have been calibrated against the {\it V}-band 
magnitude. Steward Observatory data is broken up into observing cycles separated by several months. Our observational time-frame contains 6 
Steward data cycles, and our activity periods contain from 1 to 3 Steward cycles.

We obtained 17 optical spectra from the spectroscopic monitoring program being carried out at the Instituto Nacional de Astrof\'sicia \'Optica y 
Electr\'onica (INAOE) 2.1-m telescope at OAGH in Mexico 
\citep[for details see][]{Patino-Alvarez2013a}, and the 2.1-m telescope at the Observatorio Astron\'omico Nacional at San Pedro Martir (OAN-SPM), 
Baja California, Mexico. After each object exposure, He-Ar lamp spectra were taken to allow wavelength calibration. Spectrophotometric standard stars 
were observed every night (at least two per night) in order to perform flux calibration. The spectrophotometric data reduction was carried out using the 
IRAF package\footnote{http://iraf.noao.edu/}. The image reduction process included bias and flat-field corrections, cosmic ray removal, 2D wavelength 
calibration, sky spectrum subtraction, and flux calibration. The 1D spectra were subtracted taking an aperture of 6 arcsec around the peak of the spectrum 
profile. The observational log for the observations in OAGH and OAN-SPM is shown in Table \ref{obslog}. From a visual inspection of the data, we 
determined that there is an agreement between the spectra obtained from the Steward Observatory, OAGH, and OAN-SPM.

\begin{figure}
\includegraphics[width=0.5\textwidth]{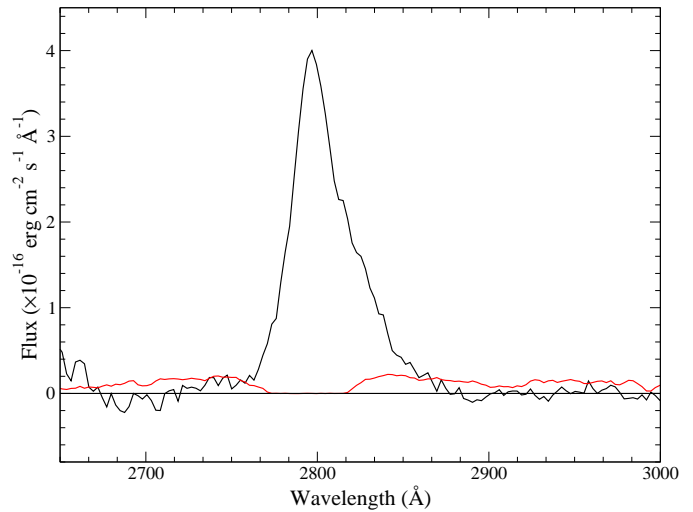}
\caption{Mean spectrum of 24 high S/N ratio spectra taken at Steward Observatory. The spectrum (in black) is in rest frame, and has the 
continuum subtracted. The fitted Fe II emission template \citep{Vestergaard2001} is shown in red. As can be seen, the UV Fe II emission is negligible.}
\label{iron}
\end{figure}

All spectra were shifted into the rest frame of the source. We applied a cosmological correction to the monochromatic flux of the form $(1 + z)^3$. 3C 279 has a high galactic latitude, therefore no corrections were necessary for galactic interstellar extinction and reddening. When we performed subtraction of the Fe II emission usually found around the Mg II $\lambda$2798~\AA~emission line, we found that Fe II UV emission in this source is insignificant. Fig.~\ref{iron} shows that, for a mean of a sample of high S/N ratio Steward Observatory spectra, the Fe II emission, represented by the red line, is negligible. Therefore, the UV-continuum at $\lambda 3000$~\AA~flux was measured directly from the rest frame spectra without Fe II subtraction. The reported error is the rms of the spectrum within the wavelength range $2900-2950$~\AA.

In order to increase the number of points for the UV-continuum light curve, we estimated the UV-continuum $\lambda 3000$~\AA~flux using a relationship between the {\it V}-band flux in mJy and the $\lambda 3000$~\AA~flux converted to mJy. We took the data where there are {\it V}-band and 
spectral observations in the same night. We have also added 5 per cent to the uncertainty in the continuum emission to take into account the flux calibration error (Paul 
Smith, Private Communication). We fit the data with a linear regression taking into account the uncertainty in both quantities using the IDL task 
FITEXY\footnote{http://idlastro.gsfc.nasa.gov/ftp/pro/math/fitexy.pro}. The linear fit obtained yields the relationship:

\begin{equation}
{\rm Flux}_{3000}=(-0.030 \pm 0.026) + (1.147 \pm 0.012)\times {\rm Flux_V}
\label{eqVC}
\end{equation}

\begin{figure}
\includegraphics[width=0.5\textwidth]{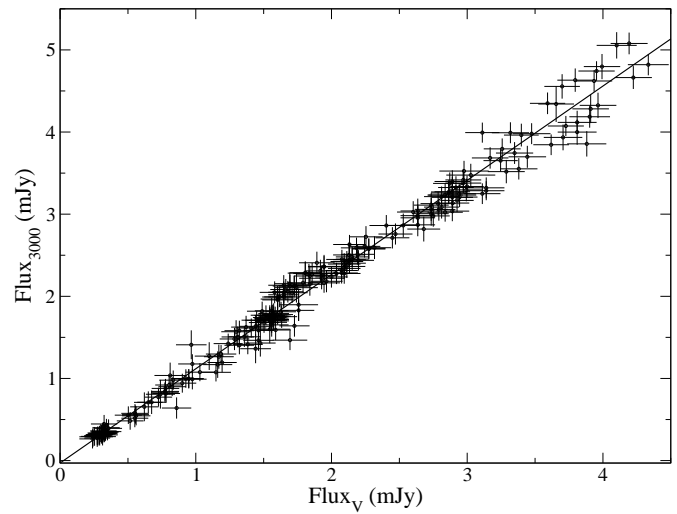}
\caption{Correlation between the {\it V}-band flux and the 3000 \AA$\,$continuum flux.}
\label{V3000}
\end{figure}

Both quantities in Eq.~\ref{eqVC} are given in mJy. Fig.~\ref{V3000} shows the flux correlation and fit to the data. After applying this transformation to the 
{\it V}-band data points, we ended up with 844 data points for UV-continuum.

\begin{table*}
\centering
\begin{minipage}{110mm}
\caption{Observational log of the spectra taken in OAGH and OAN-SPM.}
\begin{tabular}{cccccc}
\hline
UT Date & Observatory & Grating & Resolution (\AA) & Seeing & Exposure Time (s) \\
\hline
\hline
2011 Jan 30 & OAGH & 150 l/mm & 15.4 & 2.5" & 3$\times$1800 \\
2011 Feb 12 & OAGH & 150 l/mm & 15.4 & 1.8" & 3$\times$1200 \\ 
2011 Feb 13 & OAGH & 150 l/mm & 15.4 & 2.2" & 3$\times$1800 \\
2011 Mar 25 & OAGH & 150 l/mm & 15.4 & 1.9" & 3$\times$1800 \\
2011 Mar 29 & OAGH & 150 l/mm & 15.4 & 2.1" & 3$\times$1800 \\
2011 Mar 30 & OAGH & 150 l/mm & 15.4 & 2.0" & 3$\times$1800 \\
2011 Apr 01 & OAGH & 150 l/mm & 15.4 & 2.1" & 3$\times$1800 \\
2011 Apr 06 & OAGH & 150 l/mm & 15.4 & 4.0" & 3$\times$1800 \\
2011 Apr 08 & OAGH & 150 l/mm & 15.4 & 1.7" & 3$\times$1800 \\
2011 May 31 & OAGH & 150 l/mm & 15.4 & 2.2" & 3$\times$1800 \\
2011 Jun 26 & OAGH & 150 l/mm & 15.4 & 4.1" & 3$\times$1800 \\
2013 Feb 07 & OAGH & 150 l/mm & 15.4 & 3.9" & 3$\times$1800 \\
2013 Mar 14 & OAGH & 150 l/mm & 15.4 & 2.7" & 3$\times$1800 \\
2013 Apr 07 & OAGH & 300 l/mm & 15.4 & 3.1" & 3$\times$1800 \\
2013 Apr 20 & OAN-SPM & 300 l/mm & 8.8 & 1.2" & 2$\times$1800 \\
2013 May 10 & OAGH & 150 l/mm & 15.4 & 2.3" & 3$\times$1800 \\
2013 Jun 08 & OAGH & 150 l/mm & 15.4 & 3.2" & 3$\times$1200 \\
\hline
\end{tabular}
\begin{tablenotes}
\item{All observations were obtained in a 2.1m telescope using a Boller \& Chivens spectrograph located in the Cassegrain Focus. The aperture extracted for each spectrum is 2.5"$\times$6.0" in size.}
\end{tablenotes}
\label{obslog}
\end{minipage}
\end{table*}

\subsection{Millimeter}
The 1 mm data were obtained at the SMA on Mauna Kea (Hawaii) from 2008 August 19 to 2014 May 20. 3C 279 is included in an ongoing monitoring 
program at the SMA to determine the fluxes of compact extragalactic radio sources that can be used as calibrators at mm wavelengths. We obtained 365 
fluxes from this monitoring program. Data from this program are updated regularly and are available at the SMA website\footnote{http://sma1.sma.hawaii.edu/callist/callist.html}. Details of the observations and data reduction can be found in \citet{Gurwell2007}.

\subsection{X-Rays}

The X-ray data were retrieved from the public database of the Swift-XRT\footnote{\url{http://www.swift.psu.edu/monitoring/}}. 
The Swift-XRT data were processed using the most recent versions of the standard Swift tools: Swift Software version 3.9, 
FTOOLS version 6.12 and XSPEC version 12.7.1. Light curves are generated using xrtgrblc version 1.6.

Circular and annular regions are used to describe the source and background areas, respectively, and the radii of both depend on the current count rate. 
In order to handle both, piled-up observations and cases where the sources land on bad columns, PSF correction is handled using xrtlccorr \citep[Full details of the reduction procedure can be found in][]{Stroh2013}.

\subsection{Gamma-Rays}

The gamma-ray light curve from 0.1 to 300 GeV was built by using data from the Fermi Large Area Telescope (LAT). It was reduced and analyzed with the 
Fermi Science Tools v9r33p0. The Region of Interest (ROI) was selected as $15^{\circ}$ in radius, centred at the position of 3C 279. The minimisation was 
done through a maximum likelihood algorithm, and we modelled our source spectra as a log-parabola. We included all sources within $15^{\circ}$ of 3C 279, 
extracted from the 2FGL catalog \citep{Nolan2012}, with their normalisation and spectral indices kept free. The currently recommended CALDB set of 
the instrument response functions, along with the latest diffuse and isotropic background model files were applied to this analysis. In order to generate the 
spectral models and produce the gamma-ray light curve, modified versions of the user-contributed software\footnote{http://fermi.gsfc.nasa.gov/ssc/data/analysis/user/} were used. We adopted a time bin of 7 days for our light curve fluxes to increase S/N ratio and kept only the fluxes from time bins with a TS $>$ 25. Upper limits were not calculated for the bins with TS $<$ 25, because they are not suitable for cross-correlation analysis. The low number of discarded bins ensures that we do not introduce a bias due to gaps in the gamma-ray light curve.

\subsection{Polarimetry}

The polarization spectra were also taken from the Steward Observatory monitoring program. From their archive, we retrieved 322 Stokes $q$ and $u$ 
spectra and 256 total flux spectra which were used for the present analysis. The Steward Observatory monitoring program divides their observing time into cycles, which correspond to observing runs of $\sim$10 days each month.

Using the $q$ and $u$ spectra from the Steward Observatory database, we estimate the degree and position angle of polarization as a function of time. 
To avoid the noisy edges of the spectra below 5000 \AA~and above 7000 \AA, we only use data in the wavelength range of $\lambda\lambda$ 5000 - 7000 \AA~for our estimations of polarimetric measurements.

The $q$ and $u$ spectra are the wavelength-calibrated and normalised linear polarization parameters as a function of wavelength. The normalized linear 
Stokes polarization parameters are defined as $q=Q/I$ and $u=U/I$. Where Q and U are the linear Stokes polarization parameters and I the total 
intensity. In spectral form this is $q(\lambda)=Q(\lambda)/I(\lambda)$ and $u(\lambda)=U(\lambda)/I(\lambda)$.

\subsubsection{Calibrated Total Flux Spectra}

We use the calibrated total flux spectra, along with the $q$ and $u$ spectra to determine the polarized flux. There are only calibrated total flux spectra 
for a subset of the {\it q} and {\it u} spectra. There are 66 nights where the {\it V}-band magnitude of the comparison star used to calibrate the total flux 
spectra was not measured due to non-photometric conditions.

\subsubsection{Linear Degree of Polarization}
We estimate the linear degree of polarization as a function of wavelength (polarization spectrum) as

\begin{equation}
p(\lambda)=\sqrt{q(\lambda)^2 + u(\lambda)^2}.
\end{equation}

To determine if there was a wavelength dependence in the degree of polarization in the range $\lambda\lambda$5000 - 7000 \AA, we fit the polarization 
spectrum with a linear function. Using the IDL routine LINFIT, we found that the polarization spectra are statistically flat between $\lambda\lambda$5000 - 
7000 \AA; i.e. a fitting to a straight line with slope zero, results in a significant fit to the data. Hence, we can characterize the degree of polarization 
in our chosen wavelength range by a single value.

From the normalized Stokes $q$ and $u$ spectra, the linear degree of polarization is defined as \citep{Stokes1852}

\begin{equation}
P=\sqrt{q^2 + u^2}.
\end{equation}

Since the definition of the degree of polarization squares and sums $q$ and $u$ values, a Rician statistical bias is introduced into the estimate that 
causes an overestimation in the value of the degree of polarization \citep{Serkowski1958,Serkowski1962,Wardle1974, Clarke1993}. We estimate the 
corrected degree of polarization as in \cite{Serkowski1958} and \cite{Wardle1974}

\begin{equation}
P=\sqrt{<q(\lambda)>^2 + <u(\lambda)>^2 - \sigma_P}.
\label{peq}
\end{equation}

For each observation we use the median values of the $q$ and $u$ spectra to estimate the degree of polarization. $\sigma_p$ is the error in $P$ and is 
defined in equation \ref{perr}.

The ratio $P/\sigma_p$ is an estimation of the accuracy of $P$ values \citep{Simmons1985, Smith2007}. For ratio values greater than 3, the polarization 
measurements are very accurate \citep{Smith2007}. For our data set, $P/\sigma_p >> 3$, indicating our estimations are highly significant. The Rician 
statistical bias has a greater effect on the value of the degree of polarization when the ratio $P/\sigma_p$ is less than approximately 3. For over 99 per 
cent of the data set, the per cent difference between corrected and uncorrected values is less than 1 per cent. We do not report estimations of $P$ 
where $P/\sigma_p$ is less than 1, as this means the error in $P$ is larger than $P$ itself. We consider this to be an unreliable measurement with very 
low signal to noise.

We estimate the error in $P$ using root-mean-squared (rms) values of the $q$ and $u$ observations binned from $\lambda\lambda$ 5000 - 7000 \AA. We obtained the rms 
values from the Steward Observatory archive. Photon statistics is the dominant source of uncertainty in polarimetric measurements, especially of bright 
objects such as 3C 279 \citep{Impey1990, Wang1996, Gabuzda2006}. The rms values obtained from the Steward Observatory archive are a measure of 
the photon statistics of the $q$ and $u$ estimations. Therefore, they are a measure of the error of the $q$ and $u$ observations.

We estimate the error in $P$ with the rms values of $q$ and $u$ following a method suggested by Paul Smith (Private Communication). Equation 
\ref{perr} is a result of propagation of errors from binning of the data. From the propagation of errors, the $\sigma_P$ is a single rms value, in this case 
the rms of a Stokes parameter binned from $\lambda\lambda$ 5000 - 7000 \AA, multiplied by a factor of $1/\sqrt{N}$. Here, N is the number of observations. $N=500$ 
because we bin the data from 5000 - 7000 \AA$\,$ and the resolution of the spectrograph is 4\AA/pixel. 

Our adopted relationship to estimate the error in $P$ is
\begin{equation}
\sigma_P=\frac{MAX(rms_q,rms_u)}{\sqrt{N}}.
\label{perr}
\end{equation}

Simply put, we take the larger of the $q$ and $u$ rms values and divide by square root of N. We take the maximum rms value of the two Stokes 
parameters as a conservative approach to estimating the error. 

\subsubsection{Position Angle of Polarization}
\label{PA}
As in the case of $P$, we determine from a linear fit to the position angle of polarization ({\it PA}) spectra that a single value could characterize each 
observation. We estimate the {\it PA} between $\lambda\lambda$ 5000 - 7000 \AA$\,$ with the formula
\begin{equation}
PA = \frac{1}{2} \arctan{\Bigg(\frac{<u(\lambda)>}{<q(\lambda)>}\Bigg)}.
\label{paeq}
\end{equation}

In the definition of the {\it PA} based on the polarization ellipse \citep{Jones1941,Stokes1852}, there is no distinction between $n\pi$ multiplicative factors 
of the {\it PA} \citep{Adam1963}. Therefore, there is an $n\pi$ ambiguity in all {\it PA} estimations. We apply a shifting method commonly used to correct for 
the ambiguity (e.g. \citealp{Sasada2011}, \citealp{Sorcia2013} ) to our data. The shifting method applies corrections to adjacent {\it PA} values in the light 
curve and makes the assumption that variations in the {\it PA} over those adjacent points should be less than $\sim90^{\circ}$ apart.

In the literature, there is great variation in time between adjacent points, from the order of days to months. Clearly, the shorter the time between the 
adjacent points, the more viable the assumption is. Therefore, the method is most accurate for densely sampled data sets. Some authors use as little as 
five days as a limit over which to apply the shifting method \citep{Jorstad2009, Carnerero2015}. The data we use is only very densely sampled ($\sim$1 
observation/day) in small time bins($\sim$5-10 days at a time), and the average time in between time bins is three weeks. This introduces some 
uncertainty into trends that we might identify that extends over breaks in data over $\sim$1 day. However, the shifting method only changes the trends in 
the {\it PA}, not the intrinsic values themselves. The original calculated {\it PA} can be recovered by applying $n\pi$ shifts to all of the data points to lie in 
the original calculated {\it PA} range. We take the continuation of long-term trends in our data over longer cadence observations and gaps in data as 
significant, assuming no large variations in {\it PA} between time bins. This analysis is consistent with earlier works where the authors follow consistent 
trends of {\it PA} swings in their analysis over less densely sampled data \citep{Marscher2008}.

One caution to the shifting method is that it can introduce shifts where there might not be a true shift present. This happens where adjacent points are 
close to $\sim90^{\circ}$ apart and could be greater or less than $\sim90^{\circ}$ within error limits. This could create false positive shifts in the {\it PA}. 
Poor sampling can also affect the interpretation of the results, because false positives can be detected. Careful study of how the shifting method corrects 
for the $n\pi$ ambiguity in each data set is essential. The {\it PA} trends uncovered by correcting the $n\pi$ ambiguity are used to probe magnetic field 
orientation shifts over time, locations of polarized emission sites, and {\it PA} behaviour during flares (e.g. \citealp{Abdo2010b,Kiehlmann2013}).

We estimate the error in the {\it PA} in the limit of $P>>\sigma_P$, as is characteristic of our data set, as \citep{Serkowski1962,Wardle1974}
\begin{equation}
\sigma_{PA}=28.65^\circ*\frac{\sigma_P}{P}.
\label{paerreq}
\end{equation}

\begin{figure*}
\includegraphics[width=0.95\textwidth]{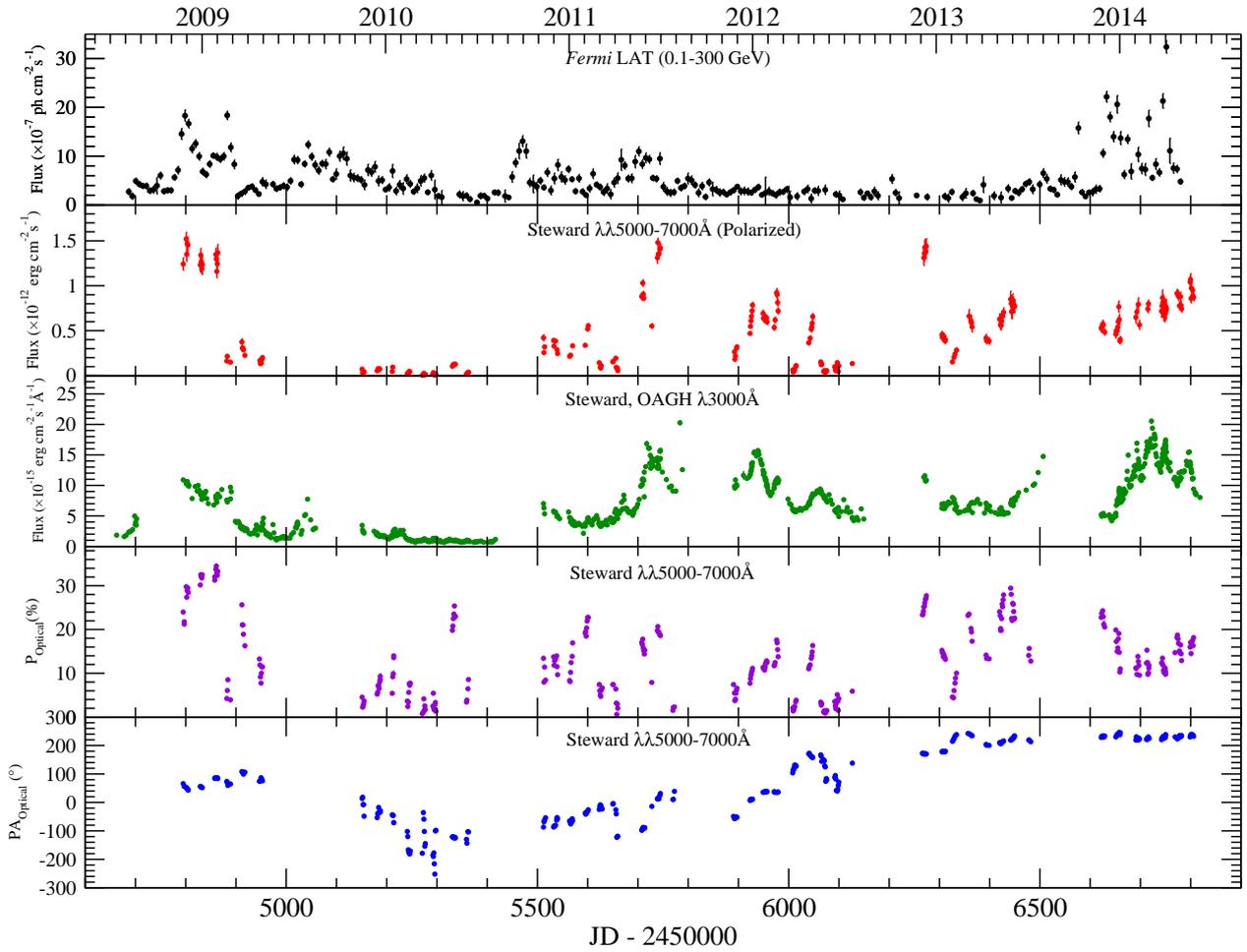}
\caption{The behaviour of the polarimetric observations ($P$, {\it PA}, Polarized Flux) compared with the gamma-rays and the UV-continuum for 3C 279. We 
labelled each panel with the band of observation and the origin of the data.}
\label{Polar}
\end{figure*}

The variability of $P$ and {\it PA}, compared to the gamma-rays and the UV continuum is shown in Fig.~\ref{Polar}. 

The Steward Observatory bins their data from $\lambda \lambda$5000 - 7000 \AA~to determine the polarimetric quantities found in the public database. 
We tested different wavelength binning and found no significant differences in the polarimetric quantities for different bin sizes. 
Our results are also consistent with the public data from the Steward Observatory.

We could not obtain a robust estimation of the error in the polarimetric quantities from the normalized Stokes spectra alone. The methodology used to 
determine the errors, which was suggested by Paul Smith (Private Communication), involved using part of the public data, the rms values of the $q$ and $u$ 
Stokes spectra. The rms values contain information about the error as they take into account the photon counting statistics for the wavelength range of 
$\lambda \lambda$ 5000 - 7000 \AA. We could not obtain rms values for other wavelength bins from the normalized Stokes spectra alone. 
So, even though we used the same wavelength range for binning the data, in the end, we performed the analysis ourselves to study the spectra 
more thoroughly.

\section{Cross Correlation Analysis}
\label{CrossAna}

We carried out the Cross-Correlation analysis using three different methods: The Interpolation Method (ICCF, \citealp{Gaskell1986}), the Discrete Cross-Correlation 
Function (DCCF, \citealp{Edelson1988}) and the Z-Transformed Discrete Correlation Function (ZDCF, \citealp{Alexander1997}). We used modified versions of these methods as described in \citet{Patino-Alvarez2013b}. The correlations between gamma-rays, X-rays, UV continuum, optical {\it V}-band, NIR {\it JHK}-bands and polarization ({\it P} and {\it PA}), were analyzed using cross-correlation analysis. The resulting lags obtained from the cross-correlation analysis between 
the different light curves are summarized in Table \ref{lags1}, and the figures are presented in Appendix~\ref{ccfigures}.

The confidence intervals (at 90 per cent) in the lags are calculated using an equation obtained by Monte Carlo simulations of different types of time series as 
explained in \cite{Patino-Alvarez_master}. The simulations show that the main contributor to the uncertainty in the lag obtained by the cross-correlation 
analysis is the sampling mean of the different light curves involved. The reason we needed simulations to obtain the uncertainty in the lag is because the 
usual error reported in the literature varies with the method used. The uncertainty in the ICCF is not well defined in the literature. One method involves 
fitting a Gaussian to the peak of the cross-correlation function and taking the standard deviation as uncertainty. However, this does not take into 
consideration the properties of the individual light curves. The DCCF uncertainty usually presented in the literature is based on the bin size selected prior 
to the analysis, which has the same problem as the ICCF. The error defined for the ZDCF by \cite{Alexander1997} is a fiducial interval. However, a fiducial 
interval is not the same as a confidence interval, a fact that is also mentioned in the paper. The significance levels are what best represent the level of certainty in 
the cross-correlation coefficient results. Our cross-correlation figures show significance levels in the correlation coefficient at 90, 95, and 99 per cent, 
also calculated from Monte Carlo simulations \citep{Emmanoulopoulos2013}, as well as the confidence intervals in the lags.

\begin{table*}
\centering
\begin{minipage}{140mm}
\caption{Cross-Correlation results for the entire time-range and the defined activity periods. All correlations are at a confidence level $>$99\%, unless otherwise indicated. We also performed the cross-correlation using $P$ and $PA$, but we found no significant correlation with any of the other light curves. }

\begin{tabular}{ccccc}
\hline
Bands & Full Time-Range & Period A & Period B & Period C \\
\hline
\hline
{\it J}-band vs. {\it H}-band & 0.0$\pm$22.3 & -0.3$\pm$24.6 & -4.2$\pm$21.7 & 0.0$\pm$5.4 \\
{\it J}-band vs. {\it K}-band & -0.1$\pm$3.2 & 0.9$\pm$3.2 & -0.3$\pm$3.6 & 1.0$\pm$1.5 \\
3000 {\rm \AA} vs. {\it V}-band & 0.0$\pm$2.0 & 0.0$\pm$2.1 & 0.0$\pm$2.1 & 0.0$\pm$1.2 \\
3000 {\rm \AA} vs. {\it J}-band & 0.2$\pm$2.7 & -0.2$\pm$2.6 & 0.0$\pm$3.2 & 0.0$\pm$1.3 \\
{\it V}-band vs. {\it J}-band & 0.0$\pm$2.7 & -0.1$\pm$2.6 & 0.8$\pm$3.2 & 0.0$\pm$1.3 \\
3000 {\rm \AA} vs. Gamma-Rays & --- & -0.7$\pm$5.0 & --- & 28.6$\pm$4.8* \\
{\it J}-band vs. X-rays & -65.1$\pm$5.8** & --- & --- & -63.6$\pm$2.2 \\
{\it V}-band vs. X-rays & -67.7$\pm$5.8** & --- & --- & -65.9$\pm$2.2 \\
3000 {\rm \AA} vs. X-rays & -92.5$\pm$5.8** & --- & --- & -63.5$\pm$2.2 \\
X-rays vs. Gamma-Rays & -0.85$\pm$5.8** & --- & --- & 1.3$\pm$3.0 \\
\hline
\multicolumn{5}{l}{* This correlation was obtained by discarding the first gamma-ray flare on period C (JD$_{245}\geq6700$).} \\
\multicolumn{5}{l}{** This correlation was obtained only in the DCCF and ZDCF methods, while the ICCF method} \\
\multicolumn{5}{l}{shows no significant correlation, which is indicative that only one part of the light curves is} \\
\multicolumn{5}{l}{dominating the correlation (spurious result), as expected when looking at the results for the} \\
\multicolumn{5}{l}{individual periods.} \\

\end{tabular}
\label{lags1}

\end{minipage}
\end{table*}

As can be seen in Table~\ref{lags1}, the {\it V}-band, the {\it J}-band and the UV Continuum emission are simultaneous. Because emission in the $
\lambda 3000$ \AA~continuum, {\it V}- and NIR {\it JHK}-bands are simultaneous; any correlation with other bands is also valid for these bands. The 
simultaneity of these bands implies the emission regions for these bands are co-spatial (e.g. \citealp{Marscher1996,Marscher1996b,Rani2013}).

On the other hand, the correlations involving the 1mm emission were analyzed using the Spearman Correlation Coefficient, 
in order to discern if the emission in the {\it V}-band and the 1 mm light curves have a common origin, as 
suggested by \cite{Aleksic2014b}. Given the high cadence of observations in the {\it V}-band, we obtained approximate {\it V}-band 
fluxes at the observation times of the 1 mm light curve via linear interpolation. We calculated the Spearman correlation coefficient for these two bands 
and found a correlation coefficient of $R=0.65$, with a probability of obtaining this correlation by chance $p<<0.01$. This low value allows us to confirm 
a significant correlation. We repeat this procedure for the rest of the bands used in this work, being careful of taking the light curve with 
the highest cadence as reference. We also made sure that only interpolated values were taken into account during the correlation 
(i.e. no extrapolation). We found a positive and significant correlation between the 1 mm emission and the following bands: 
3000 \AA~continuum ($R=0.45$), {\it J}-band ($R=0.45$), and {\it K}-band ($R=0.52$). 
We did not find a significant correlation between 1 mm and the following bands: 
Gamma-rays ($R=-0.31$), {\it H}-band ($R=0.35$), {\it P} ($R=-0.31$), {\it PA} ($R=0.20$), and X-rays ($R=0.18$). 
This is an indication that the UV, optical and NIR continuum are dominated by non-thermal emission. We suspect that the reason for the lack of 
significant correlation between the 1 mm emission and the {\it H}-band, is due to the low number of observations on the {\it H}-band compared to the 1 mm light curve.


\section{Variability and Activity Periods}
\label{Periods}

Based on the behaviour of the gamma-rays with respect to the UV, optical and NIR, we separated the entire time range into three different activity periods: A flaring period in multiple bands, with counterparts in gamma-rays; a flaring period in multiple bands with no counterparts in gamma-rays; and another flaring period in multiple bands with apparent counterparts in gamma-rays. We also 
performed cross-correlation analysis on each individual activity period, for all the light curves. For the polarimetric quantities we define the continuum levels of the $P$ 
as the per centage that best describes the level that is returned to, after flaring activity. The continuum level of the {\it PA} is the angle that the fluctuations 
rotate about or return to, after some varying time-frame or amplitude swing. Hereafter, all the dates we present are in the format $\rm{JD}_{245}=
\rm{JD}-2'450'000$. 

\subsection{Activity Period A}

\begin{figure}
\centering
\includegraphics[width=0.5\textwidth]{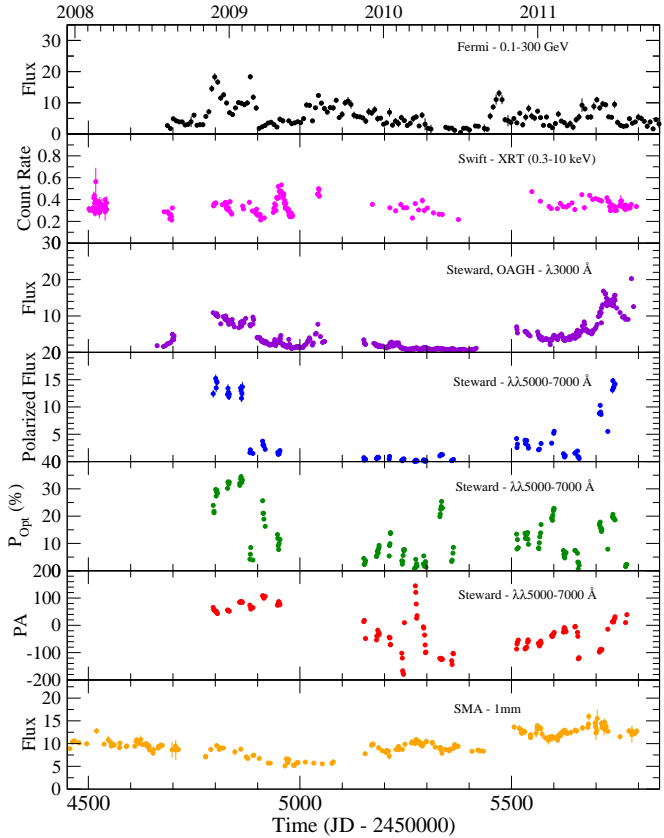}
\caption{Activity period A. The gamma-rays, UV, optical and NIR are well correlated and simultaneous. The units are the same as in Fig.~\ref{curves}.}
\label{PeriodA}
\end{figure}

This period ranges from $\rm{JD}_{245}=4650-5850$ (see Fig.~\ref{PeriodA}). During this period we observe multiple flares in the gamma-ray emission, 
as well as counterparts in the optical ({\it V}-band), UV continuum, and NIR emission ({\it J}-, {\it H}- and {\it K}-bands). In the 1 mm light curve we 
observe a response to each of these flares, however, the amplitude of the 1 mm flares is not as high as in the other wavelengths. The cross-correlation for 
this period shows a delay of 0.7$\pm$5.0 days between the UV spectral continuum and the gamma-ray emission. This is consistent with zero delay, 
which implies co-spatial emission regions.

After studying the polarimetric behaviour over each Steward cycle, we found that the spread in $P$ in the first Steward cycle is the largest for the entire data set. The 
maximum and minimum values for $P$ are 34.5 per cent and 1.6 per cent, respectively. This corresponds to a spread of 32.9 per cent, and the drop from 
the maximum to the minimum occurs over 29 days. This behaviour occurs during flaring activity in other wavebands, as discussed in 
Section \ref{Periods}. The {\it PA} behaviour during this time exhibits 4 small scale swings ($\sim20^{\circ}$).

We observed five clear flares in $P$ (in this context a flare is a local maximum in the $P$ light curve). Flaring time-scales of $P$ range from 40 to 85 days 
and the change of $P$ during flaring was as high as an order of magnitude increase.

The $P$ flare around JD$_{245}=$ 5330 is superseded by two apparent large swings in {\it PA} with a delay of approximately 50 days: $\sim180^{\circ}$ 
southward and $\sim180^{\circ}$ northward, then $\sim270^{\circ}$ southward and $\sim180^{\circ}$ northward. In a recent paper on the polarimetric 
behaviour of 3C 279 \citep{Kiehlmann2016}, this {\it PA} behaviour is not observed. The cadence of data in that paper is much higher, therefore we 
believe these swings could be artifacts resulting from poor sampling in that portion of the light curve. This exemplifies the reason why care must be taken with the shifting method and the dangers of shifting over large time gaps. In the \cite{Kiehlmann2016} paper, a southward rotation of the PA is reported over this time period, which is consistent with our data.

The polarized flux follows the trend of $P$ taking into account the continuum flux level. At the beginning of the activity period, $P$ is high and the 
continuum is decreasing, thus the polarized flux is high, but decreasing. When the $P$ drops, so does the polarized flux. The {\it PA} during this beginning 
period is oscillating around an increasing {\it PA} continuum value. There is a small amplitude flare in the continuum and a large amplitude flare in the 
gamma-rays as $P$ drops to a minimum, around JD$_{245}=$ 4900, as such, the polarized flux is at a minimum.

The third Steward cycle in this activity period has high levels of $P$ variability associated with several flares in the UV continuum. The UV-mm flares have 
been previously discussed, but the $P$ during the largest flare in this cycle is near the maximum of the polarized flux over the entire time-frame, 
suggesting simultaneous flaring in $P$. Preceding this flare is the beginning of a large swing in {\it PA} that occurs over approximately 200 days. Over the 
cycle, there is a general northward rotation, but at ~JD$_{245}=$ 5650 it rotates sharply southward. The extent of the swing is not known for sure, 
because there is an observational gap in the middle of the swing. However, after the gap, the {\it PA} is rotating sharply northward, returning to the 
continuum {\it PA} level. The return to the continuum level happens to coincide with the maximum of the largest UV flare. Such a large long-term swing in 
{\it PA} during a large flare in the UV suggests that a feature is moving along a helical magnetic field, as proposed by \citet{Kiehlmann2016}. 

In Activity Period A, the continuum $P$ level from JD$_{245}=4800-5800$ is low, $<10$ per cent, until JD$_{245} = 5300$, where it increases until 
JD$_{245} = 5500$. It drops back down to a low level for the rest of the activity period. We found a beginning trend of a northward rotation of the 
continuum {\it PA} followed by a swing to a southward rotation at approximately JD$_{245} = 4900$, which coincides with a flare in $P$, but no concurrent 
multiwavelength counterparts. There is a northward rotation of the continuum {\it PA} in Steward cycle 3, but the time of the swing occurs during an 
observational gap. At the very end of the third steward cycle, we observe a swing to southward rotation, which also happens to coincide with a flare in $P$ 
and no concurrent multiwavelength counterparts.

\subsection{Activity Period B}

\begin{figure}
\centering
\includegraphics[width=0.5\textwidth]{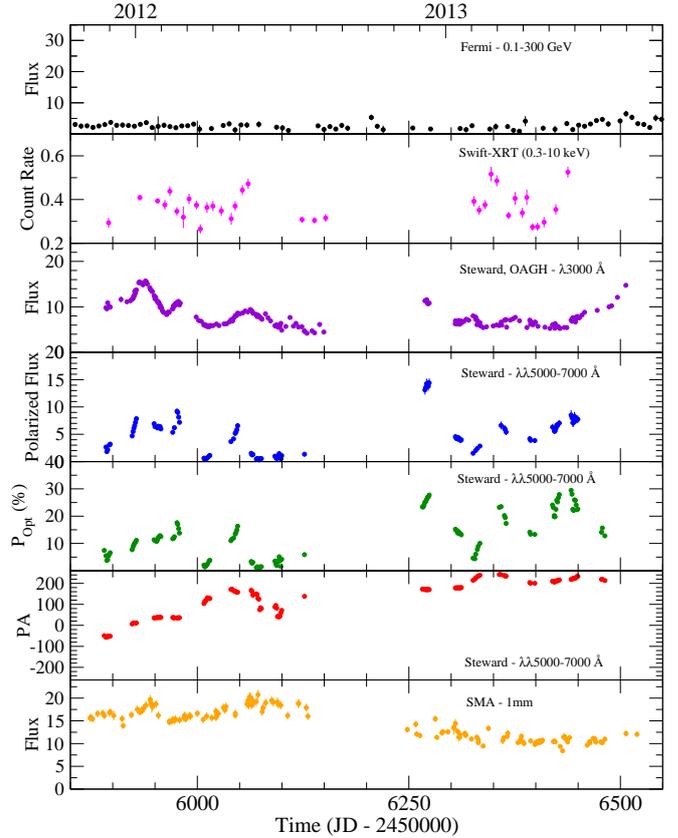}
\caption{Activity period B. We do not observe a counterpart in gamma-rays to flares in the UV, optical and NIR. The units are the same as in Fig.~\ref
{curves}. }
\label{orphan}
\end{figure}

This period ranges from $\rm{JD}_{245}=5850-6400$ (see Fig.~\ref{orphan}). During this time period, we observe multiple flares in the optical {\it V}-band, with clear counterparts in the UV spectral continuum and NIR bands. We also observe an increase in the 1 mm emission corresponding with each of 
these flares. The highest levels of 1 mm emission over the entire time-frame of our observations occurs during this activity period. There are increases in 
polarization degree coincident with these flares; this might indicate that these flares have a non-thermal origin. However, there are no counterparts to any 
of these flares in the gamma-rays. This kind of behaviour has only been reported once by \citep{Chatterjee2013} on the source PKS 0208-512. In the 
previous case, the reported period of low gamma-ray flux is $\sim$150 days, while the period we observe in 3C 279 is $\sim$500 days.

We analyze the polarimetric behaviour over the two Steward cycles that make up this time period. The {\it P} is highly variable during this period, and we find correlations with the high levels of activity in the other emission bands over this time-frame. There is a clear general northward rotation of the {\it PA} at the beginning, flattening out at the end. There are two high amplitude swings in the {\it PA} on the order of at least $\sim100^{\circ}$. This period contains the minimum value of $P$ for all of the activity periods, 1.4 per cent at JD$_{245}=$ 6008. Due to the gaps in $P$ data, we observed few clear flares in $P$, but we compared local minima and estimated flaring time-scales that range from 8 to 83 days. We were able to confirm the existence of flares by noting the concurrent rising and falling trends in the $P$ light curve.

Similarly to activity period A, we observed flares that had an order of magnitude increases and decreases in $P$. The second Steward cycle in this period shows some of the highest amplitude $P$ flares over the time-frame, while the other emission bands show low amplitude flares. There is also a jump in the continuum $P$ level during this time period at approximately JD$_{245}=$ 4600. This can be explained by an increase in the Lorentz factor, since the gamma-rays base level changes, which indicates a change in the dominant emission process \citep{Ghisellini2001}. Due to the high amplitude flares of $P$, the polarized flux increases as a general trend. 

The continuum $P$ level is low, $<5$ per cent, until approximately JD$_{245} = 6400$, where it jumps to approximately $10$ per cent. The continuum 
level of {\it PA} rotates northward throughout the activity period. We note that the swing to northward rotation from the southward rotation observed at the 
end of activity period A must have occurred during the observational gap between periods. At the end of Steward cycle 5, the {\it PA} becomes 
roughly constant at approximately $\sim220^{\circ}$. The point where the {\it PA} becomes constant is coincident with the jump in the base level
of the $P$. This could be due to a strengthening magnetic field forcing a well ordered magnetic field topology.

\subsection{Activity Period C}
\label{APC}

\begin{figure}
\centering
\includegraphics[width=0.5\textwidth]{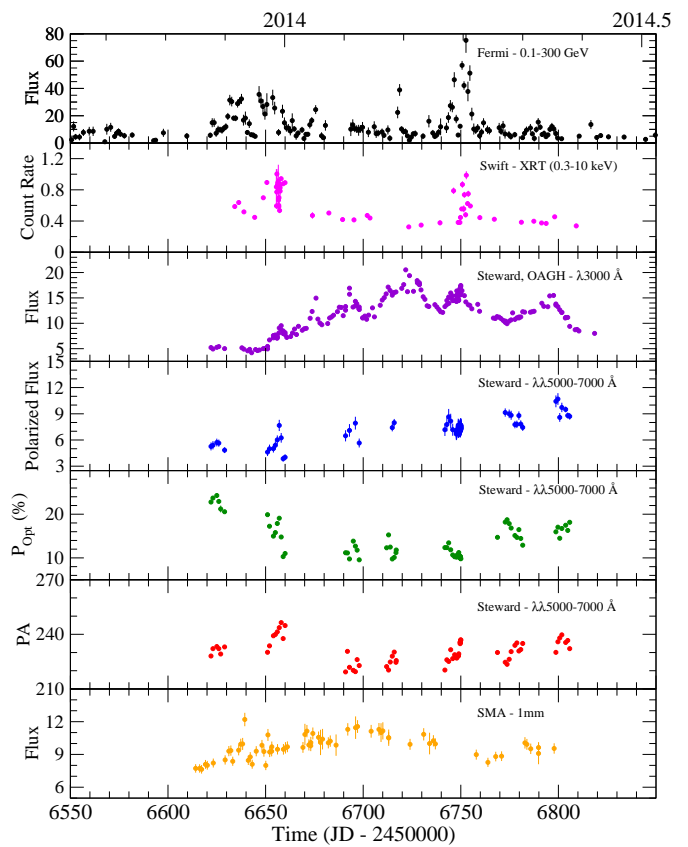}
\caption{Activity period C. We found a delay between gamma-rays and the UV continuum. This delay is different than the one found in Period A. The units 
are the same as in Fig.~ \ref{curves}.}
\label{delay-time}
\end{figure}

This period ranges from $\rm{JD}_{245}=6400-6850$ (see Fig.~\ref{delay-time}). During this time period, we observe the highest levels of gamma-ray 
emission in our time-frame of study. At the start of this time period we observe a very intense flare in the gamma-rays with a clear counterpart in 
the 1 mm emission, and high levels of degree of polarization, however, we do not see any response in the wavelength range from UV to NIR. This gamma-ray flare was first reported by \cite{Hayashida2015}, where via SED modelling, they conclude that it is mainly driven by External Inverse Compton \citep[EIC,][]{Sikora1994}. \citet{Dotson2015} report a similar behaviour for the source PKS 1510-089 during two gamma-ray 
flares in 2009. \cite{Dotson2015} suggest that the GeV variability during these flares is due to an increase of the external photon field.

After the gamma-ray flare in 3C 279 without a counterpart in UV, optical or NIR, we subsequently observed multiple flares in the UV, optical and NIR 
bands, with possible counterparts in the gamma-rays. The cross-correlation analysis between the UV continuum and the gamma-rays showed a delay of 
-67.8$\pm$5.0 days, with the gamma-rays leading the UV continuum. However, when we analysed only the time-frame after the intense flare at the start 
of this period, we obtained a delay of 28.6$\pm$4.8 days. 

The cross-correlation analysis also shows a delay between the X-rays and the NIR, optical V, and UV bands, at $\sim -64$ days; meaning the X-rays precede the other bands. The cross-correlation between the X-rays and gamma-rays for this period, shows a delay of 1.3$\pm$3.0 days (consistent with zero).

Several works that perform SED modelling of 3C 279 \citep[e.g.][]{Abdo2010b, Hayashida2012, Dermer2014, Aleksic2014a, Hayashida2015} point towards inverse Compton being the dominant source of swift XRT X-rays (0.3-10 keV). The question is: What is the dominant source of seed photons? Is it the jet \citep[SSC, e.g.][]{Bottcher2013}, or an external source \citep[EIC, e.g.][]{Pian1999,Bottacini2016}?

It is well established that to produce X-rays from 0.3 to 10 keV, via Inverse Compton, the more likely seed photons are at millimeter and radio wavelengths \citep[e.g.][]{Croston2005,Worrall2016}. On the other hand, it is expected that light curves from the seed photon frequency and the up-scattered frequency, to have similar behaviours; i.e. a correlation between the light curves. If SSC was the dominant component, we would expect a correlation between the 1 mm emission (dominated by jet synchrotron) and the X-rays, which we did not find. This reduces the possibility (but does not eliminate) that the dominant emission process for X-rays is SSC.

We also have to consider the fact that we find a correlation with a delay of $1.3 \pm 3.0$ days (consistent with no delay) between the X-rays and Gamma-rays light curves for Period C. Such correlation would only be expected in the case of both seed photon fields having a common origin (and therefore being correlated). Since part of the conclusion of this paper, as well as \cite{Hayashida2015}, is that the gamma-ray emission during Period C is dominated by External Inverse Compton, and we cannot help but point to a source external to the jet as the responsible for the main source of seed photons for X-ray production during Period C.

In $P$ we observe a dip over a time-scale of $\sim$183 days, with the minimum at the middle of the time period. The continuum level of $P$ starts at 
approximately the same level as the end of activity period B, then begins to increase after JD$_{245} = 6700$. The $P$ drops from 20 per cent to 
5 per cent during the dip which occurs over the first 30 days of the cycle, while the recovery occurs over approximately 150 days. In the observed 
emission bands, there is large amplitude flaring activity throughout the period. Thusly, even though $P$ dips at the beginning of the cycle and 
then recovers, the polarized flux increases over the activity period. 

The continuum level of the {\it PA} is nearly constant around a value of approximately $\sim230^{\circ}$. We observed two small amplitude swings in 
{\it PA} (on the order of $\sim$10 degrees), first from northward to southward movement, and then back to northward. The swing back to northward
occurs at the minimum $P$ value of the dip. The standard deviation from the mean value of the {\it PA} is 6.4 degrees, which implies these swings 
are deviations from a constant {\it PA} continuum. The polarized flux increases for the duration of the activity period; even though the $P$ dips 
at the same time the UV continuum flares, so that the rising UV flux coincides with falling $P$ and falling UV flux coincides with rising $P$. We will 
discuss this and multiwavelength behaviour correlations further in Section \ref{resultsanddiscussion}.

\subsection{Overall behaviour of the Continuum Level of the Polarimetric Quantities}
\label{PolLong}

Overall, the continuum level of $P$ remains lower than $5$ per cent, except for JD$_{245} = 5300-5500$ and from JD$_{245} = 6400$ until the end of the 
observational time-frame, where it is approximately $10$ per cent. The continuum level of the {\it PA} is characterized by four swings. There is a 
predominantly northward rotation throughout the time period, with the southward rotations lasting 400 and 100 days. The constant trend at the end of the 
observational time-frame is approximately 400 days long. Two of the swings in {\it PA} occur simultaneously with flares in $P$, while the other inferred two 
swings occur during observational gaps. The beginning of the constant trend in the continuum level of the {\it PA} coincides with the jump in the $P$ 
continuum at JD$_{245} = 6400$.

\subsection{Polarization Behaviour on Short Time-Scales}
\label{PolBehav}

Micro-variability occurs on the short time-scale bins of around 10 days that make up each Steward observation cycle. We analyze the micro-variability 
over the entire data set, as there are no observed trends in behaviour that are more common or rare in any specific activity period. There is a wide range 
of behaviour and we do not observe consistently similar correlated behaviour between $P$ and {\it PA}. We observe variability in $P$ on the order of a 
few days with changes in $P$ from 3 to 10 per cent. We also observed {\it PA} swings lasting a few days with changes that ranged from $\sim$3 to 10 
degrees.

We observe that micro-variability is present during our entire time-frame, both in the active and inactive states of UV continuum and other emission 
bands. It seems that $P$ and {\it PA} behave chaotically or randomly on this timescale. This leads us to speculate that stochastic processes such as 
turbulence could be responsible for the small amplitude variability \citep{Jones1985, Marscher2014, Kiehlmann2016}.

\subsection{Gamma-Ray Spectral Index}

\begin{figure}
\centering
\includegraphics[width=0.5\textwidth]{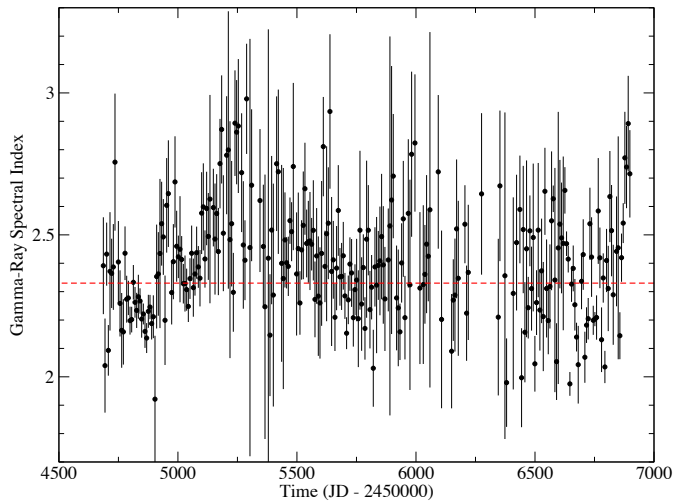}
\caption{Gamma-ray spectral index over the entire observation period.}
\label{GSI}
\end{figure}

We created a plot of the behaviour of the gamma-ray spectral index as a function of time over the entire time-frame of study (see Fig.~\ref{GSI}). We use 
time bins of 7 days; for each time bin, we fit 3C 279 spectra as a power-law of the form

\begin{equation}
\frac{dN}{dE}=N_0\Bigg( \frac{E}{E_0} \Bigg)^{-\gamma}
\end{equation}

where $N_0$ is the prefactor, $\gamma$ is the gamma-ray spectral index, and $E_0$ is the energy scale. 

We test the variability of the gamma-ray spectral index by fitting a constant value to the calculated spectral indices over our entire observational period. 
Under the assumption that the gamma-ray spectral index can be represented by a constant value, the best fit shows a chi-square value of $\chi^2=646$. Taking into 
account the degrees of freedom of the fit (N$_{\rm{dof}}=277$), the probability of getting this $\chi^2$ value by chance (estimated with IDL routine 
MPCHITEST \citep{Markwardt2009}) is $\rm{P}<1\times10^{-31}$. Such a small probability corresponds to a variability signal of over 11-sigma, thus we 
rejected the hypothesis that the gamma-ray spectral index did not vary over the last six years.

The variability of the gamma-ray spectral index implies changes of the energy distribution of the electron population responsible for the Inverse Compton 
effect. The softening of the spectrum may imply either cooling of the pre-existing electron population (by synchrotron emission); or the injection of a less energetic electron population to the pre-existing one. The hardening of the spectrum implies an injection of an energetic electron population (i.e. more energetic than the pre-existing one). This result is important in the context of a variable SED since the synchrotron spectrum of the source will also be affected.

\begin{figure}
\centering
\includegraphics[width=0.5\textwidth]{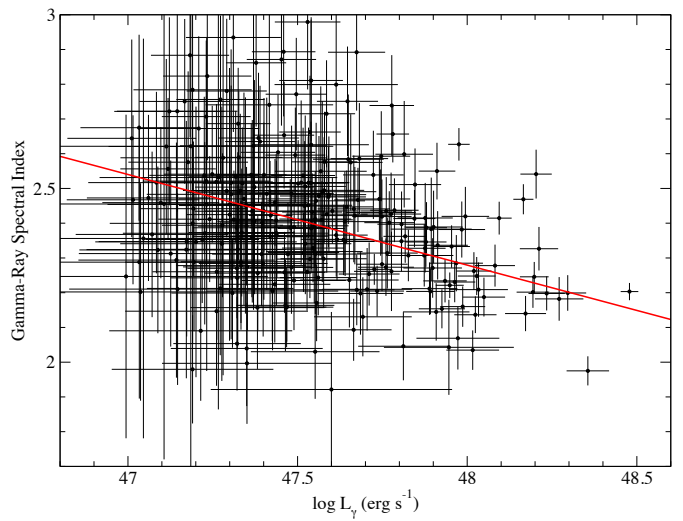}
\caption{Gamma-ray luminosity vs. the gamma-ray power-law spectral index. The red line represents the result of the linear fitting performed using the IDL task FITEXY.}
\label{Lum-Index}
\end{figure}

Using a sample of 451 blazars, \cite{Fan2012} showed that there is an anti-correlation between the gamma-ray luminosity and spectral index averaged over two years.
Because of this, we decided to explore the variability behaviour of 3C 279 in the gamma-ray spectral index vs. gamma-ray luminosity plane. 
We calculate the gamma-ray luminosity using equations 1 and 2 from \citet{Ghisellini2009}. 
By applying the Spearman Rank Correlation test we found that there is a significant anti-correlation with a correlation coefficient of -0.28, and a probability 
of getting this result by chance of $\sim10^{-6}$. We also used the IDL routine FITEXY to confirm the anti-correlation. We found a statistically significant 
linear anti-correlation (see Fig.~\ref{Lum-Index}), with a probability of obtaining the $\chi^2$ value by chance $<< 0.01$. This confirms the result from the Spearman Rank 
Correlation test.

\section{Results and Discussion}
\label{resultsanddiscussion}

We analyze the multiwavelength variability of 3C 279 over a time-frame of six years, using light curves of multiple emission bands from 1 mm up to gamma-rays. We summarize our results as follows.

\begin{enumerate}

\item We identify the simultaneity of the UV $\lambda3000$ \AA~continuum emission, the optical {\it V}-band, and the NIR {\it J}-, {\it H}- and 
{\it K}-bands, using three cross-correlation analysis methods as described in Section~\ref{CrossAna}. This correlation allowed us to speculate that
the emission from the middle UV range to the NIR are emitted from the same region. We propose that this emission region is the jet itself or a 
moving feature along the jet via synchrotron emission. The variable and high $P$ found in the observations discards the accretion disk as the main emission source (big blue bump) and further supports the hypothesis of these bands being dominated by synchrotron emission.

\item We find a significant correlation between the {\it V}-band and the 1 mm emission for the entire time range of the observations used in this paper, 
with a Spearman correlation coefficient of 0.65, and a probability of obtaining this correlation by chance of P$<<0.01$. This indicates that the optical {\it V}-band emission should be dominated by non-thermal emission from the jet. The correlation found between the UV continuum, {\it V}-band, and NIR bands, 
strongly suggests that the emission from these bands is dominated by non-thermal emission. We conclude that the jet dominates the emission 
at these wavelengths.

\item The cross-correlation analysis on the gamma-rays and the UV continuum (therefore, all the way to the NIR) for period A shows a delay of 0.7$\pm
$5.0 days, which is consistent with no delay. Similarly to the lower energy shape of the SED, the high energy SED shape of a blazar is usually described 
as a downward facing parabola, produced mainly by Inverse Compton scattering (IC) \citep{Bloom1996}. The wavelength range of the high energy SED 
component is typically from the extreme ultraviolet to TeV energies, however, this is very dependent on the model of the high-energy component. The 
Synchrotron Self-Compton (SSC, \citealp{Bloom1996}) and External inverse Compton (EIC, \citealp{Sikora1994}) models are both able to reproduce the 
SED of blazar-type objects; therefore it is likely that both models are potentially viable. The SED of 3C 279 has been modeled using both EIC models and 
SSC models together to explain the behaviour of different parts of the SED \citep{Hayashida2012}. Due to the correlation, we find between the {\it V}-band and 1 mm emission, we attribute the UV continuum mainly to emission from the jet. Since our cross-correlation analysis is consistent with no 
delay between the gamma-ray emission and the UV, we postulate that they are co-spatial. A correlation between jet emission and gamma-ray 
emission supports the hypothesis that the primary high-energy component is Inverse Compton emission arising from seed photons in the jet. 
Therefore, the SSC model better describes our data. It is worth clarifying that we are not concluding that the gamma-ray emission can be explained solely by SSC, as it is likely that we have a contribution from both, SSC and EIC; our results only show that the dominant contribution to the gamma-ray emission is the SSC component.

It is important to take into account that even when there are EIC models that show significant correlations and short delays between the optical and gamma-ray bands, these models have the optical total flux dominated by a source external to the jet (not just the seed photons being dominated by the external source, there is a difference between the two scenarios). The rise of polarised flux during this period indicates that the optical total flux is dominated by synchrotron emission. In the cases where optical total flux is dominated by the same source of seed photons for gamma-rays, a correlation is of course expected. The short delays usually suggest that the gamma-ray emission region is in the vicinity of the black hole (e.g. inside the BLR, with a typical radius of $10^{15}$ - $10^{17}$ cm, consistent with the models where the shortest delays are obtained). In this case, the short delay makes total sense.

With this in mind, we make an estimation of the shortest expected delay for EIC in 3C 279, as the light travel time for the BLR size. \cite{Bottcher2016} calculates the radius of the BLR for 3C 279 as $2.3\times10^{17}$ cm, which translates to 88.8 light days. Taking into account the cosmological time dilation we have an expected time delay of 136.4 days. Even assuming that the calculation of \cite{Bottcher2016} overestimates the radius by one order of magnitude, this still leaves us with a minimum delay for EIC that is too large to be compatible with the observed delay between gamma-rays and UV/Optical/NIR, discarding the possibility of EIC being the dominant component of gamma-ray emission for Period A.

\item We report period B as an anomalous activity period in 3C 279 observed from JD$_{245}\sim$ 5850 to 6400. In this period we observe multiple flares in the $\lambda3000$ \AA~continuum with counterparts in the optical {\it V}- and NIR bands, as well as the highest flux levels of 1 mm emission during our entire 
time-range. However, there is no counterpart in the gamma-ray emission to any of these flares. This was previously reported by 
\citet{Patino-Alvarez2014}. This kind of behaviour has been reported only for one other source. \citet{Chatterjee2013} reported an anomalous flare for the 
source PKS 0208-512, which was proposed to be caused by a change in the magnetic field in the emitting region without any change in doppler factor or the number of emitting electrons.

A possible explanation for this behaviour is a change in opacity in the jet to gamma-ray emission. An ejection of a component into the jet is one of the ways flares are created, this could increase the bulk Lorentz factor of the gamma-ray emission region. The increase in the bulk Lorentz 
factor of the jet leads to an increment in the interaction cross-section of the electron-positron pair production in the gamma-ray emitting region. This would lead to an increased rate of gamma-ray absorption that would naturally lead to a lack of observed gamma-ray emission. We consider this to be the most likely scenario given the evidence that many flares in AGN with relativistic jets have been linked with ejected components into the jet 
\citep{Arshakian2010,Leon-Tavares2010,Leon-Tavares2013}. In Appendix \ref{app:cross_sections}, we present theoretical calculations of the cross-sections for the inverse Compton scattering, and for the electron-positron pair production; and how the cross-sections change with increasing Lorentz factor. In Appendix~\ref{comparison}, we mention a few works whose conclusions closely relate to the conclusion of this work regarding the absorption of gamma-rays via electron-positron pair production \citep{Protheroe1986,Mastichiadis1991,Zdziarski1993,Petry2000}.

Another possibility is the absorption of gamma-rays by the broad line region. There are multiple works that, via modelling, show that high energy photons surrounded by the intense radiation field of the BLR are prone to $\gamma\gamma$ absorption \citep[e.g.,][and references therein]{Donea2003, Reimer2007, Liu2008, Sitarek2008, Bottcher2009b, Tavecchio2009, Bottcher2016, Abolmasov2017}. Despite the difference in physical parameters and assumptions used in the works mentioned before, the conclusion is nonetheless the same, the BLR is an efficient $\gamma\gamma$ absorber. This has also been considered by several studies as a counter-argument for the gamma-ray emission zone being located inside the BLR in FSRQ \citep[e.g.,][]{Tavecchio2011}.

\item At the start of activity period C, from JD$_{245}\sim$ 6600 to 6650, we find a flare in the gamma-rays, as well as in the 1 mm and $P$; however, we 
do not find any counterpart in the spectral range from UV to NIR. \cite{Hayashida2015} described the behaviour of the gamma-ray flares occurring in this time period and modelled the SED of 3C 279 during the different flares with a leptonic model (Synchrotron + SSC + EIC). It is clear from the SED models that the EIC dominates the emission over the SSC at the energies of Fermi-LAT. Such behaviour has been reported before for the source PKS 1510-089 \citep{Dotson2015}, where they suggest that the GeV variability is due to an increase of the external photon field.

In the time range from JD$_{245}\sim$ 6700 to 6850, the cross-correlation analysis between the UV continuum and the gamma-rays shows a delay of 28.6$\pm$4.8 days. The finding of this delay implies that the dominant seed photon source for gamma-rays and the dominant source of UV continuum (synchrotron from the jet), are not co-spatial, as was the case for Period A. This strongly suggests that the dominant mechanism of gamma-ray emission during this activity period is EIC. We also found a delay consistent with zero between the X-rays and Gamma-rays, which as explained in Section~\ref{APC}, also implies a dominance of an EIC process as the gamma-ray emission mechanism. Both of our results match the conclusions reached by \cite{Hayashida2015}.

Furthermore, the conclusion that the gamma-rays are dominated by EIC during this period is the natural progression of our explanation of what happens in Period B. There, the increase in the bulk Lorentz Factor of the jet would also increase the cooling of the electrons in the jet (the large flares observed in other bands during Period B). As a result, this allows external sources of seed photons to dominate the interaction for the Inverse Compton emission.

\citet{Ramakrishnan2016} studied multiwavelength light curves of 3C 279 between 2012.5 and late 2015. They report no significant correlation 
($\geq3\sigma$) between the gamma-rays and the optical R-band, as well as the R-band and radio at 37 and 95 GHz. This apparent discrepancy 
between their results and ours can be reconciled by considering the different time frames used in the two studies: \citet{Ramakrishnan2016} used a time 
consisting of parts of periods B and C. Cross-Correlating time periods where the dominant emission mechanism changes, leads to a diminished 
significance of the correlation coefficient.

\item In general, we can see from the data that polarimetric variability is common in the active and inactive states of the multiwavelength emission of 3C 
279. We do not observe any trends of $P$ variability amplitude or timescale in any particular activity state or time-frame. It seems that during some time 
periods, $P$ behaves chaotically or randomly, and this is seen during the quiescent periods of emission in the other bands.

We present two main results from our polarimetric observations. Firstly, the behaviour of activity period C is of special interest and secondly, a general 
result that different timescales of variability can distinguish stochastic versus deterministic variability behaviour.

The behaviour of the $PA$ and $P$ during Period C is unique for the observational time-frame. $P$ shows a dip at the start of this period, followed by a slow steady increase. The $PA$ is roughly constant during the entire time period, which implies that the magnetic field topology is unchanging during this high level of flaring activity in the other bands (at least in the optical emitting region). Given the gamma-ray activity observed during this period, which is often associated with magnetic field topology variability \citep{Marscher2010}, this behaviour indicates that the source of seed photons for gamma-rays is external to the jet, as concluded previously using other observational evidence. The changes in polarization degree along with the flaring in the NIR-UV can be explained by transverse shocks in the jet, with a well ordered magnetic field, that does not change the magnetic field topology. Rather, energy is injected by the acceleration of particles across the shocks \citep{Marscher1985,Lyutikov2004,Nalewajko2012}.

A chaotic or random 
behaviour is most easily explained by invoking a highly variable magnetic field topology and strength in different regions of the jet, most likely from 
turbulence and instabilities \citep{Jones1985, Ferrari1998, Marscher2014, Kiehlmann2016} or moving components \citep{Blinov2016}. Also, during high 
activity periods where non-thermal emission is involved, the 'active' or deterministic $P$ behaviour is superimposed on a 'quiescent' behaviour which is 
predominantly stochastic \citep[e.g.][]{Marscher2014,Kiehlmann2016}. The data show this effect during flaring events where the behaviour in $P$ and {\it 
PA} is more pronounced, indicating the dominance of a deterministic component which we deem to be directly related to synchrotron emission. While there is little evidence for different time-scales of variability, \cite{Kiehlmann2016} also report that the changes in the direction of the polarization are due to deterministic processes. They find that a smooth 360$^{\circ}$ swing observed in the data, is not consistent with stochastic processes; leading to the conclusion that deterministic processes are mainly responsible for this behaviour.


\item We test the gamma-ray power-law spectral index for variability by fitting a constant value. The result of the fit shows a $\chi^2=646.0$ with 
N$_{\rm{dof}}=277$, the probability of getting this $\chi^2$ value by chance is $\rm{P}\sim1\times10^{-31}$, which means that the gamma-ray power-law 
spectral index in 3C 279 has been variable with a significance of over 11-$\sigma$ over the past six years. We believe our results conclusively show that the 
gamma-ray power-law spectral index has varied over the studied time-frame. \citet{Fan2012} previously reported that spectral index variability is 
associated with gamma-ray flux variability. Our data supports this, as the source was very variable in gamma-ray flux over the time-frame of the study. 
\citet{Vercellone2010} found a trend which suggests that a harder gamma-ray SED is observed when an object is brighter, from a study of the source 3C~454.3, using AGILE data. \citet{Fan2012} reported as well that the spectral index flattened (i.e. a harder spectrum) with higher gamma-ray luminosity 
for a sample of 451 blazars, using the average of the first two years of Fermi observations found in the 2FGL Catalog \citep{Nolan2012}. 
The trend we find between the gamma-ray spectral index and gamma-ray flux variability over the time-frame of study matches the ones 
mentioned above (see Fig.~\ref{Lum-Index}). The physical interpretation of spectral index changes is related to the cooling dynamics 
of emitting particles and the accretion regimes of the object \citep{Abdo2010a,Ghisellini2009}. However, \citet{Abdo2010a} reported 
that the spectral index does not generally change with gamma-ray flux variability for a large sample of {\it Fermi}-LAT sources. They also report a weak trend in some sources, where the spectral index decreases when the source is brighter. They mention that their 
observation of nearly constant spectral index does not exclude fine variability over short periods of time (such as during a flare). 
We have shown that the gamma-ray spectral index of 3C 279 not only varies during our observational time-frame (see Fig. \ref{GSI}), 
including periods of flaring activity, but also that there is a statistically significant harder-when-brighter trend (see Fig. \ref{Lum-Index}).

\end{enumerate}


\section{Conclusions}

In this paper, we also discuss the observed simultaneity of the continuum emission from UV to NIR. This implies that the emission from these bands is co-spatial during the entire time-frame of our study. We also discuss our conclusion that this continuum is produced by synchrotron emission. The main 
supporting evidence of this is the dominance of a single component over a wide wavelength range. Synchrotron emission is one process capable of such 
behaviour and is the most probable of the emission processes that are inherent to AGN. The non-thermal origin of the emission implies the jet is the 
source of the emission. It follows that synchrotron emission dominates the part of the spectrum that we conclude is co-spatial. 

The flaring we report in activity period A ($\rm{JD}_{245}=4650-5850$) in the gamma-ray band is dominated by Synchrotron Self-Compton emission. This is supported by finding no delay between the gamma-rays and UV continuum. The lack of gamma-ray flaring variability during activity period B ($\rm{JD}_{245}=5850-6400$) cannot be explained by a decrease of jet activity. This is supported by the flaring behaviour observed in UV, optical, NIR and 1 mm bands. We propose the mechanism that causes gamma-ray absorption is electron-positron pair production, arising from an increase in the bulk Lorentz factor of the jet in the emitting region. The gamma-ray flaring in activity period C ($\rm{JD}_{245}=6400-6850$) is dominated by the External Inverse Compton mechanism. This is supported by our finding of a non-zero delay between the gamma-rays and UV continuum, which is an outcome expected from the EIC process.

Finally, we proved the existence of significant variability of the gamma-ray spectral index over the entire time-frame of observations.

Our analysis suggests that the dominant emission mechanism for the gamma-rays changes with time, however, there is also the possibility that the gamma-ray emission zone 
changes with time. During activity period A the synchrotron emission is simultaneous with the gamma-ray emission, which pinpoints the location of the 
gamma-ray emission zone in the jet, near the electron acceleration zone. The data we have over activity period B does not allow us to make such a 
distinction. In Period C we find a delay between optical/UV emission and the gamma-rays, where the optical/UV emission is mainly produced by synchrotron emission. This 
suggests a possible change of the location of the gamma-ray emission zone, with respect to activity period A. It is a possibility that the gamma-rays are 
being emitted closer to the black hole, while illuminated by seed photons from an external source (e.g. the Broad Line Region).

A further study that will help clarify the source of seed photons is to characterize the variability of the flux and profile of the Mg$\,${\sc ii} $\lambda
$2798 \AA~emission line. \citet{Punsly2013} reported a change in the flux and in the profile asymmetry for 3C 279, using three spectra. In a 
forthcoming paper, we will present a variability analysis of the Mg\,{\sc ii} $\lambda2798$ \AA~emission line using approximately 270 spectra 
observed at the Steward and Guillermo Haro Observatories.


\section*{Acknowledgments}

We thank the anonymous referee for their positive, constructive and helpful comments. We are thankful to P. Smith for his help on the polarimetry analysis. This work was supported by CONACyT research grants 151494 and 280789 (M\'exico). V. P.-A. 
acknowledges support from the CONACyT program for Ph.D. studies. V.P.-A. and V.C. are grateful to UTSA for their hospitality during their stay. S.F. 
acknowledges support from the University of Texas at San Antonio (UTSA) and the Vaughan family, support from the NSF grant 0904421, as well as the 
UTSA Mexico Center Research Fellowship funded by the Carlos and Malu Alvarez Fund. This work has received computational support from 
Computational System Biology Core, funded by the National Institute on Minority Health and Health Disparities (G12MD007591) from the National 
Institutes of Health. Data from the Steward Observatory spectro-polarimetric monitoring project were used; this program is supported by Fermi Guest 
Investigator grants NNX08AW56G, NNX09AU10G, and NNX12AO93G. 1 mm flux density light curve data from the Submillimeter Array was provided by 
Mark A. Gurwell. The Submillimeter Array is a joint project between the Smithsonian Astrophysical Observatory and the Academia Sinica Institute of 
Astronomy and Astrophysics and is funded by the Smithsonian Institution and the Academia Sinica.

\bibliographystyle{mnras} 

\bibliography{3C279_refs} 

\clearpage


\onecolumn

\appendix

\section{Cross-Correlation Figures}
\label{ccfigures}
\clearpage

\begin{figure}
\begin{center}
\includegraphics[width=0.48\textwidth]{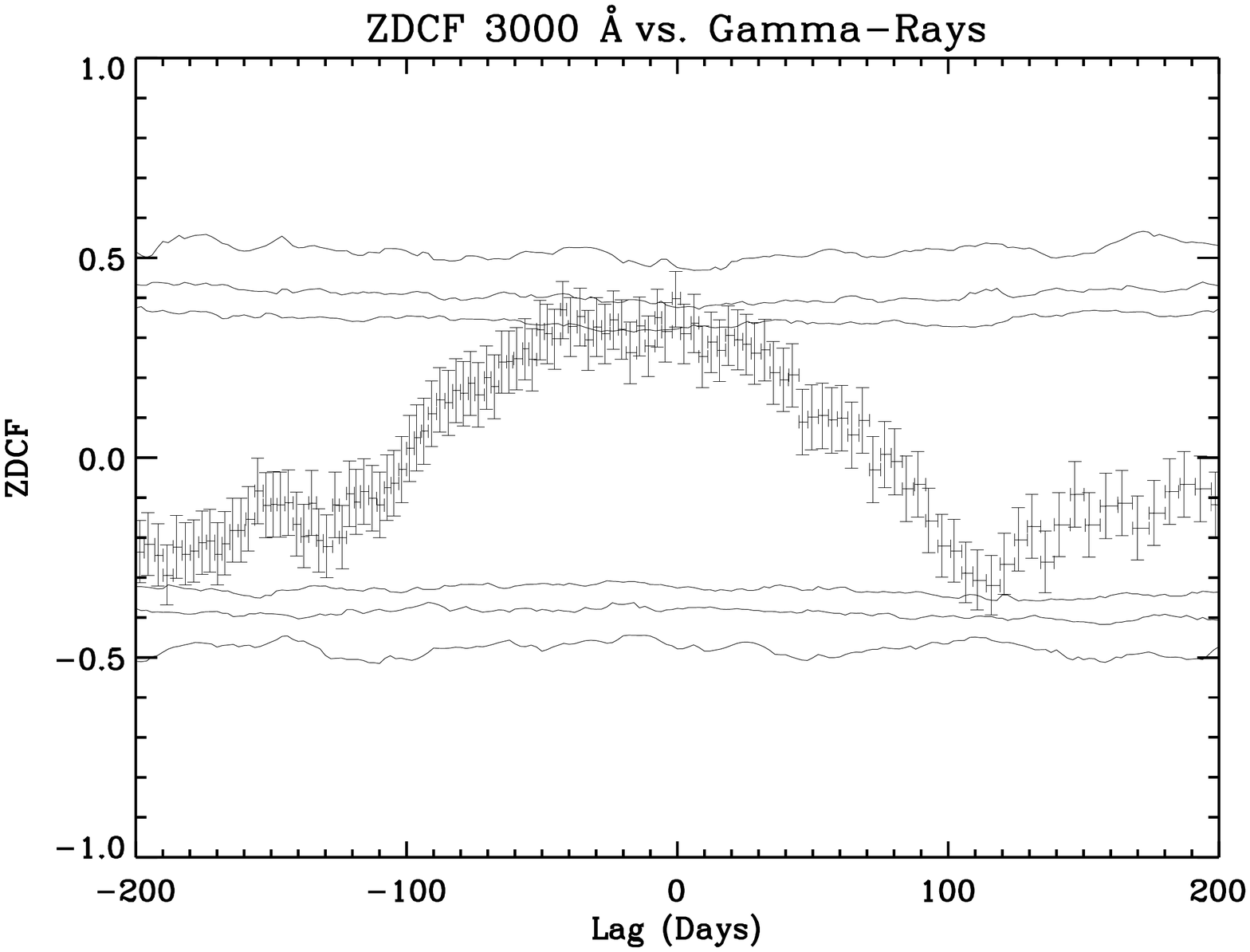}
\includegraphics[width=0.48\textwidth]{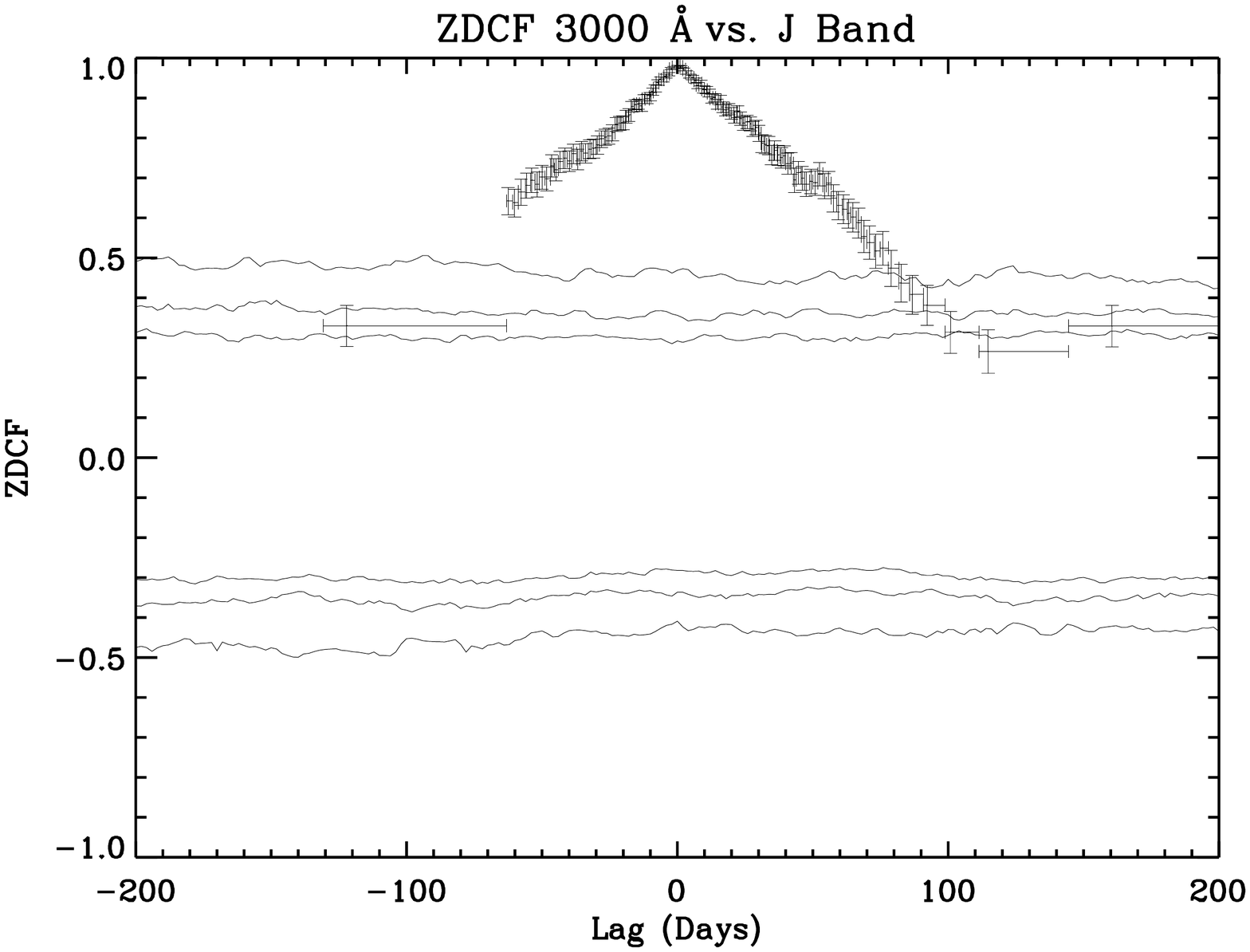}
\includegraphics[width=0.48\textwidth]{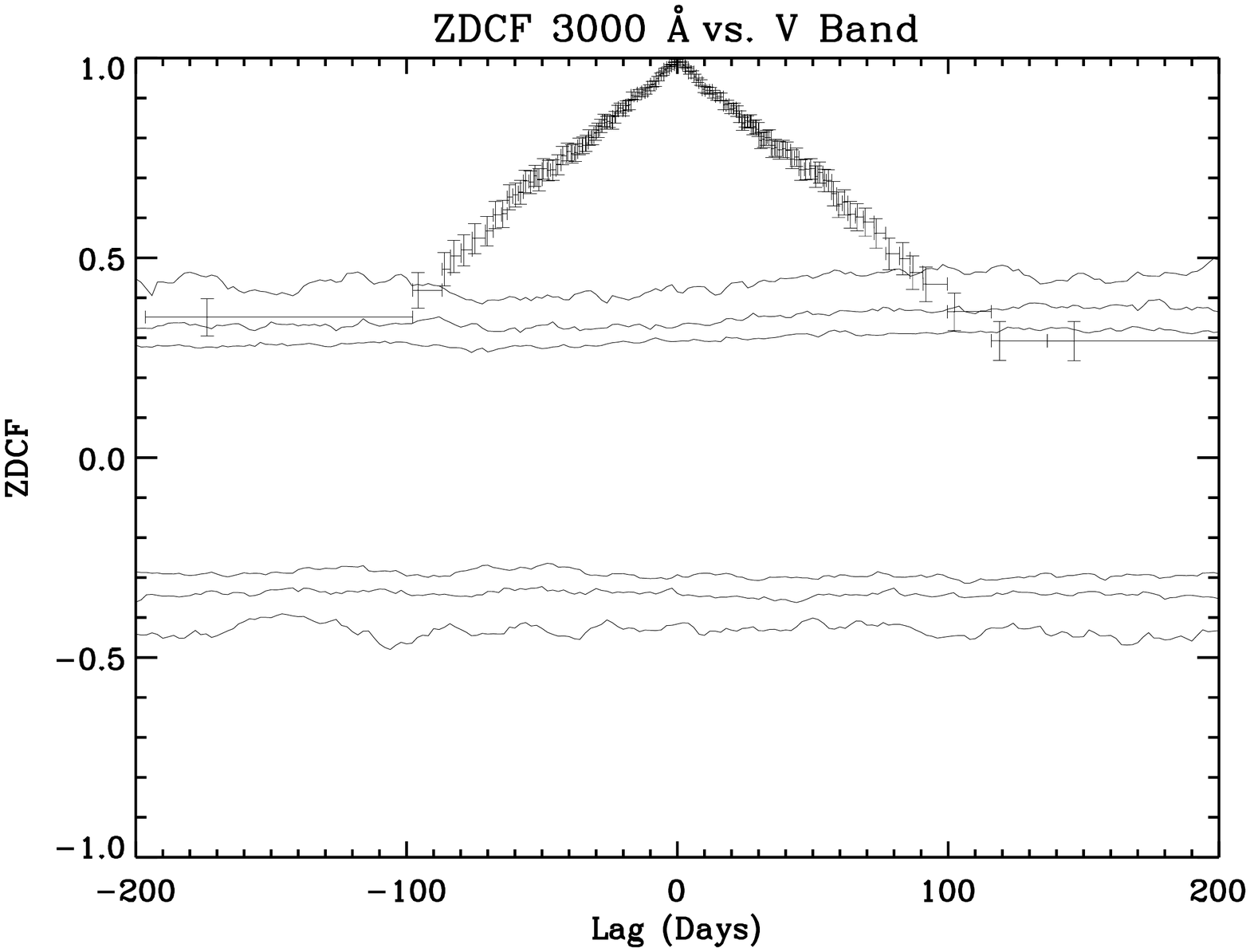}
\includegraphics[width=0.48\textwidth]{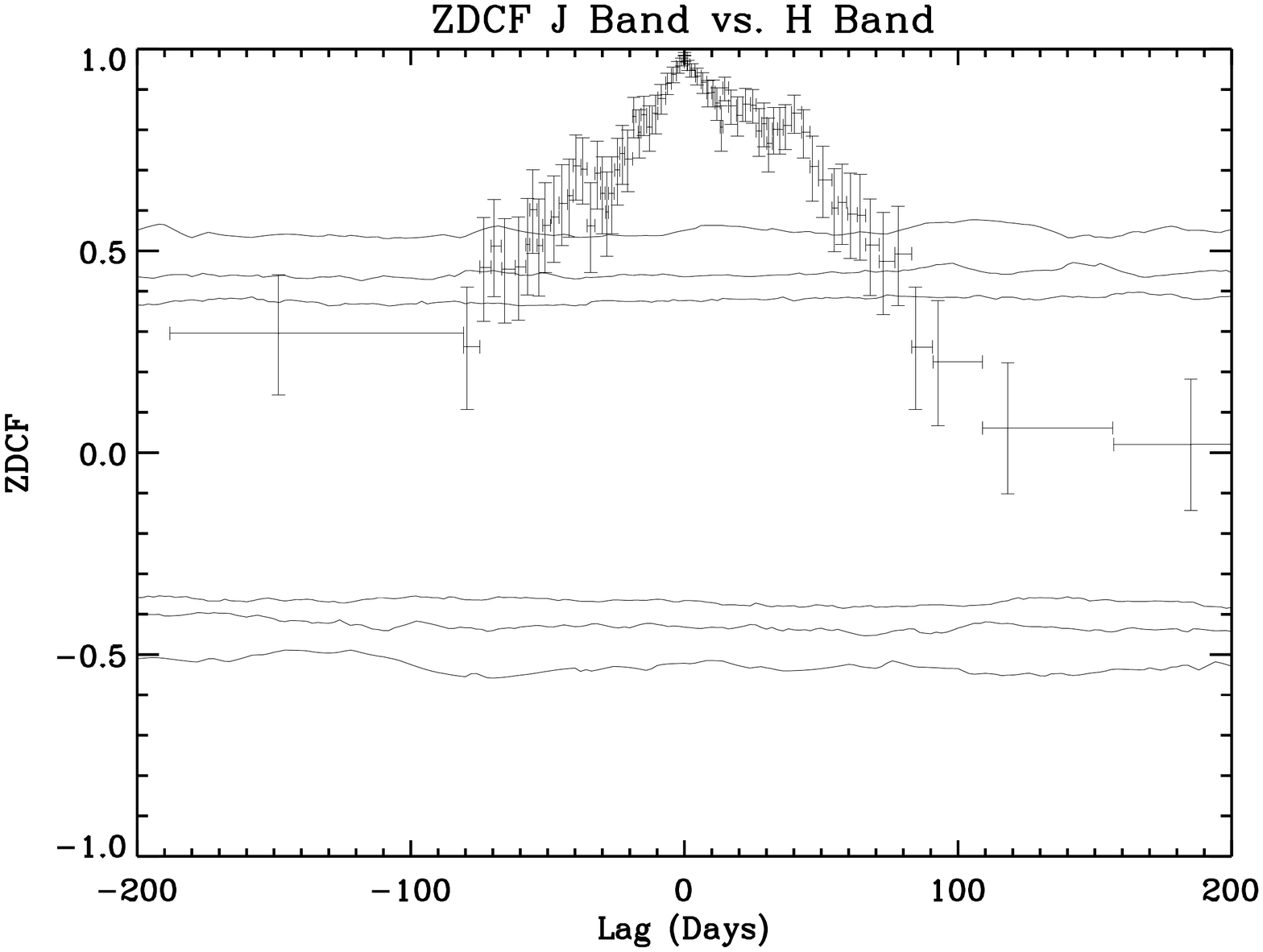}
\includegraphics[width=0.48\textwidth]{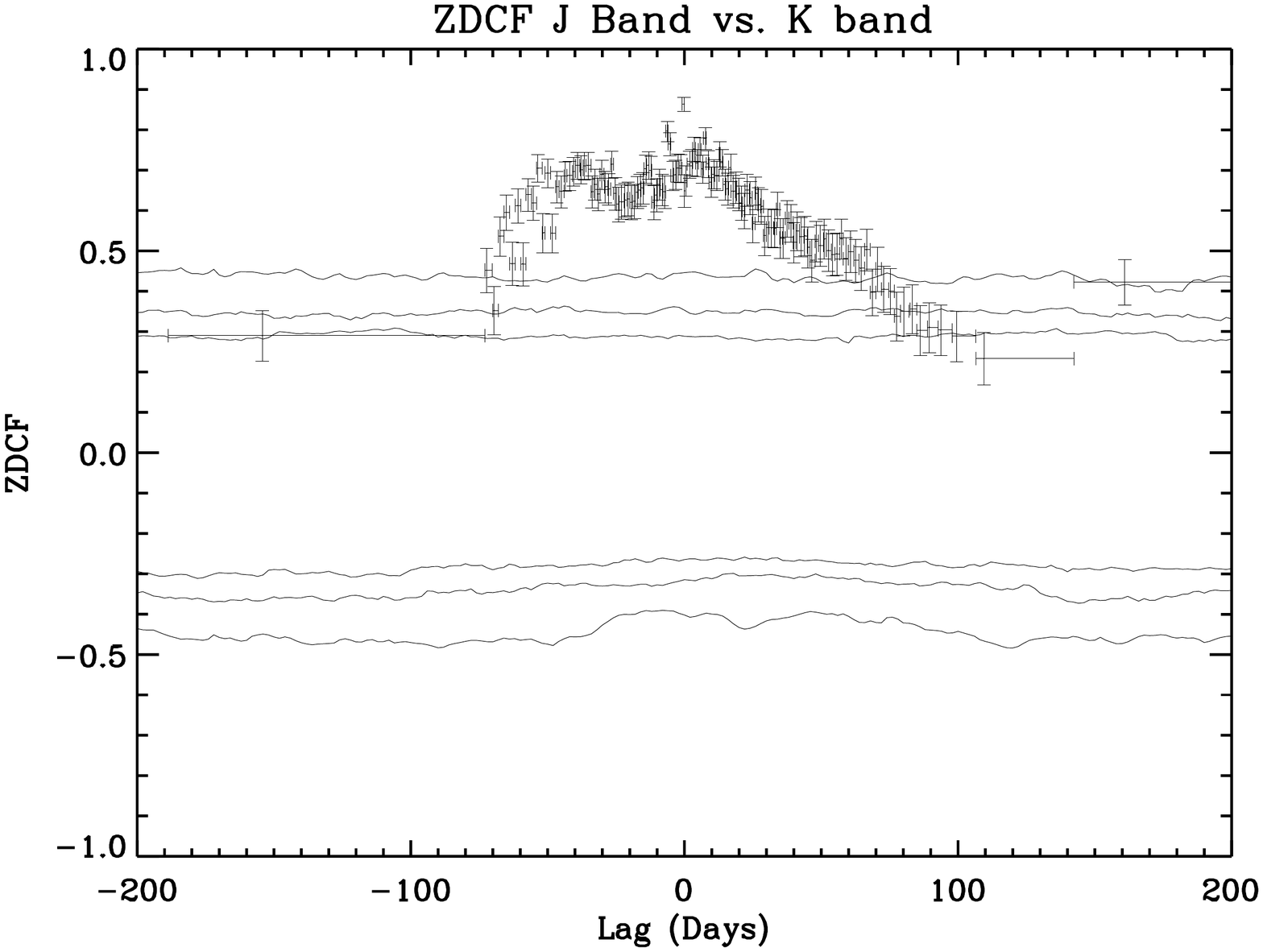}
\includegraphics[width=0.48\textwidth]{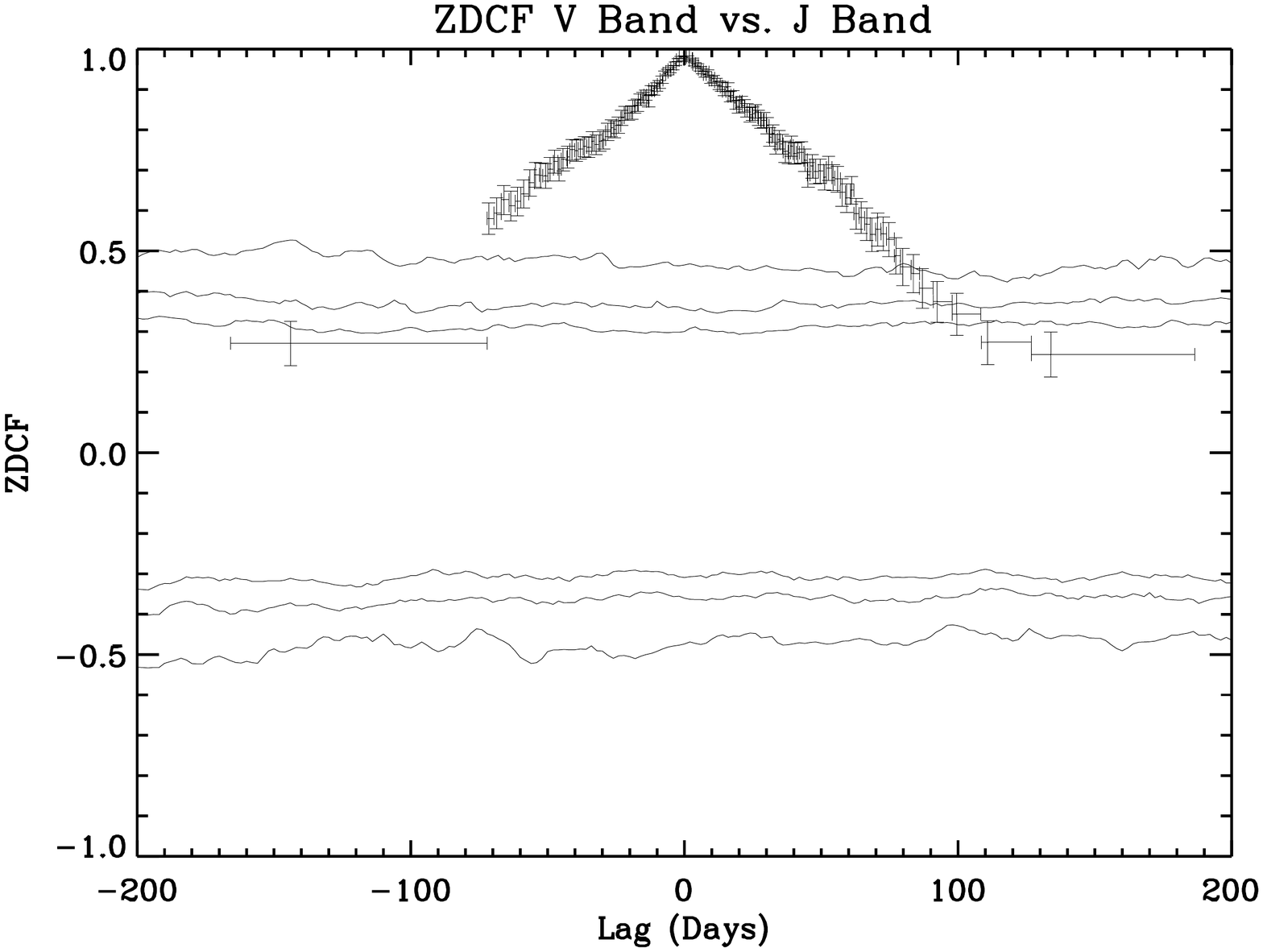}
\end{center}
\caption{Cross-correlation functions obtained for the full time-frame of study presented in this work. Bands involved and method used are shown in each 
panel. Confidence levels at 90\% , 95\% and 99\% are drawn.}
\label{CC_Full}
\end{figure}

\begin{figure}
\begin{center}
\includegraphics[width=0.48\textwidth]{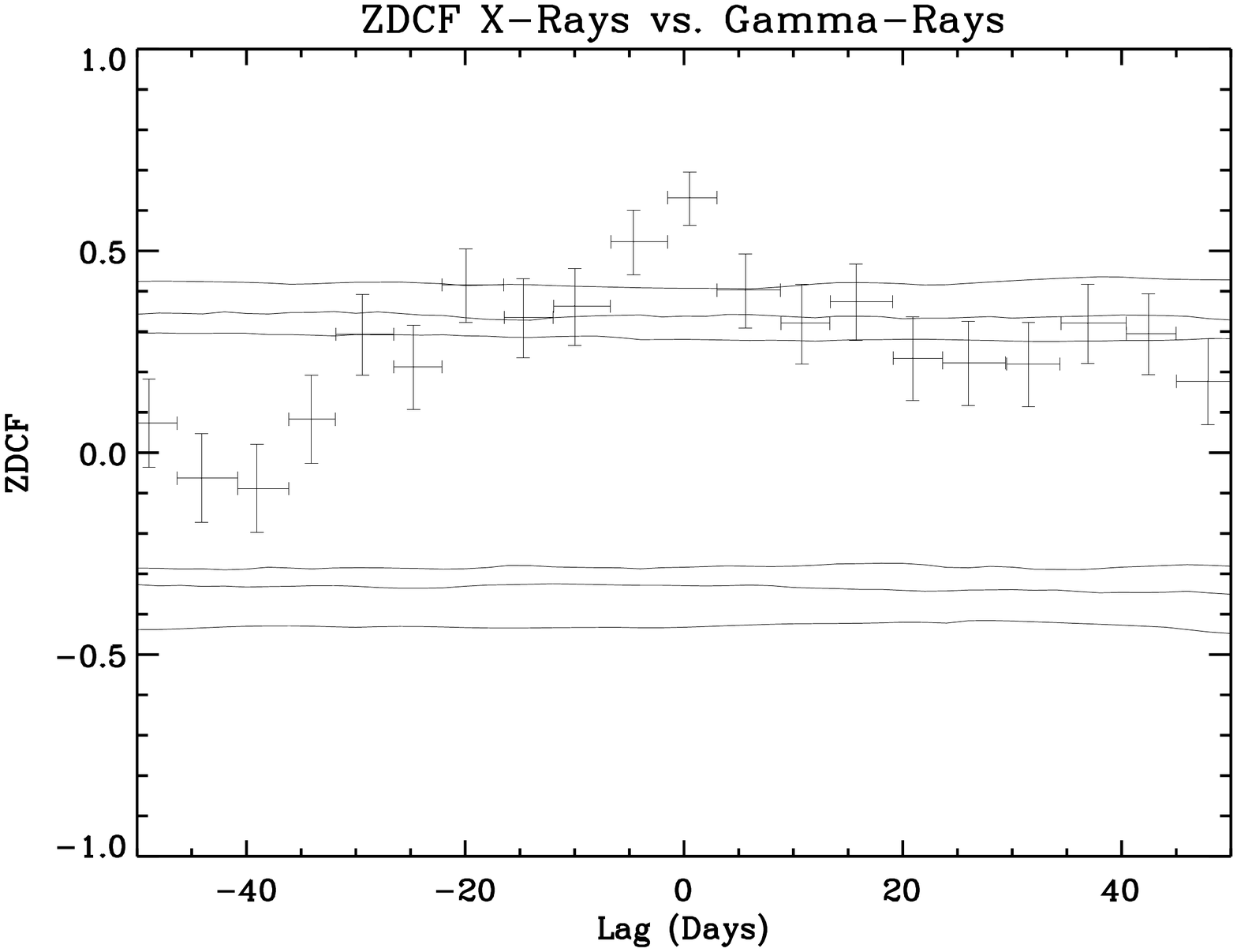}
\includegraphics[width=0.48\textwidth]{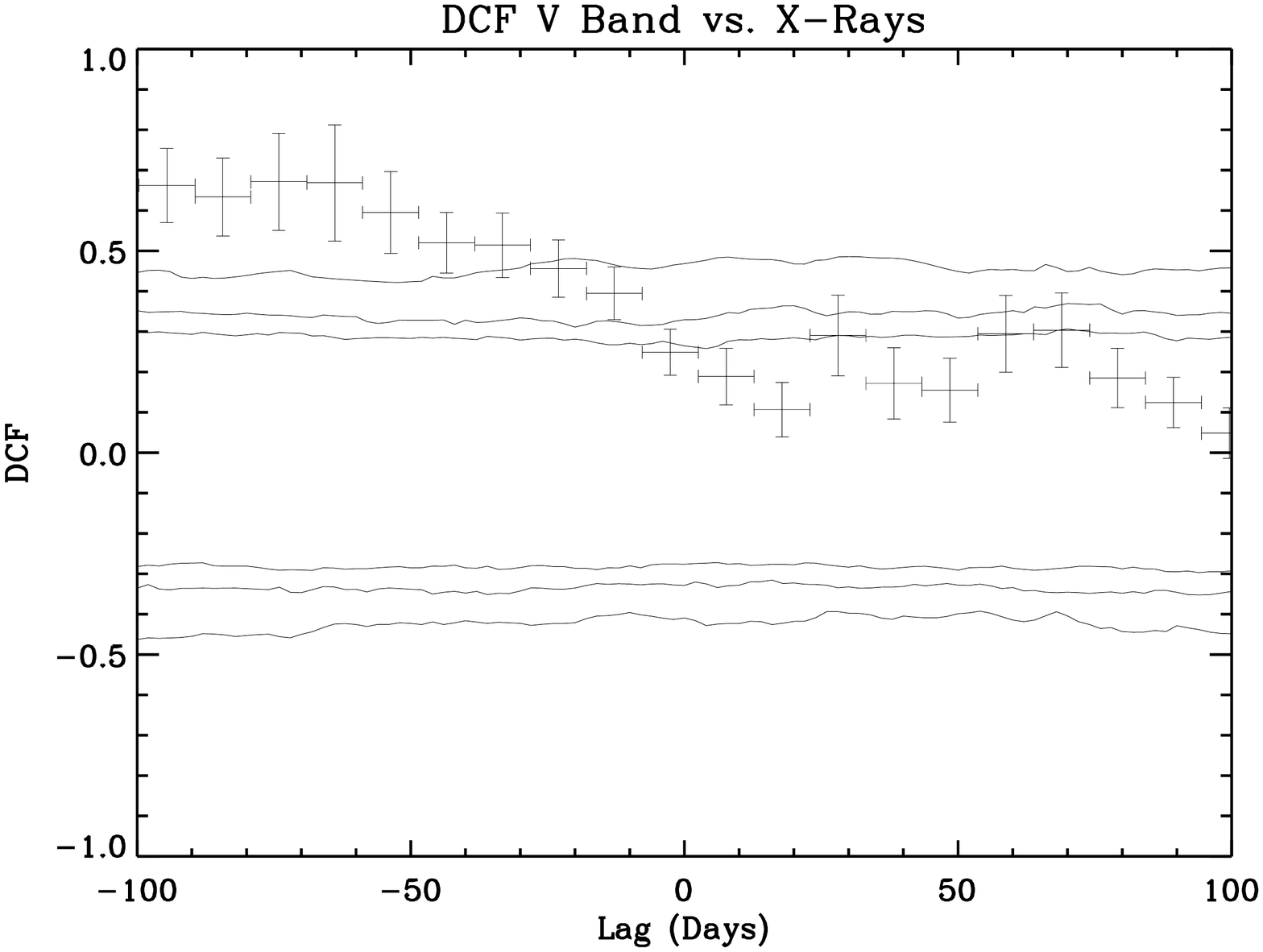}
\includegraphics[width=0.48\textwidth]{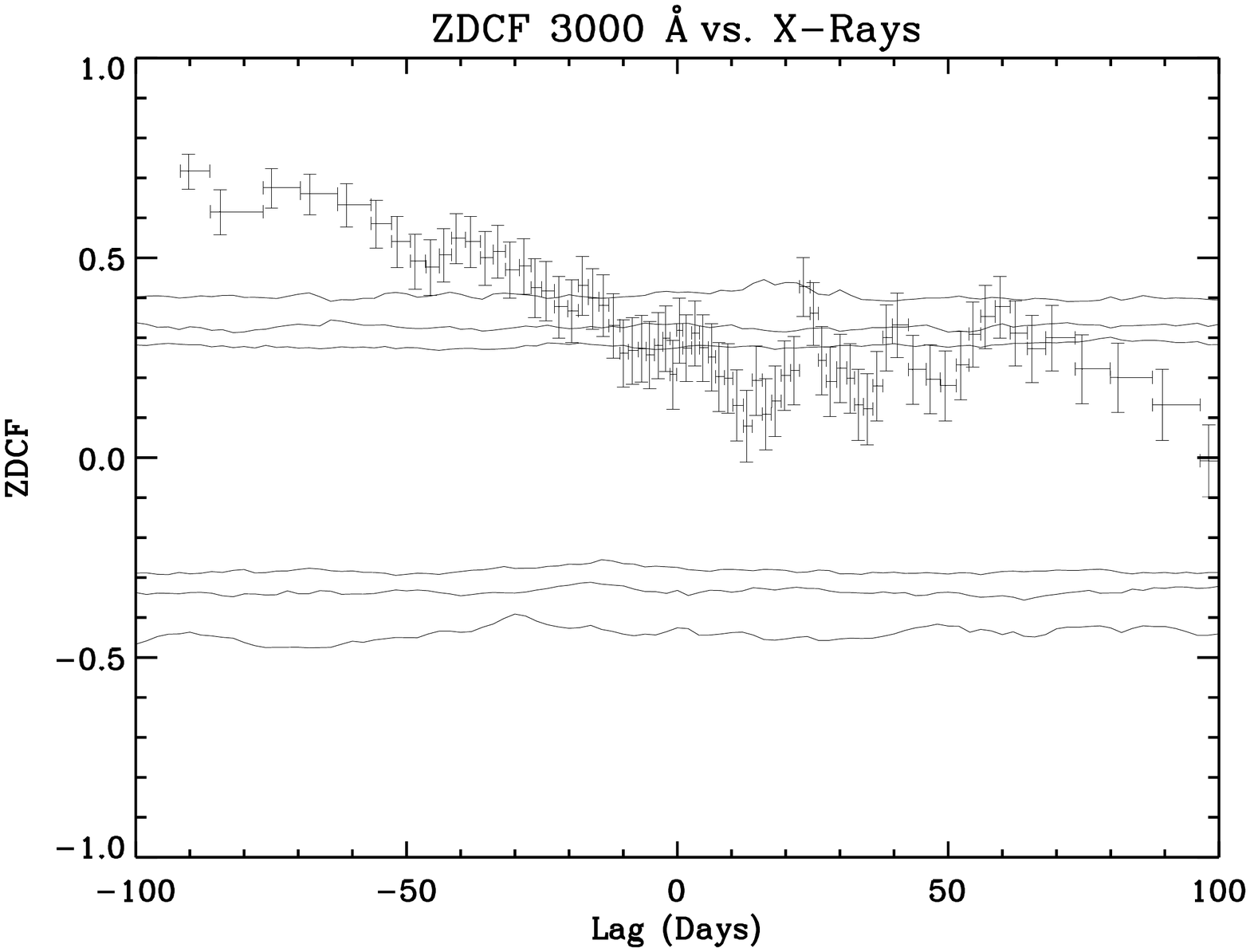}
\includegraphics[width=0.48\textwidth]{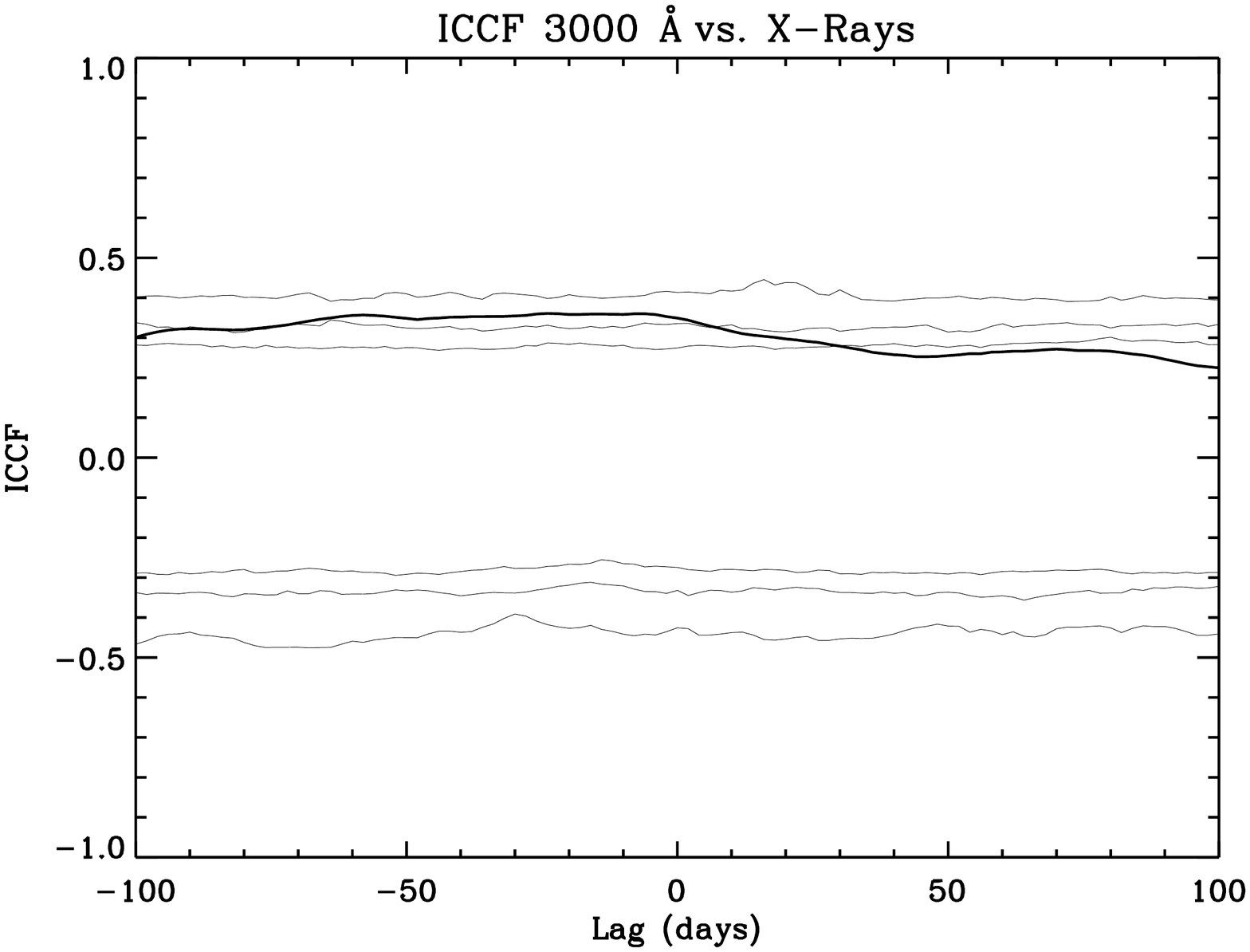}
\includegraphics[width=0.48\textwidth]{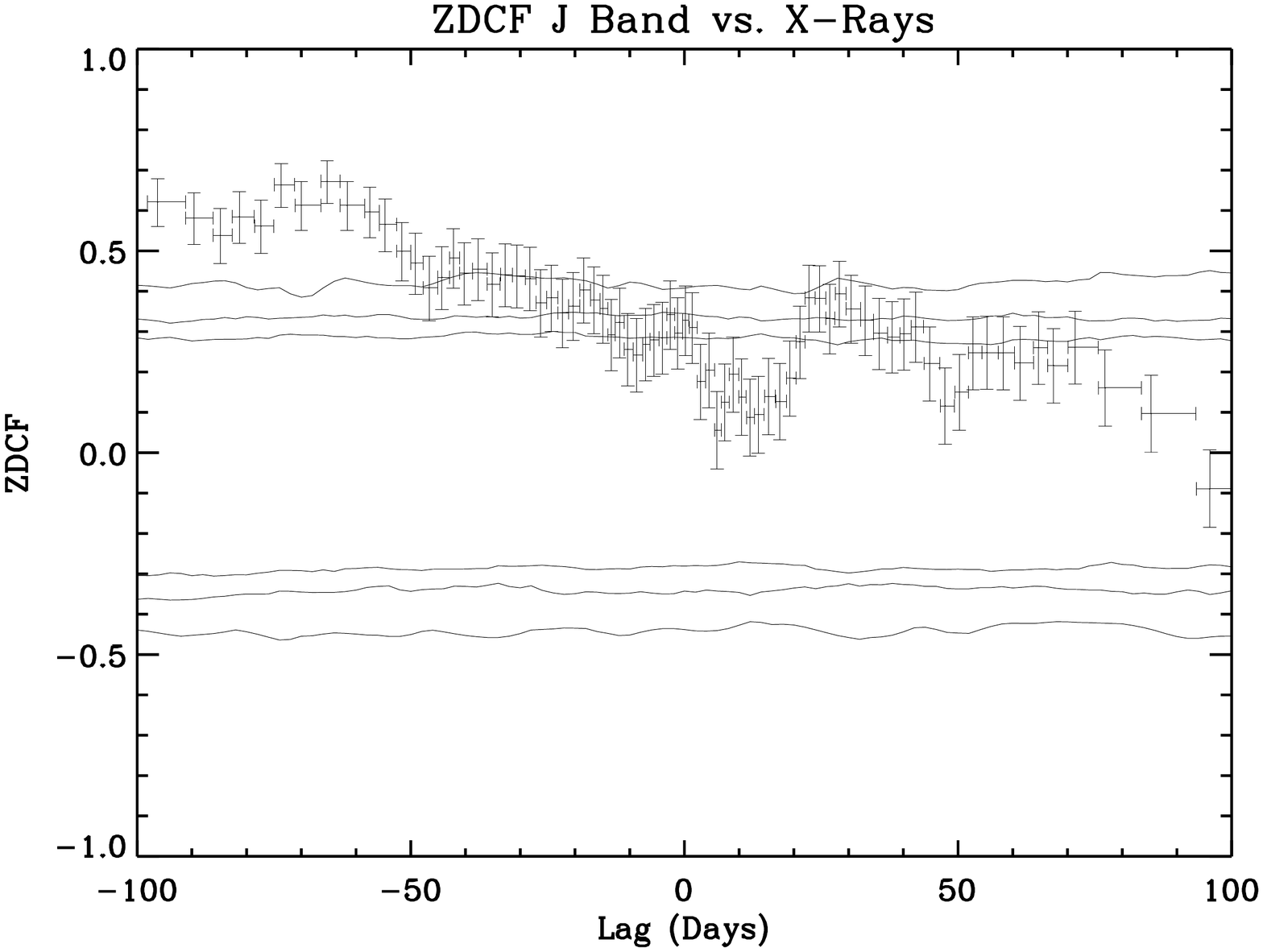}
\end{center}
\caption{Continuation of Fig.~\ref{CC_Full}. Middle panels show the discrepancy in results between the ICCF and the ZDCF methods (is the same case for the DCCF), that is mentioned in Table~\ref{lags1}.}
\label{CC_Full2}
\end{figure}


\begin{figure*}
\begin{center}
\includegraphics[width=0.48\textwidth]{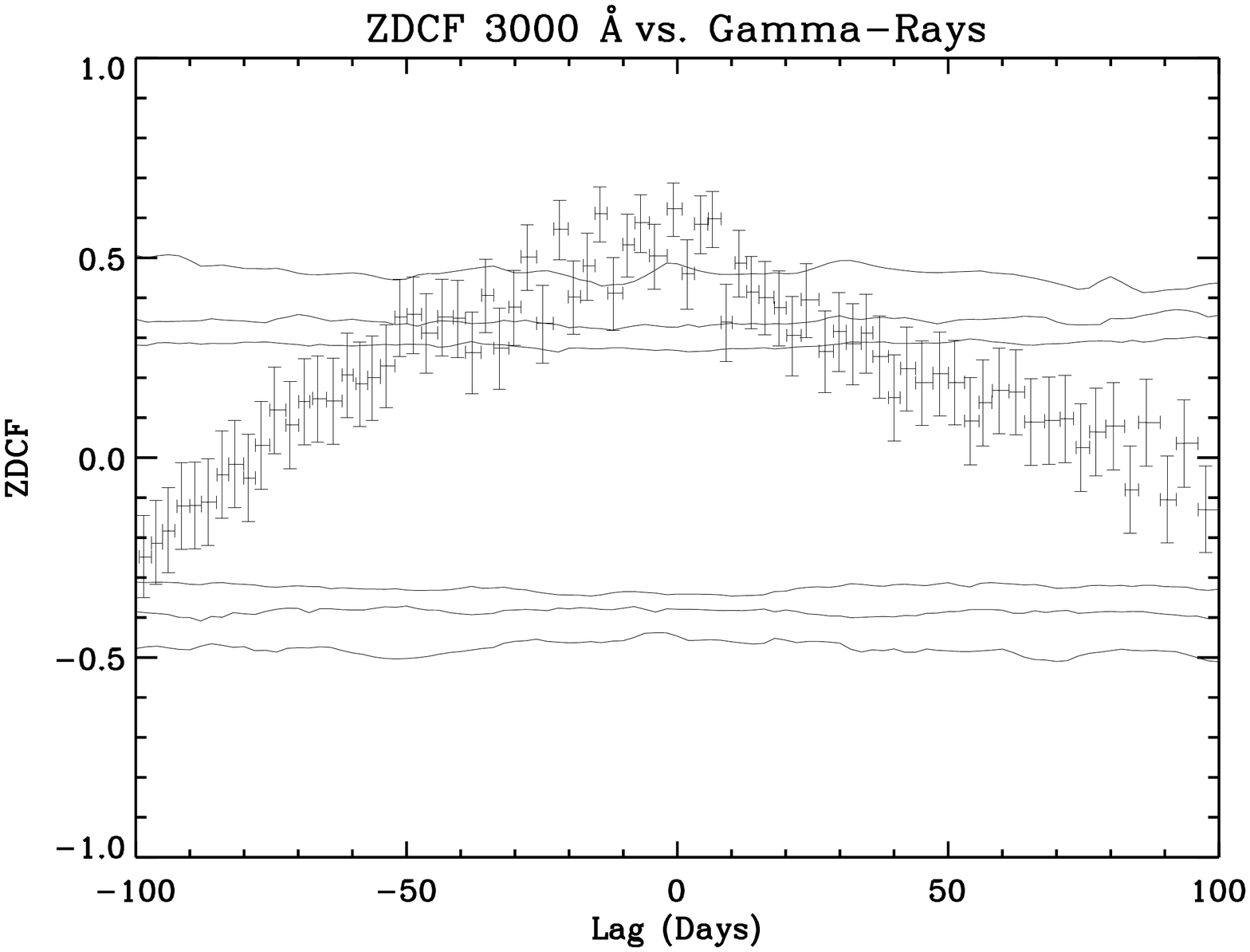}
\includegraphics[width=0.48\textwidth]{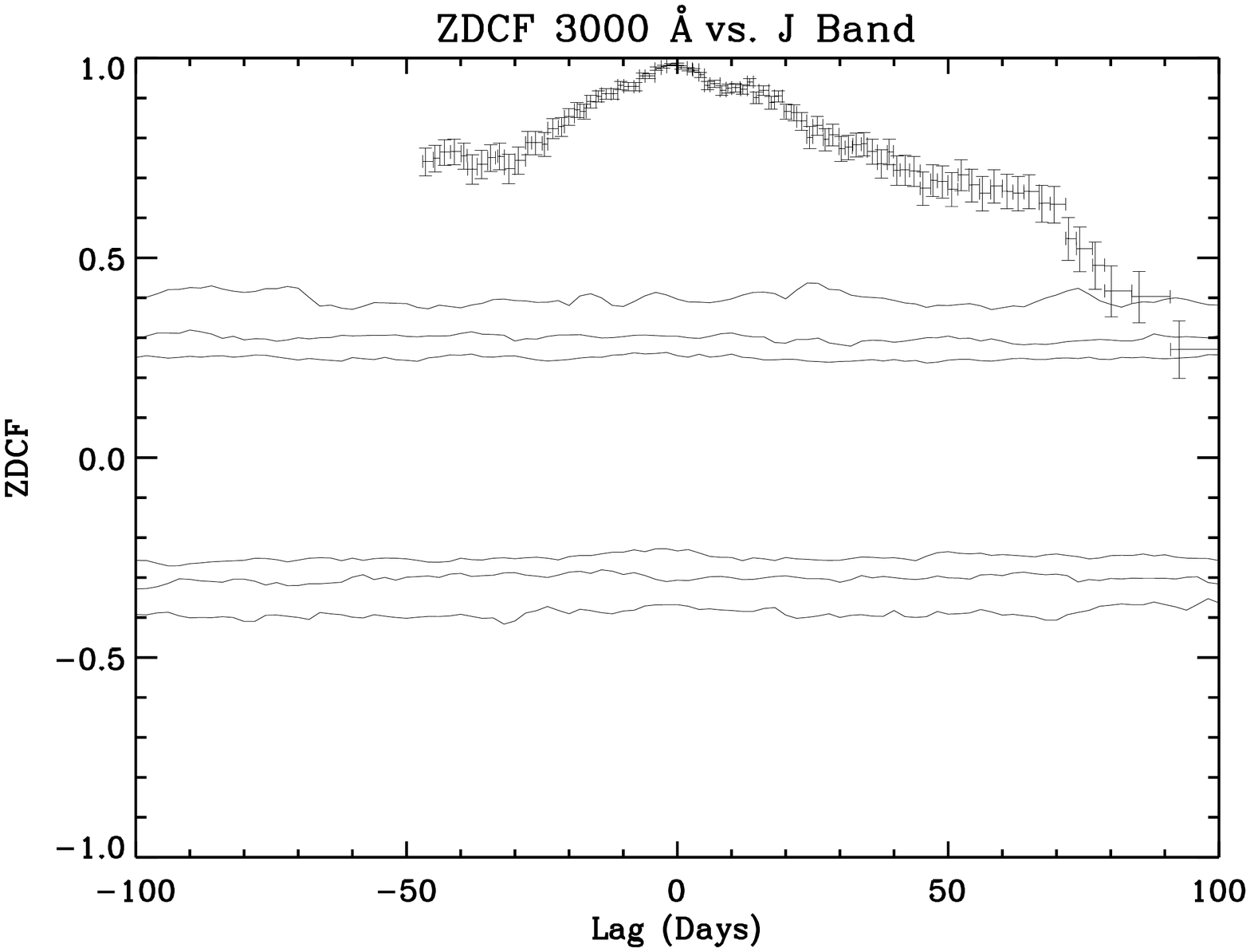}
\includegraphics[width=0.48\textwidth]{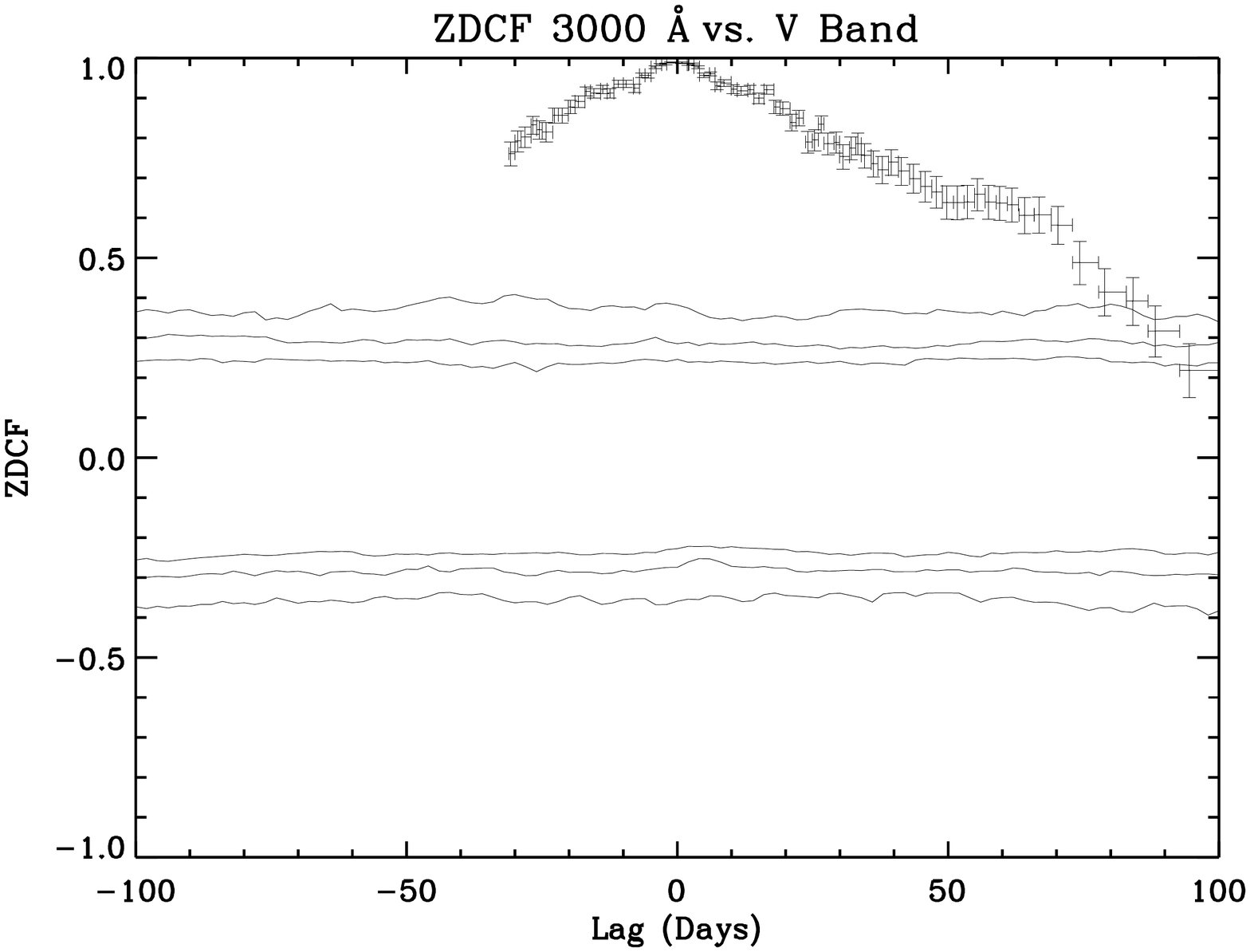}
\includegraphics[width=0.48\textwidth]{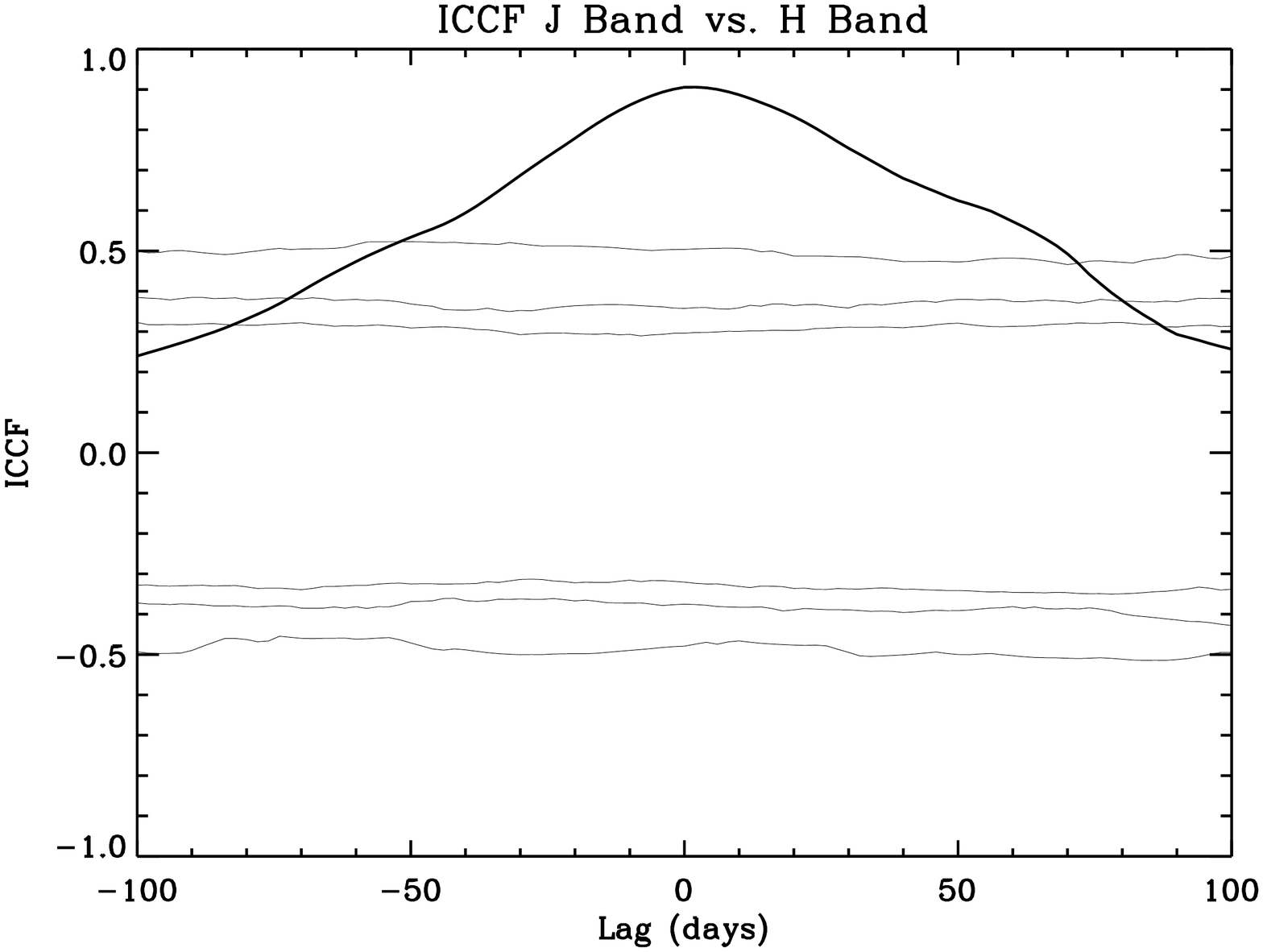}
\includegraphics[width=0.48\textwidth]{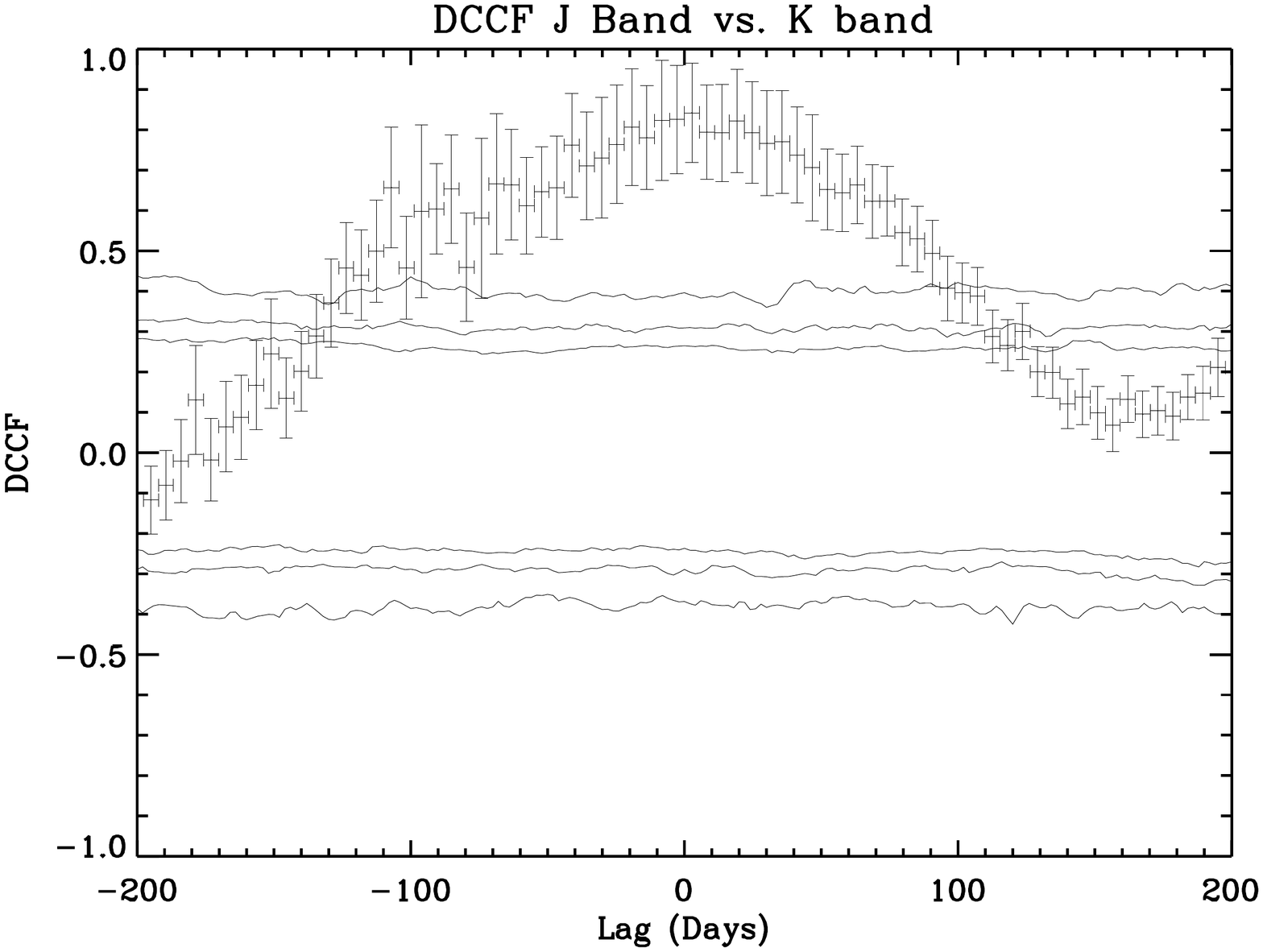}
\includegraphics[width=0.48\textwidth]{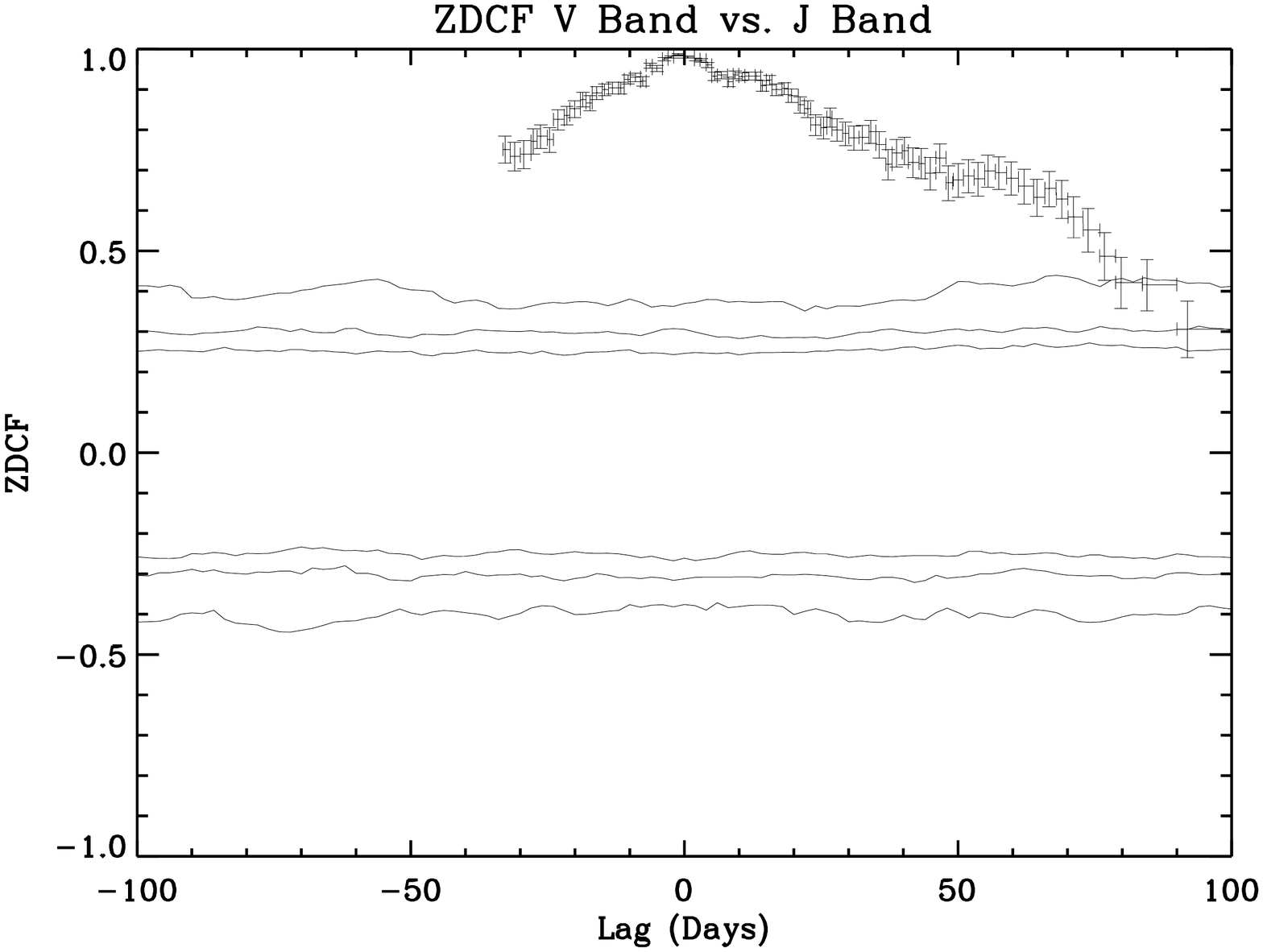}
\caption{Same as Fig.~\ref{CC_Full} for the Period A ($\rm{JD}_{245}=4650-5850$). }
\label{CC_PA}
\end{center}
\end{figure*}

\begin{figure}
\begin{center}
\includegraphics[width=0.48\textwidth]{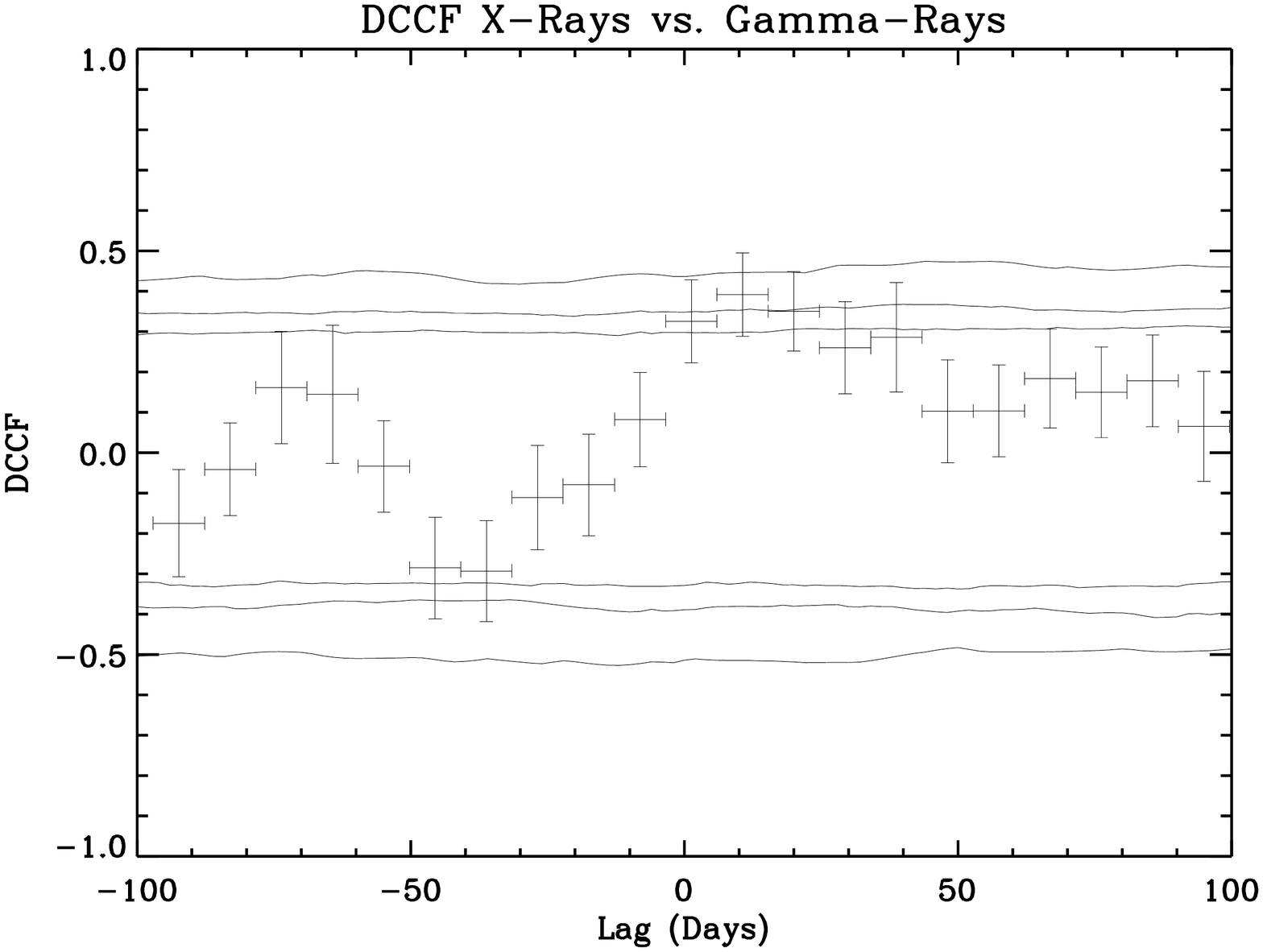}
\includegraphics[width=0.48\textwidth]{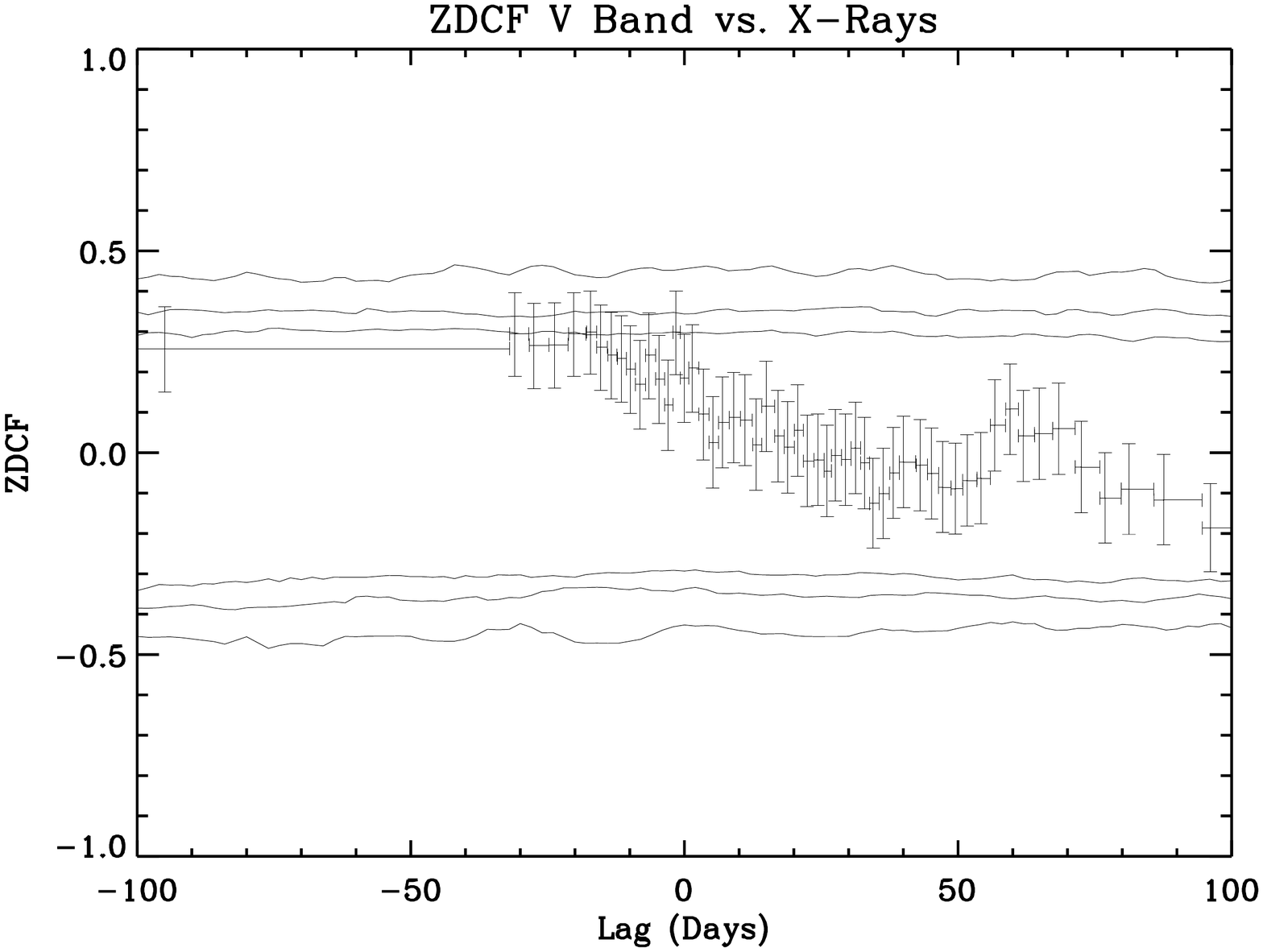}
\includegraphics[width=0.48\textwidth]{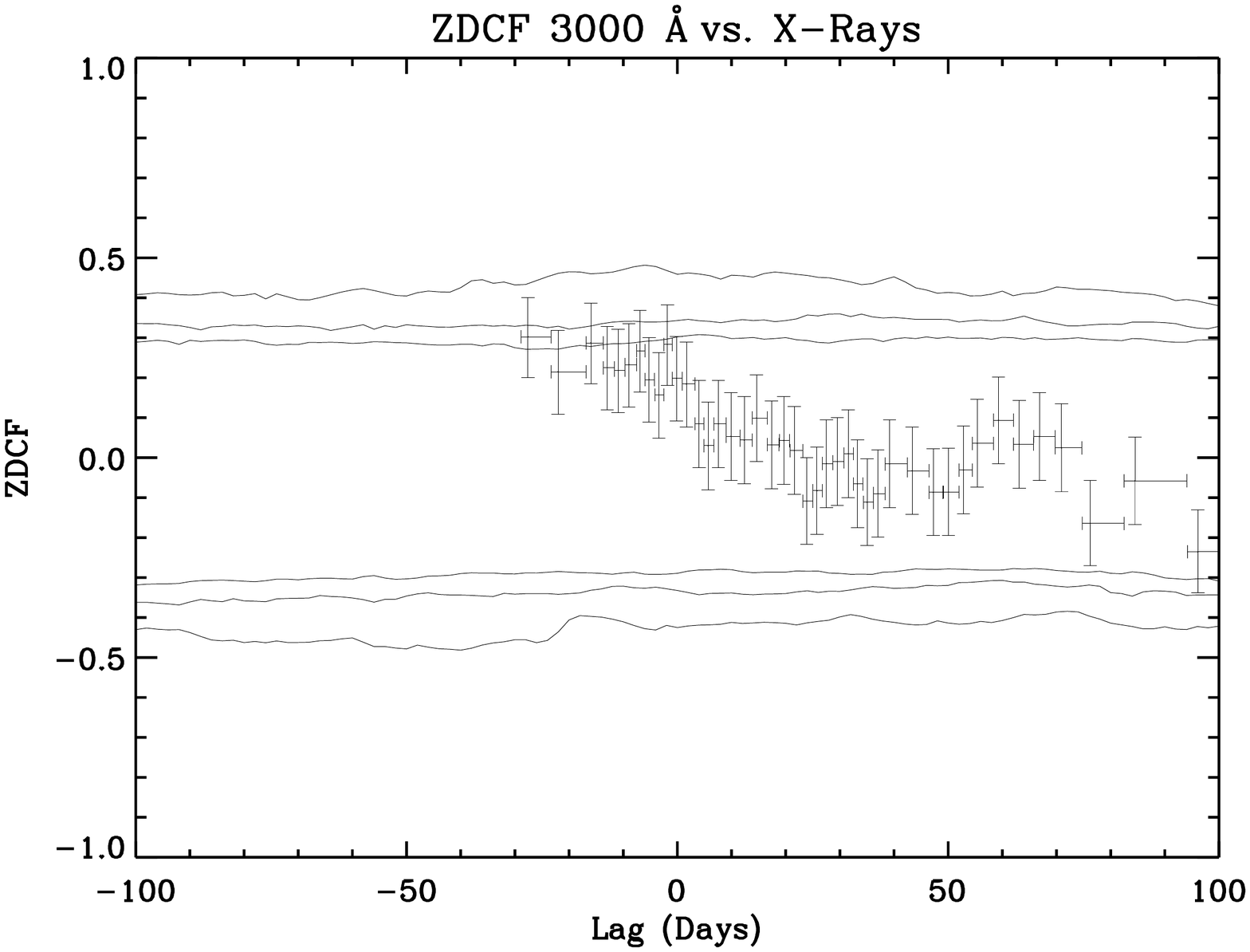}
\includegraphics[width=0.48\textwidth]{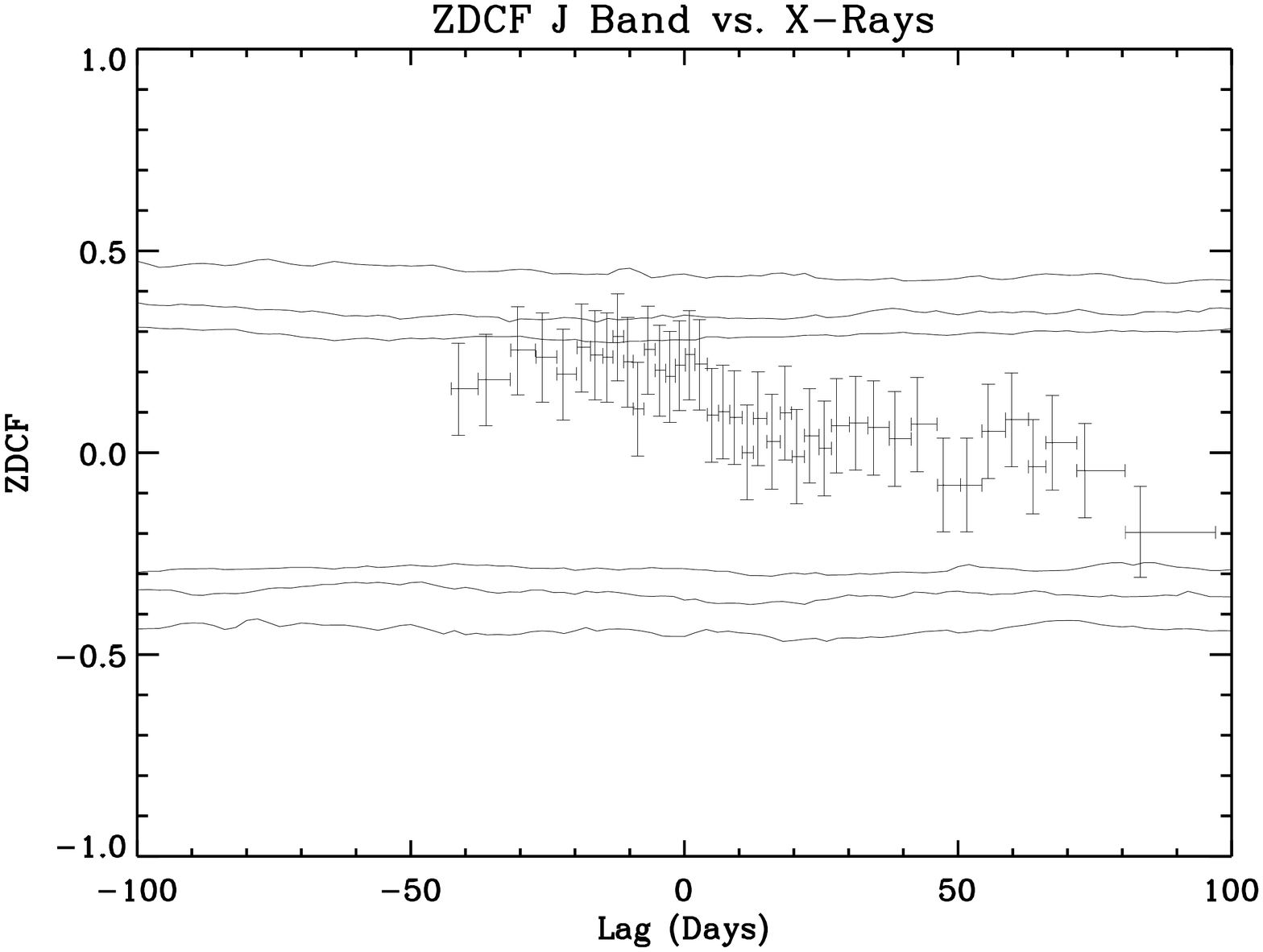}
\end{center}
\caption{Continuation of Fig.~\ref{CC_PA}.}
\label{CC_PA2}
\end{figure}


\begin{figure*}
\begin{center}
\includegraphics[width=0.48\textwidth]{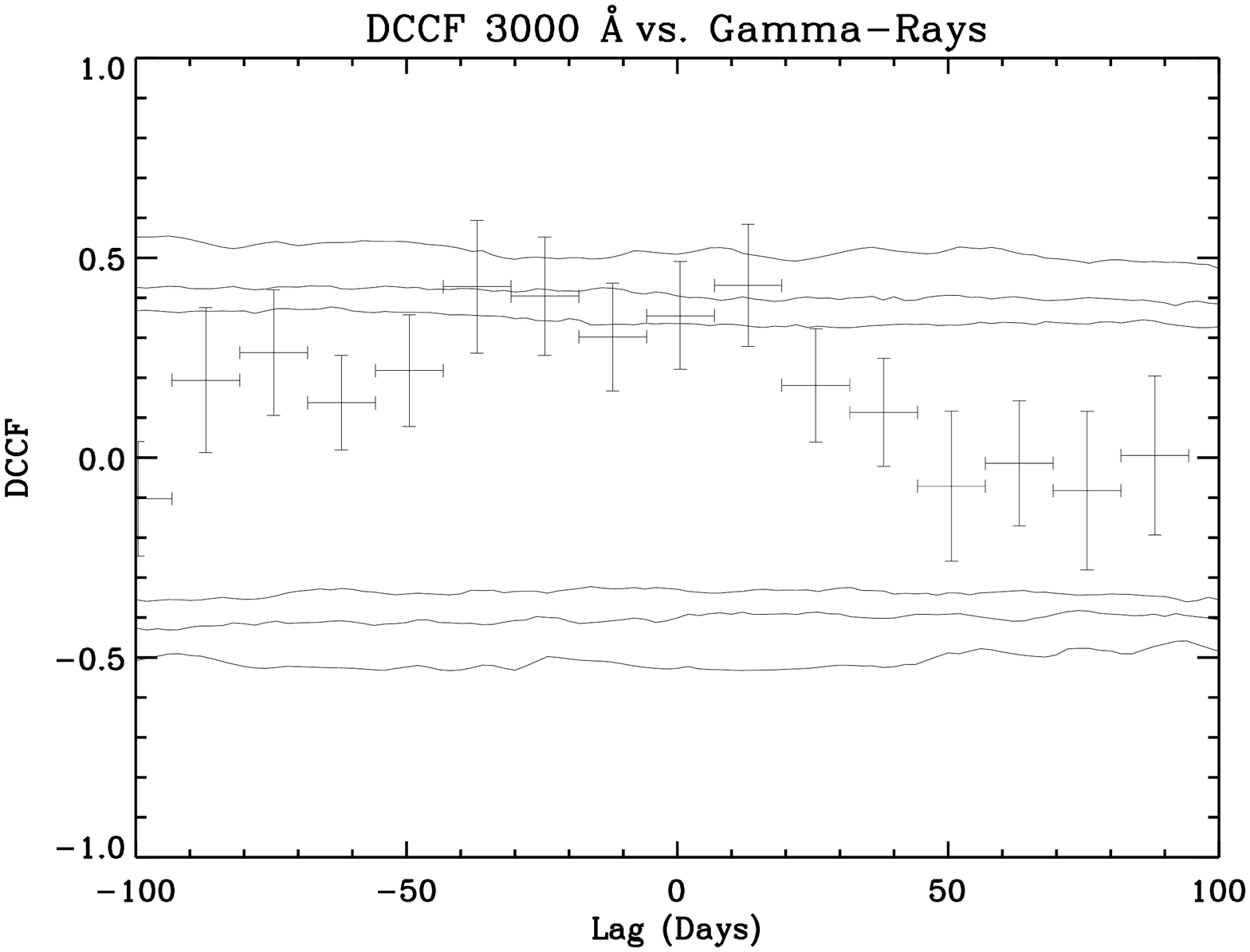}
\includegraphics[width=0.48\textwidth]{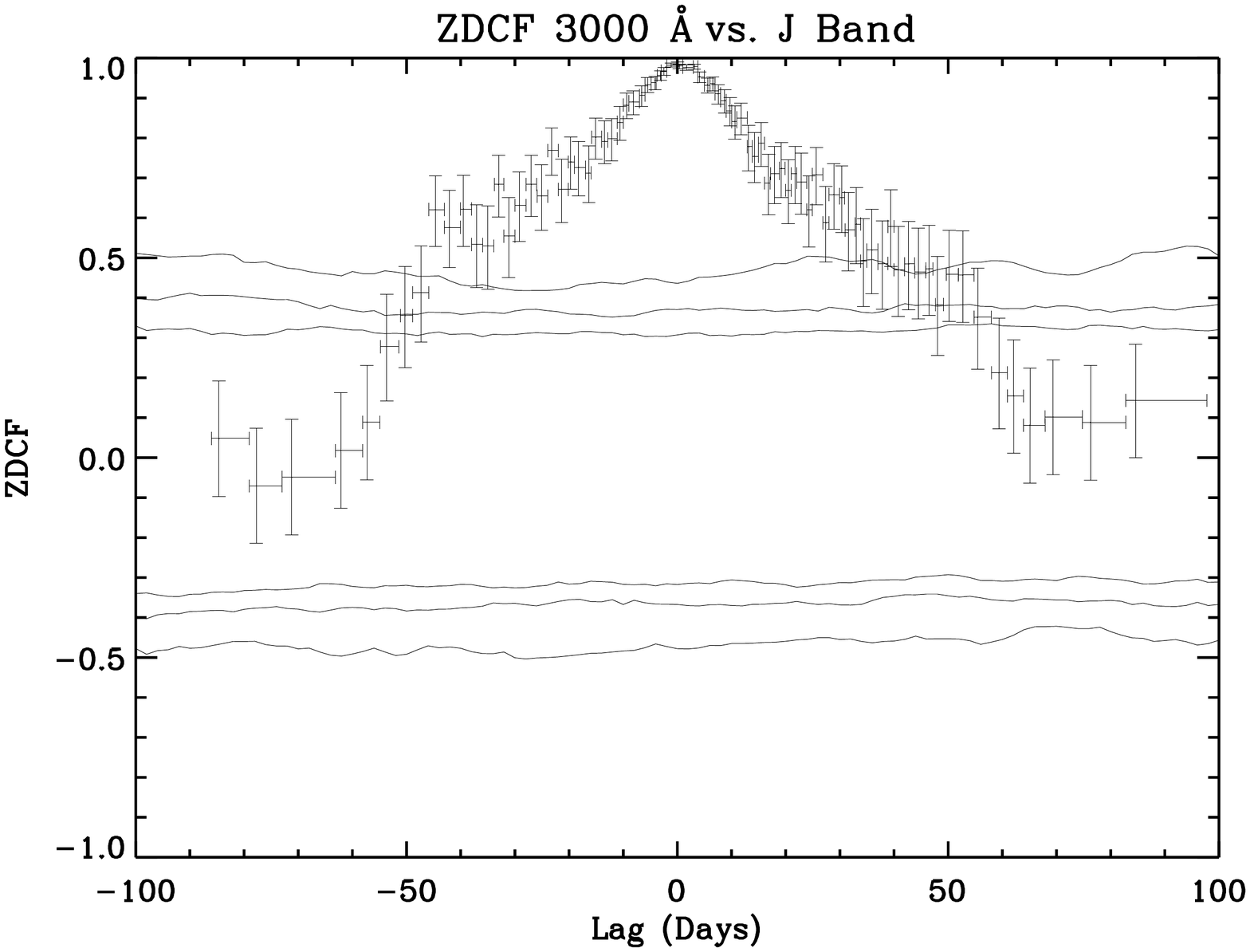}
\includegraphics[width=0.48\textwidth]{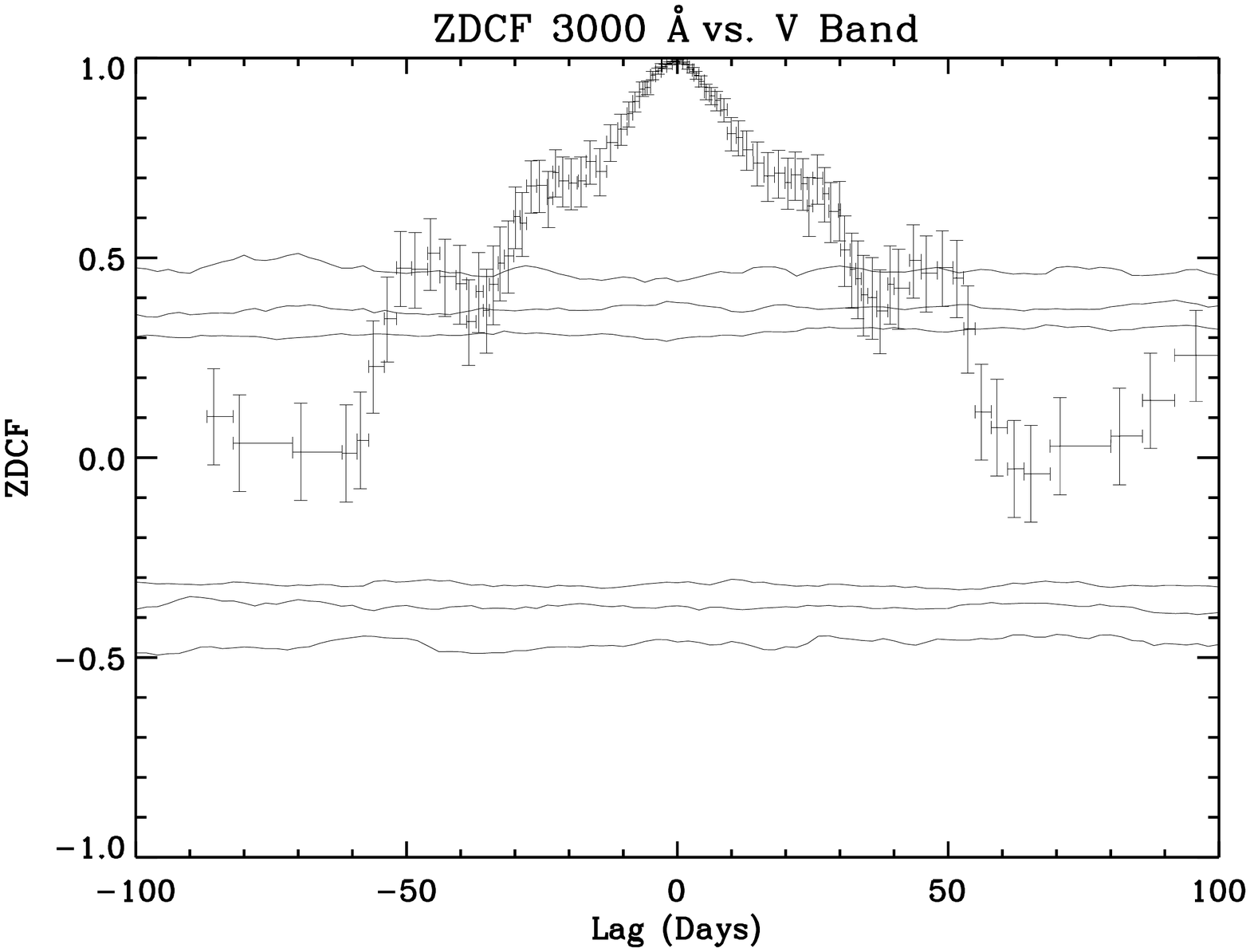}
\includegraphics[width=0.48\textwidth]{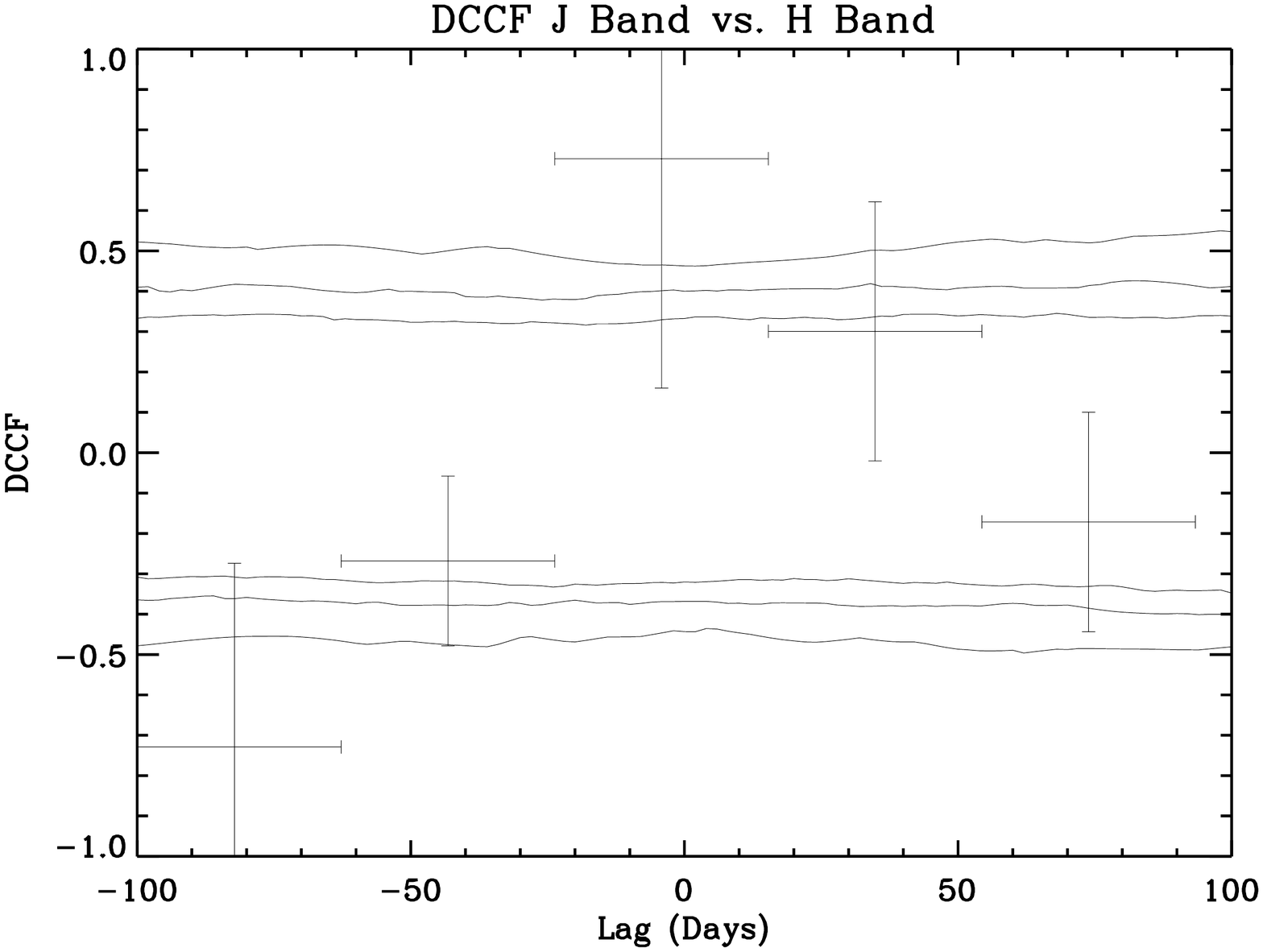}
\includegraphics[width=0.48\textwidth]{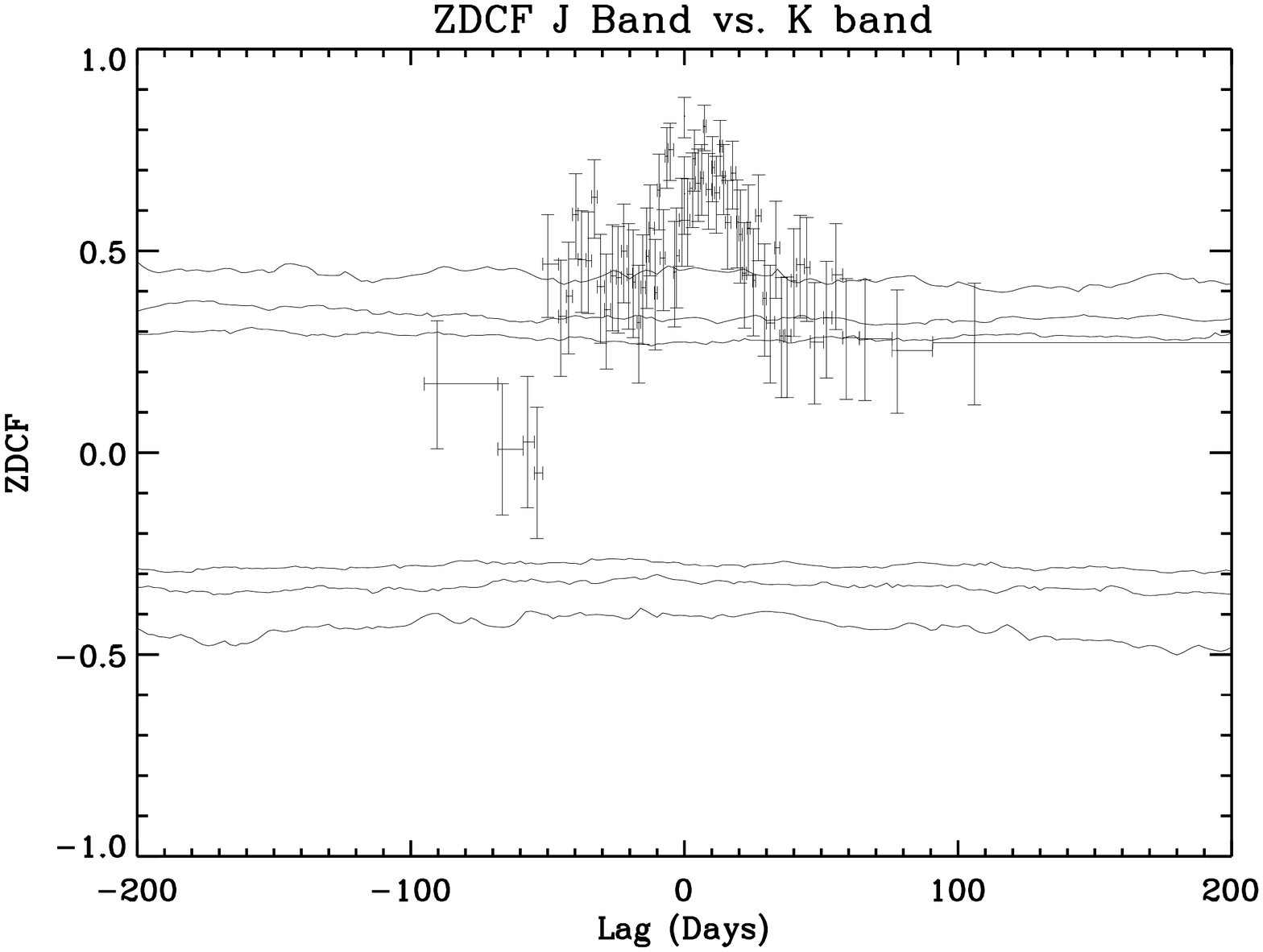}
\includegraphics[width=0.48\textwidth]{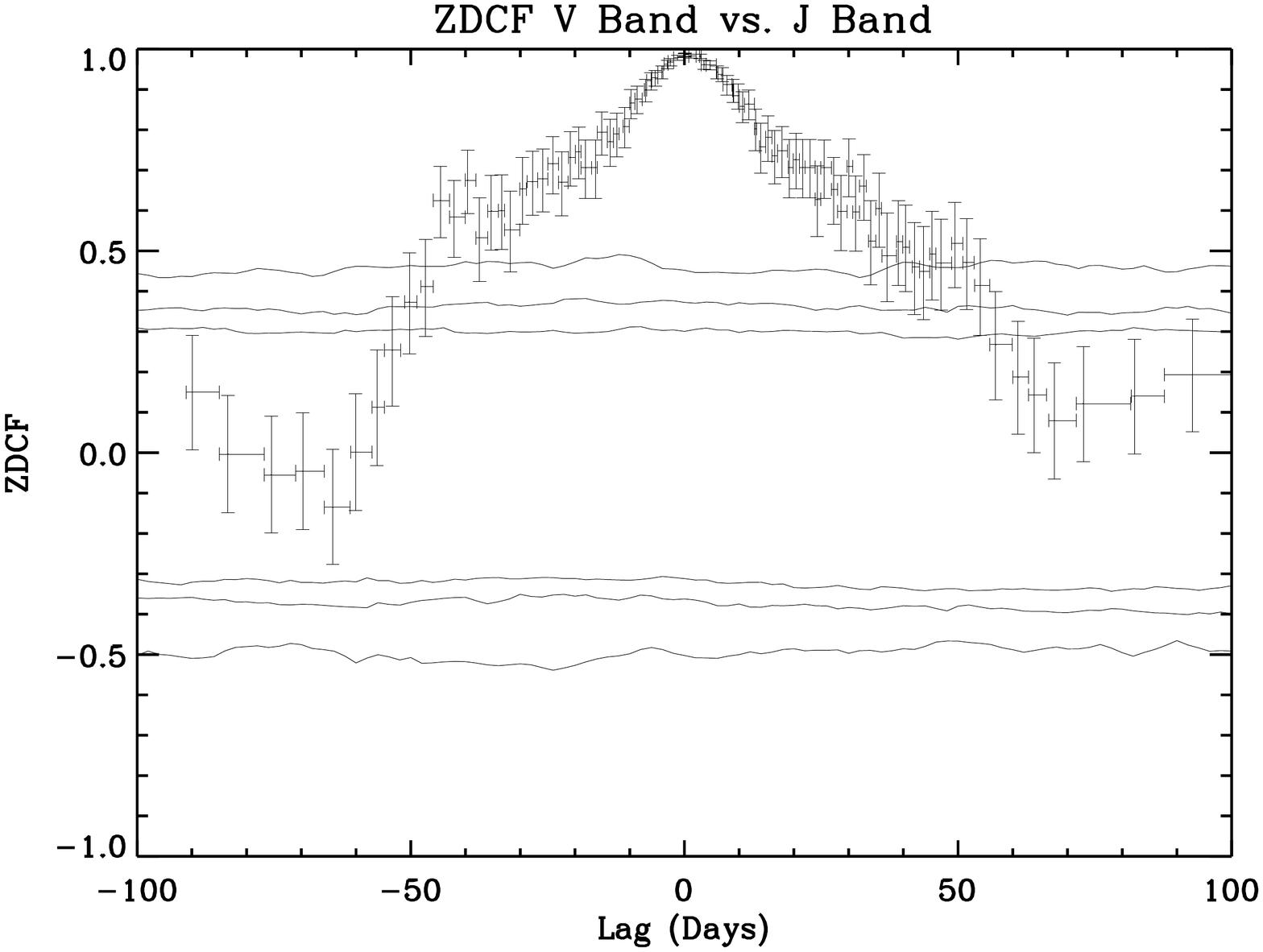}
\caption{Same as Fig.~\ref{CC_Full} for the Period B ($\rm{JD}_{245}=5850-6400$).}
\label{CC_PB}
\end{center}
\end{figure*}

\begin{figure}
\begin{center}
\includegraphics[width=0.48\textwidth]{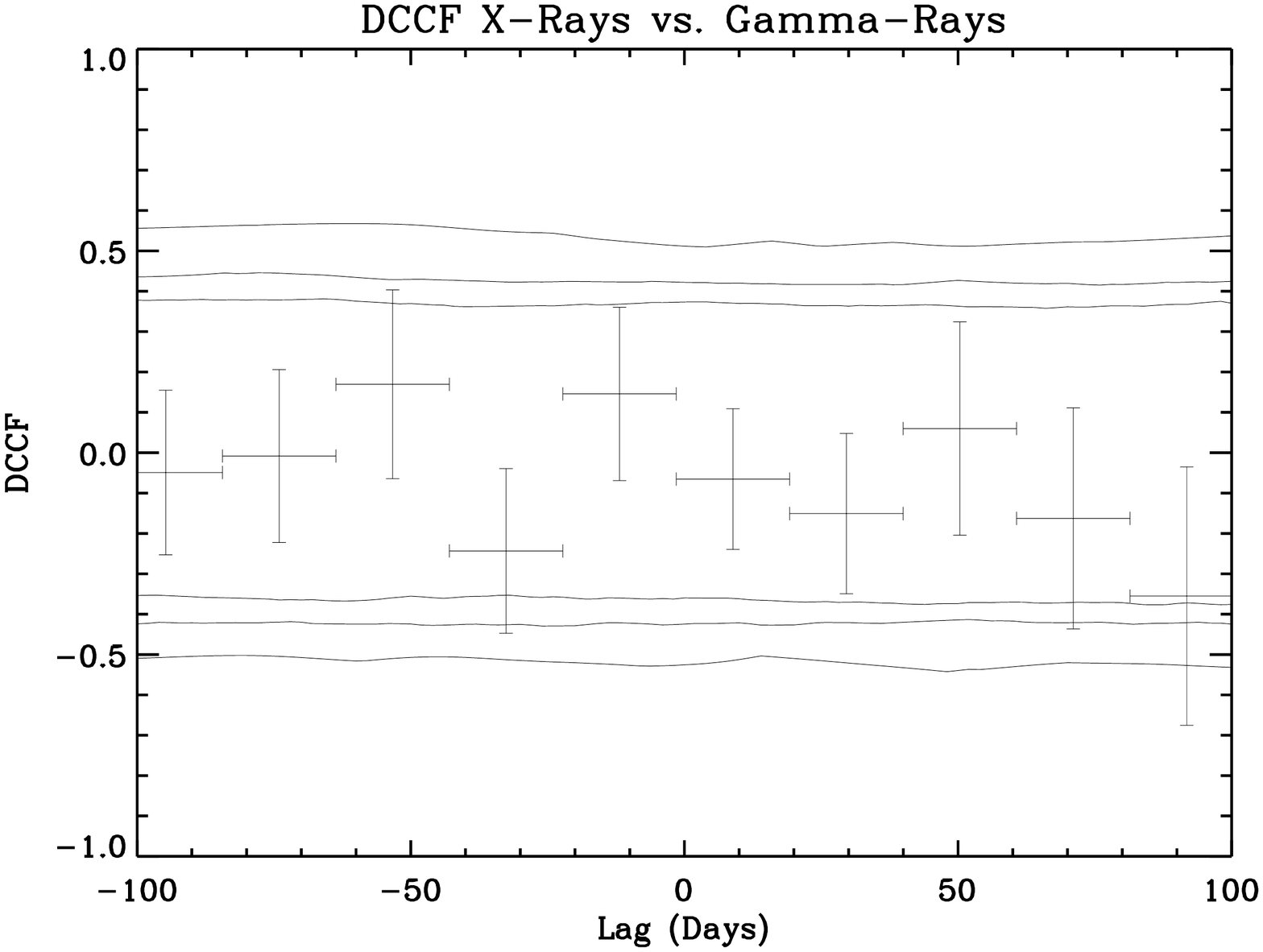}
\includegraphics[width=0.48\textwidth]{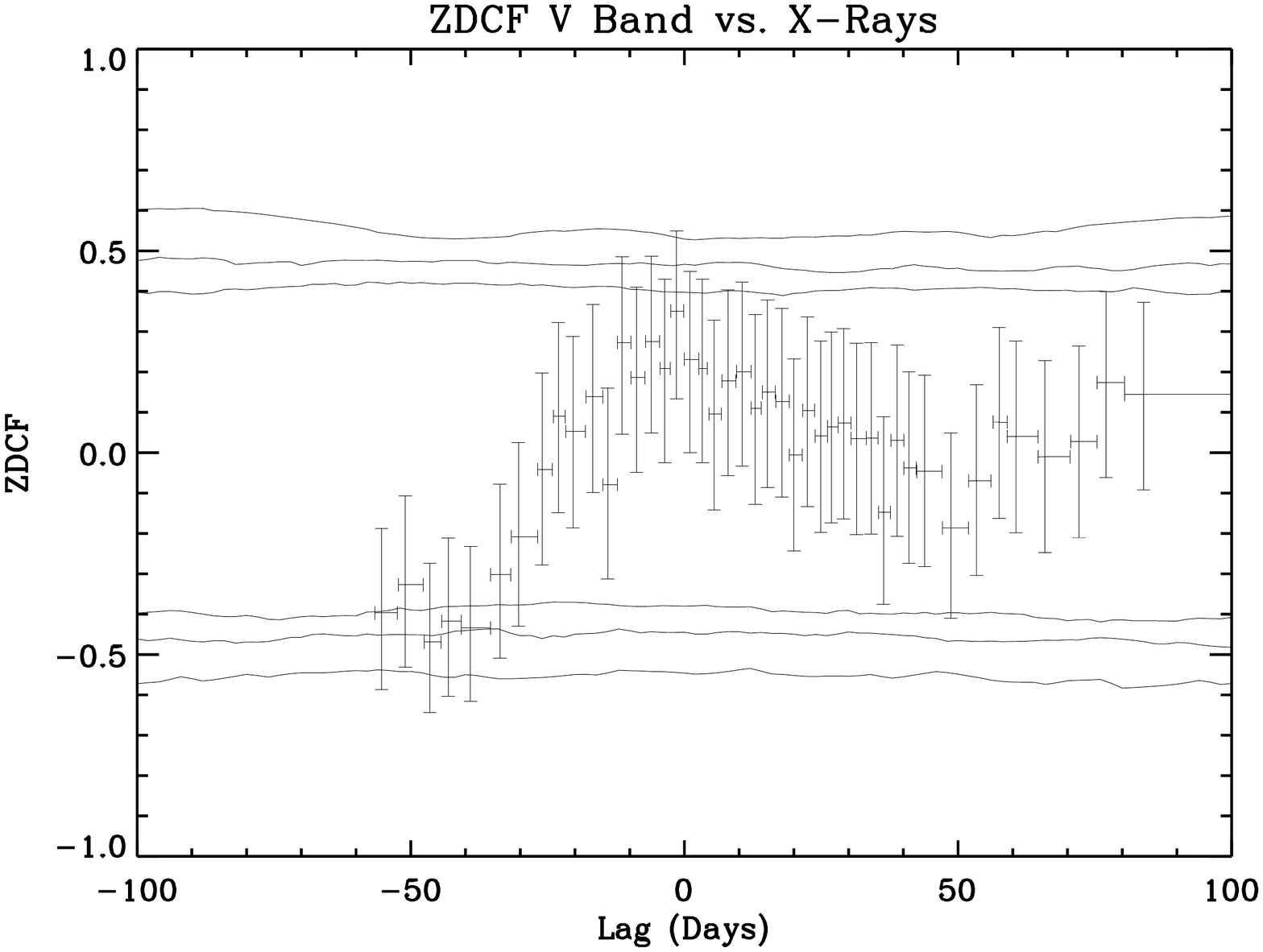}
\includegraphics[width=0.48\textwidth]{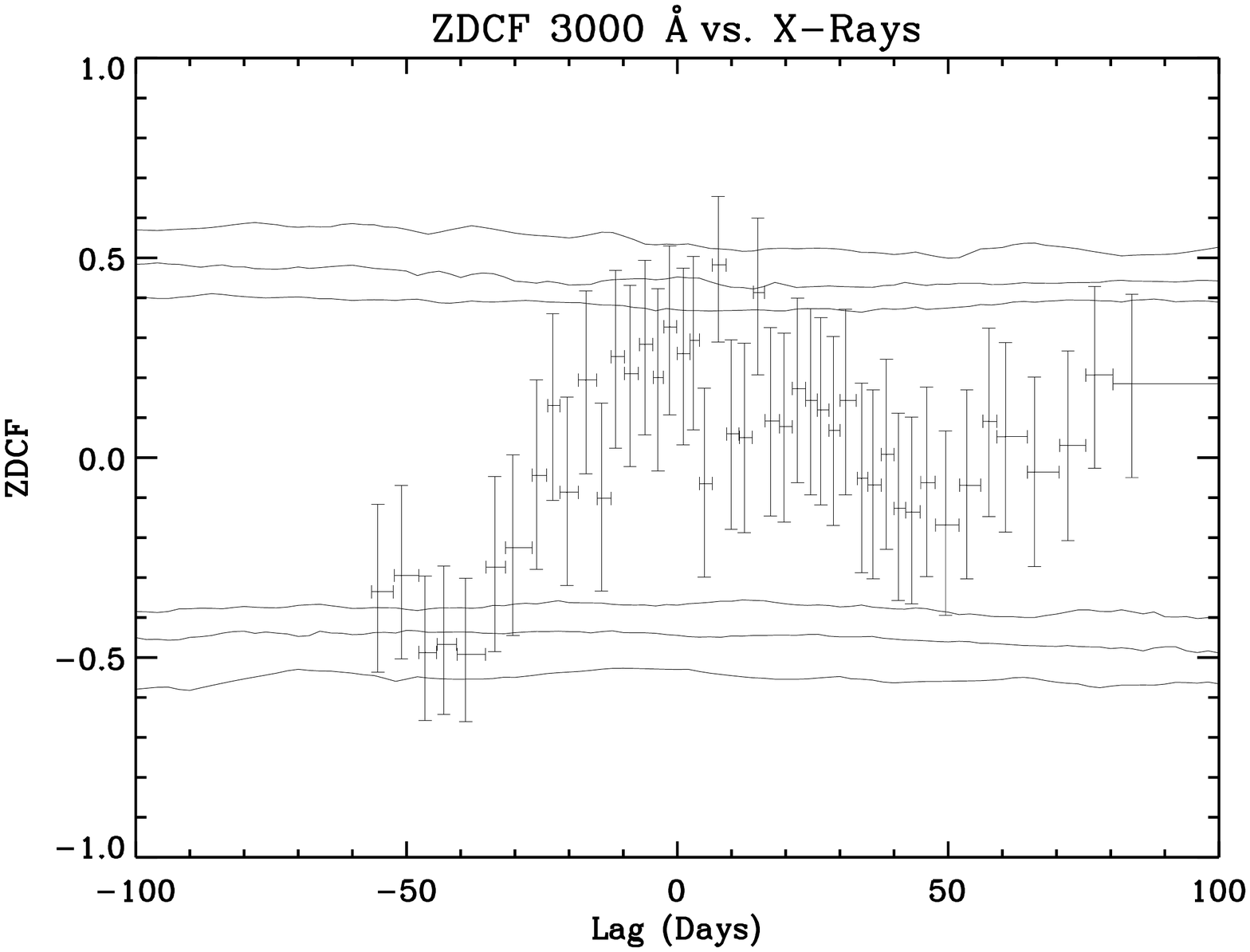}
\includegraphics[width=0.48\textwidth]{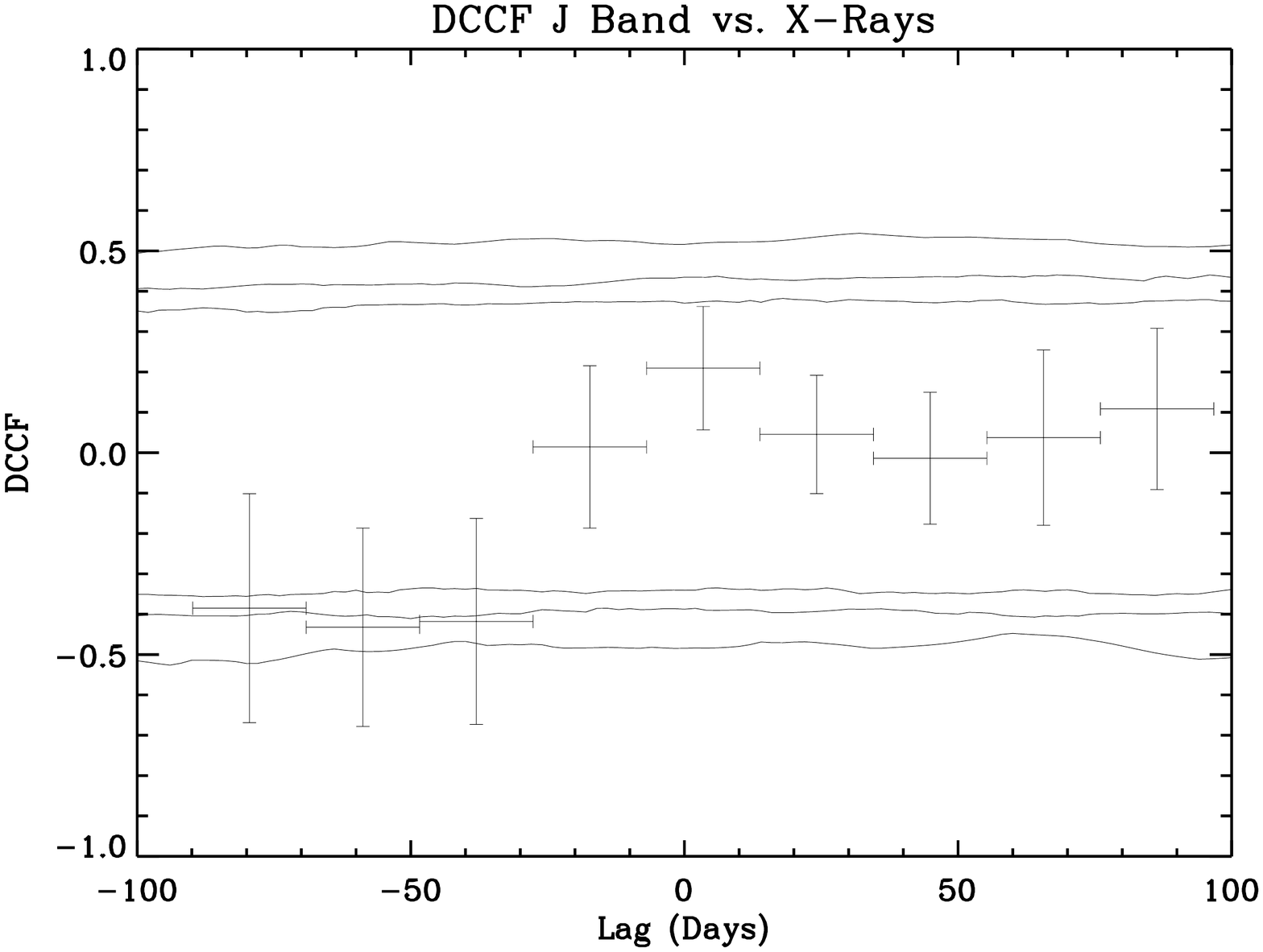}
\end{center}
\caption{Continuation of Fig.~\ref{CC_PB}.}
\label{CC_PB2}
\end{figure}


\begin{figure*}
\begin{center}
\includegraphics[width=0.48\textwidth]{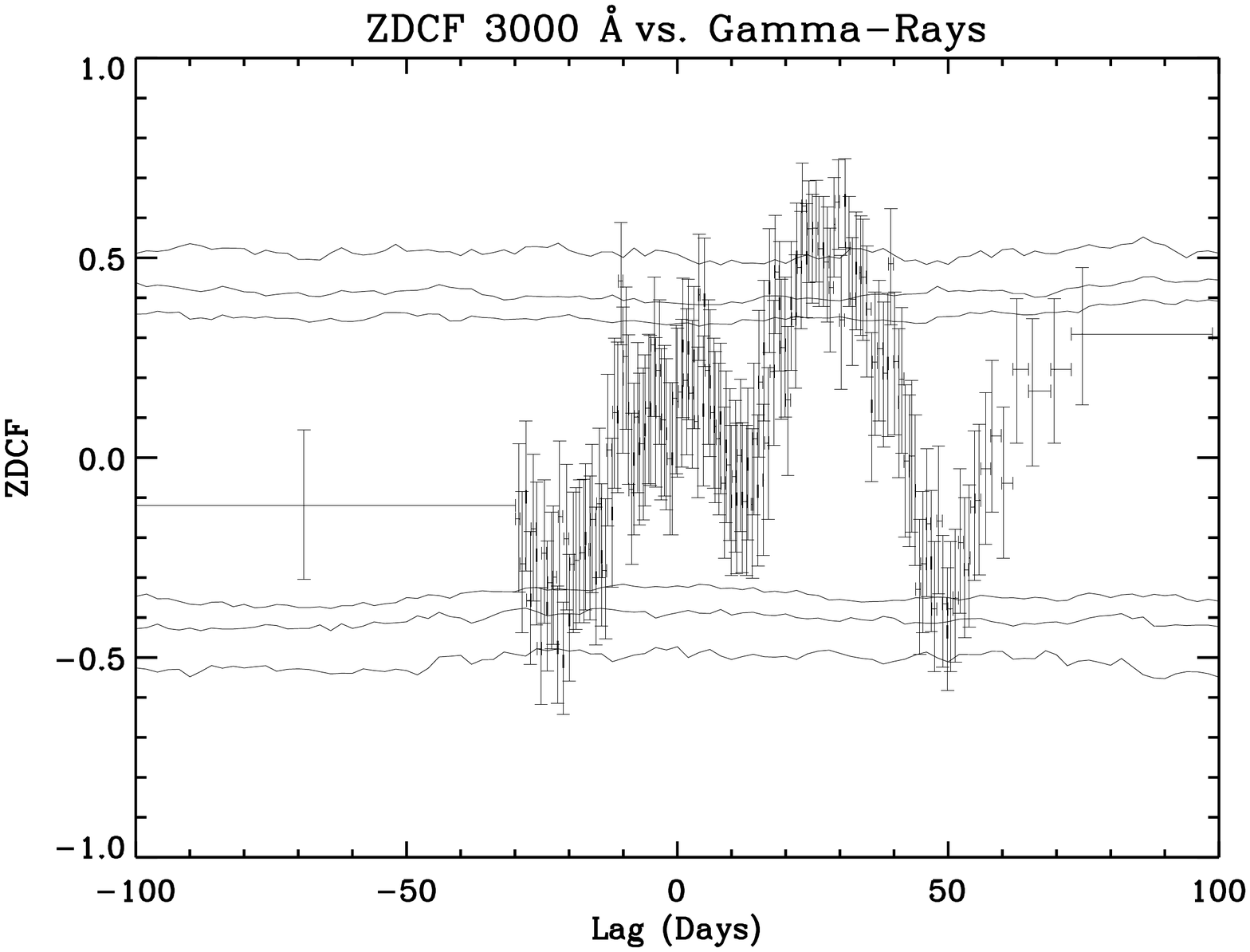}
\includegraphics[width=0.48\textwidth]{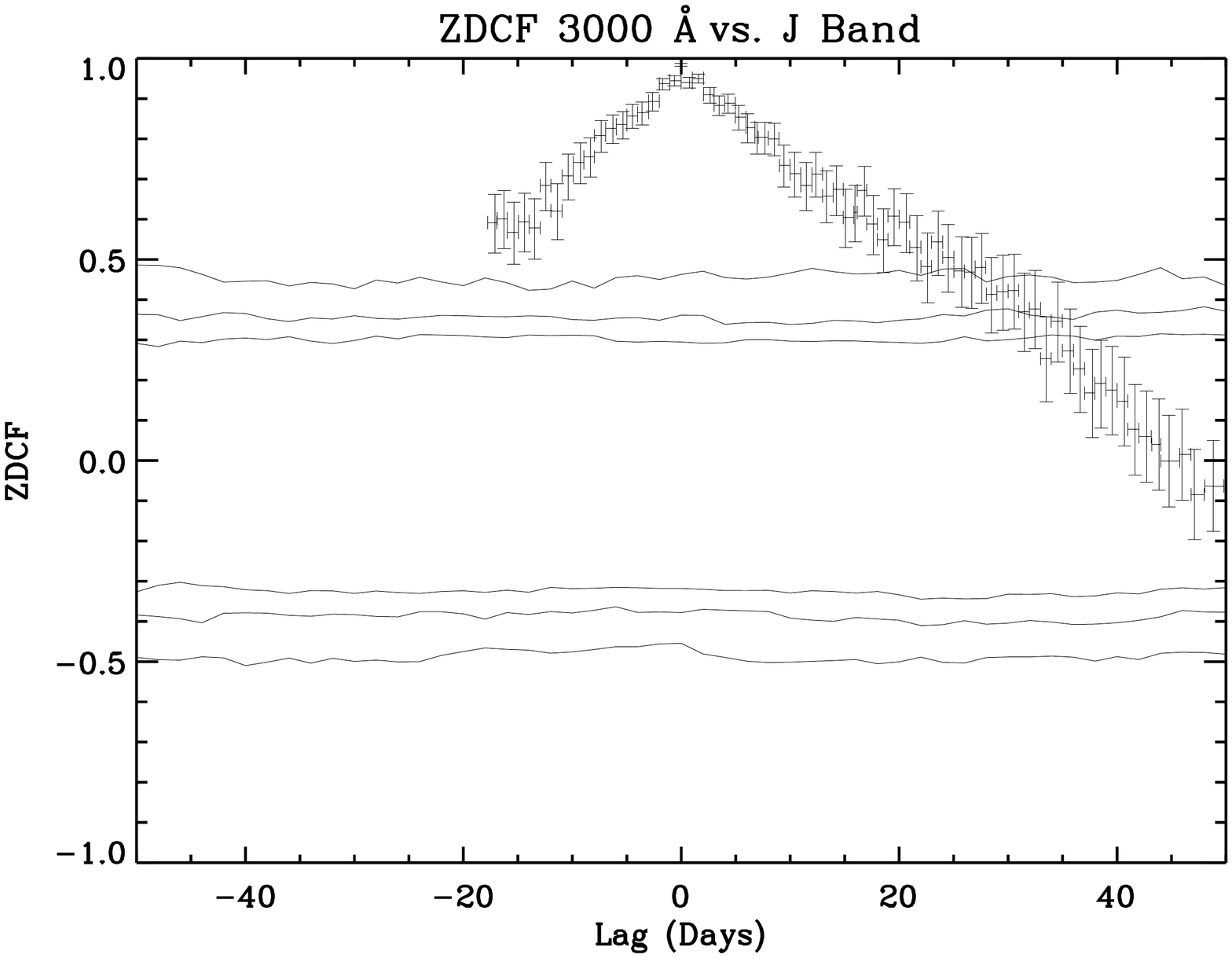}
\includegraphics[width=0.48\textwidth]{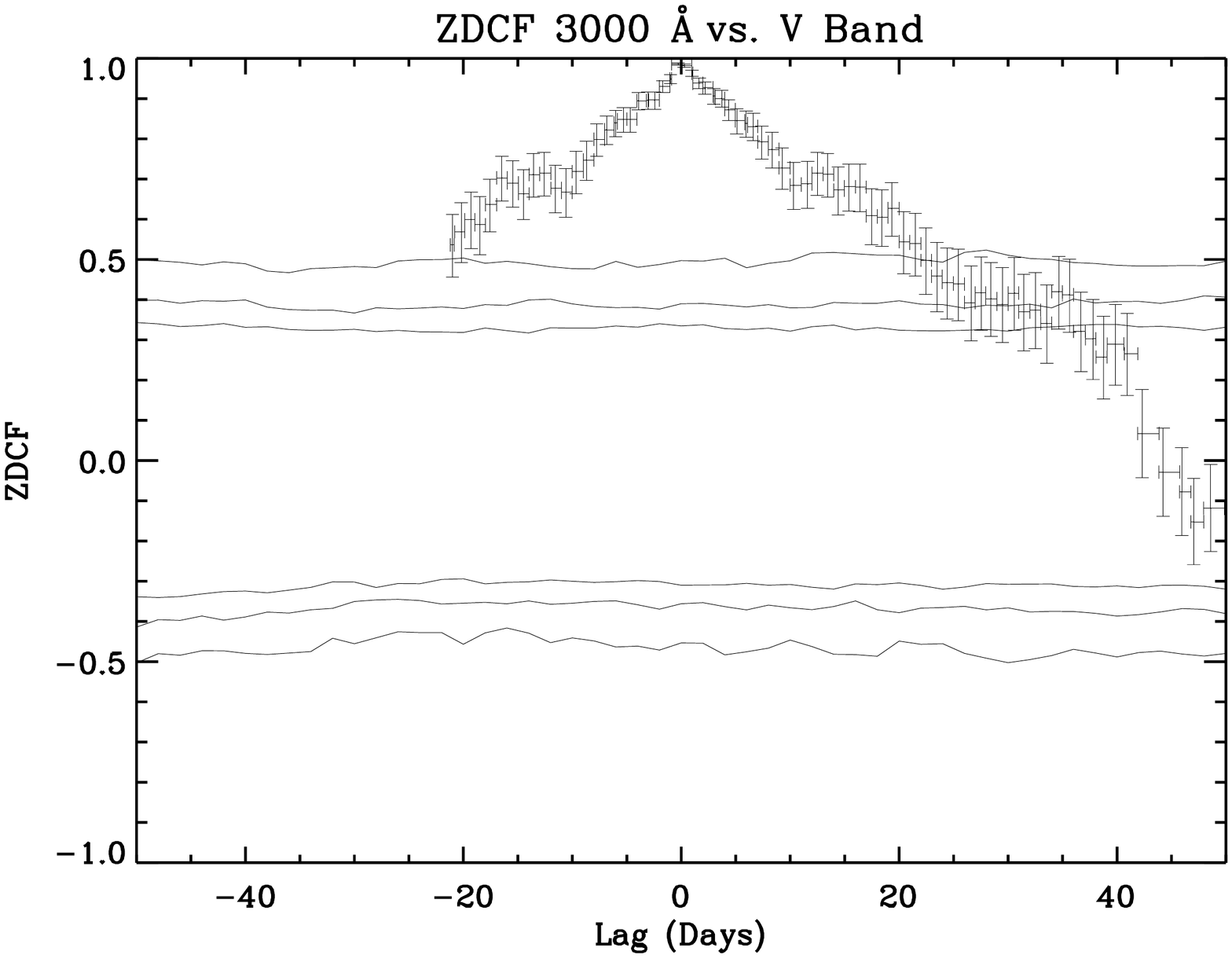}
\includegraphics[width=0.48\textwidth]{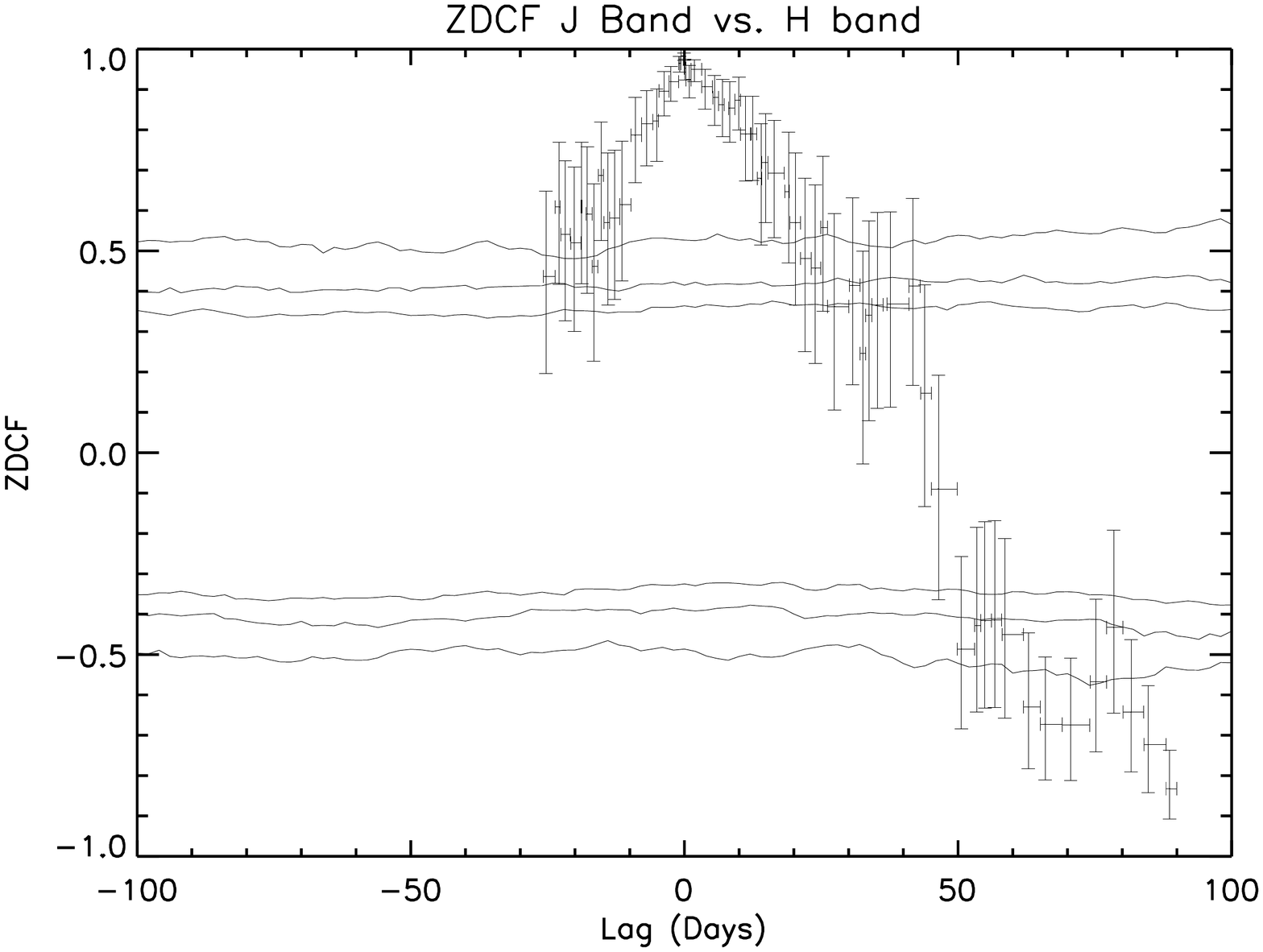}
\includegraphics[width=0.48\textwidth]{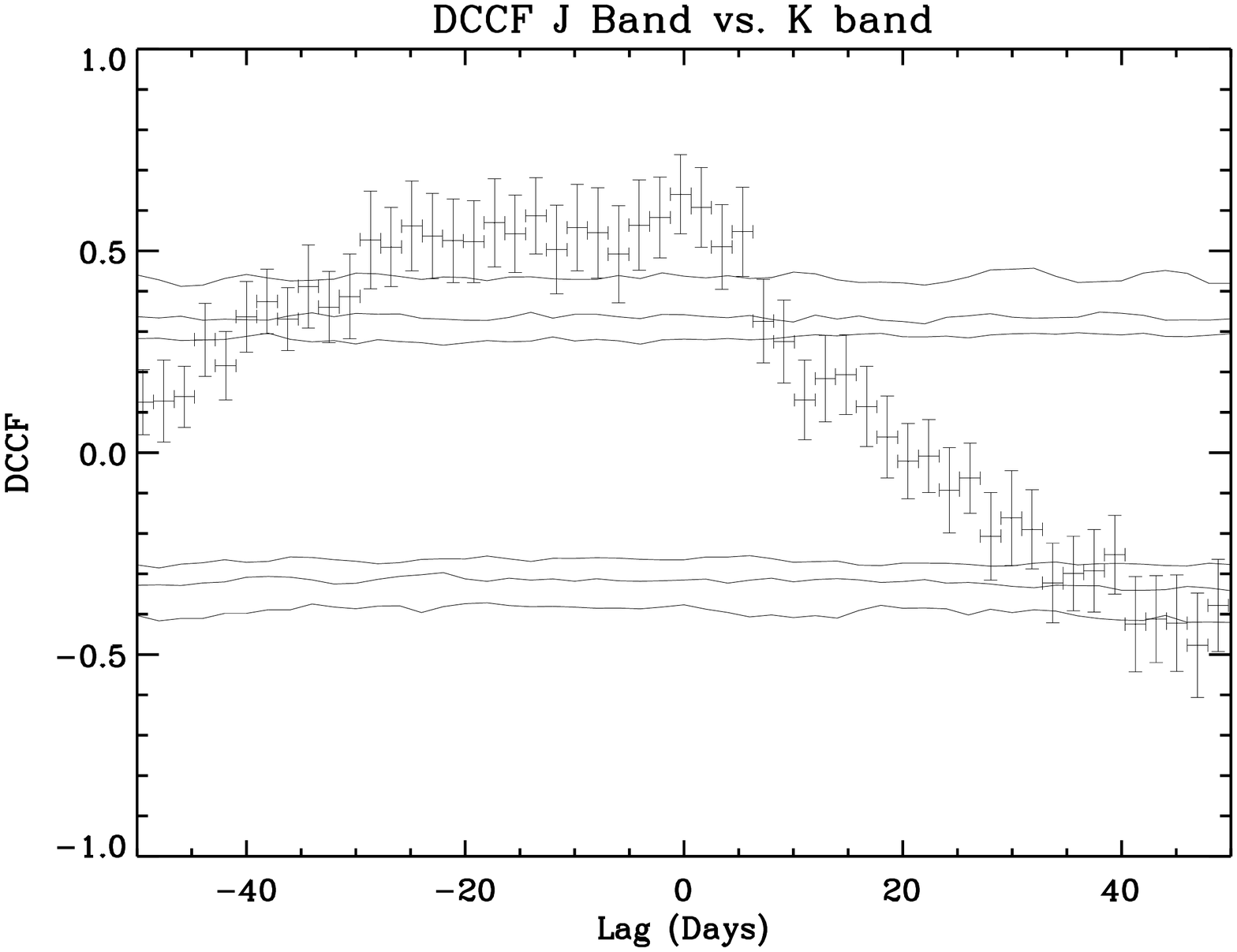}
\includegraphics[width=0.48\textwidth]{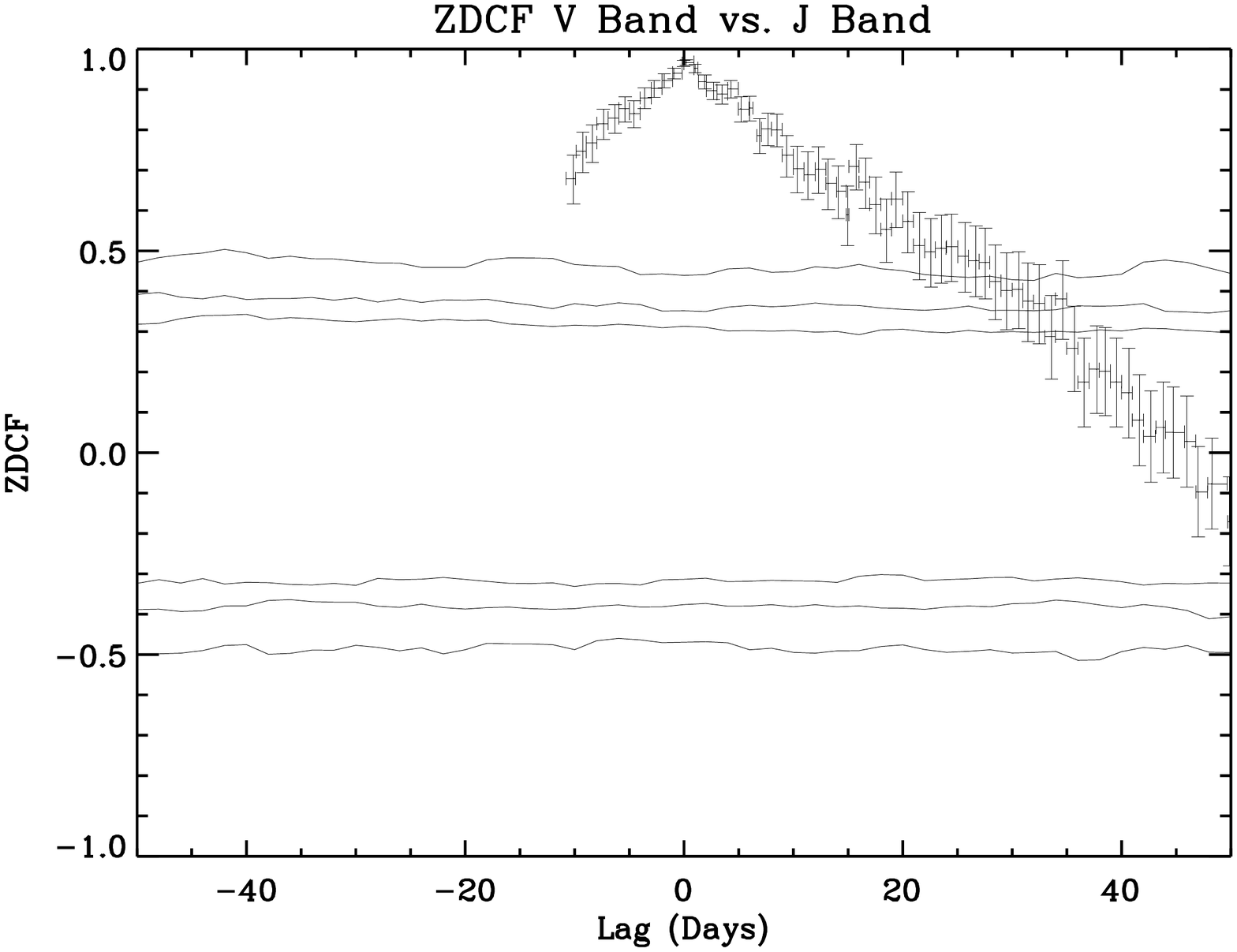}
\caption{Same as Fig.~\ref{CC_Full} for the Period C ($\rm{JD}_{245}=6400-6850$). The Cross-Correlation function shown for 3000 \AA~vs. gamma-rays corresponds to the period of JD$_{245}\geq6700$.}
\label{CC_PC}
\end{center}
\end{figure*}

\begin{figure}
\begin{center}
\includegraphics[width=0.48\textwidth]{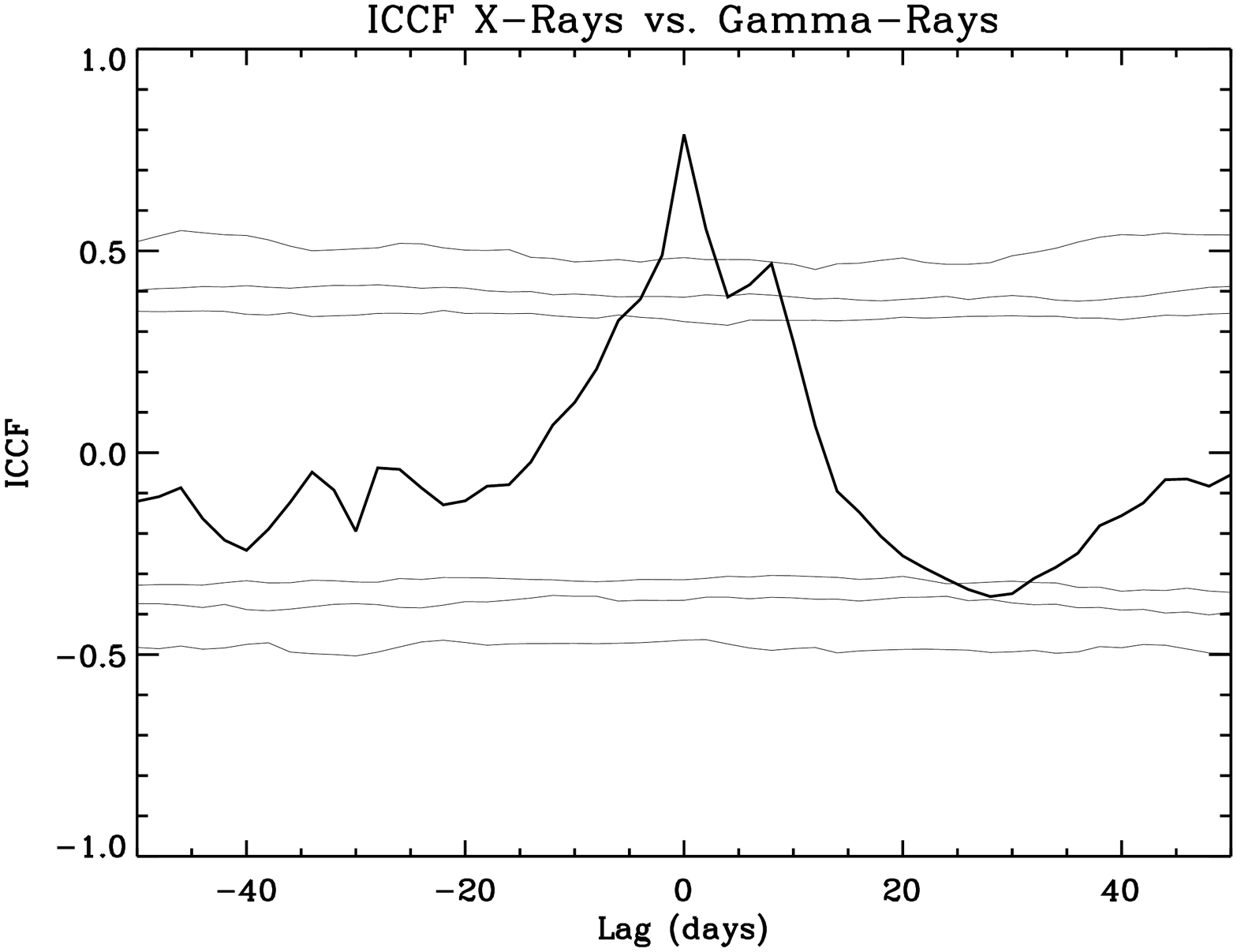}
\includegraphics[width=0.48\textwidth]{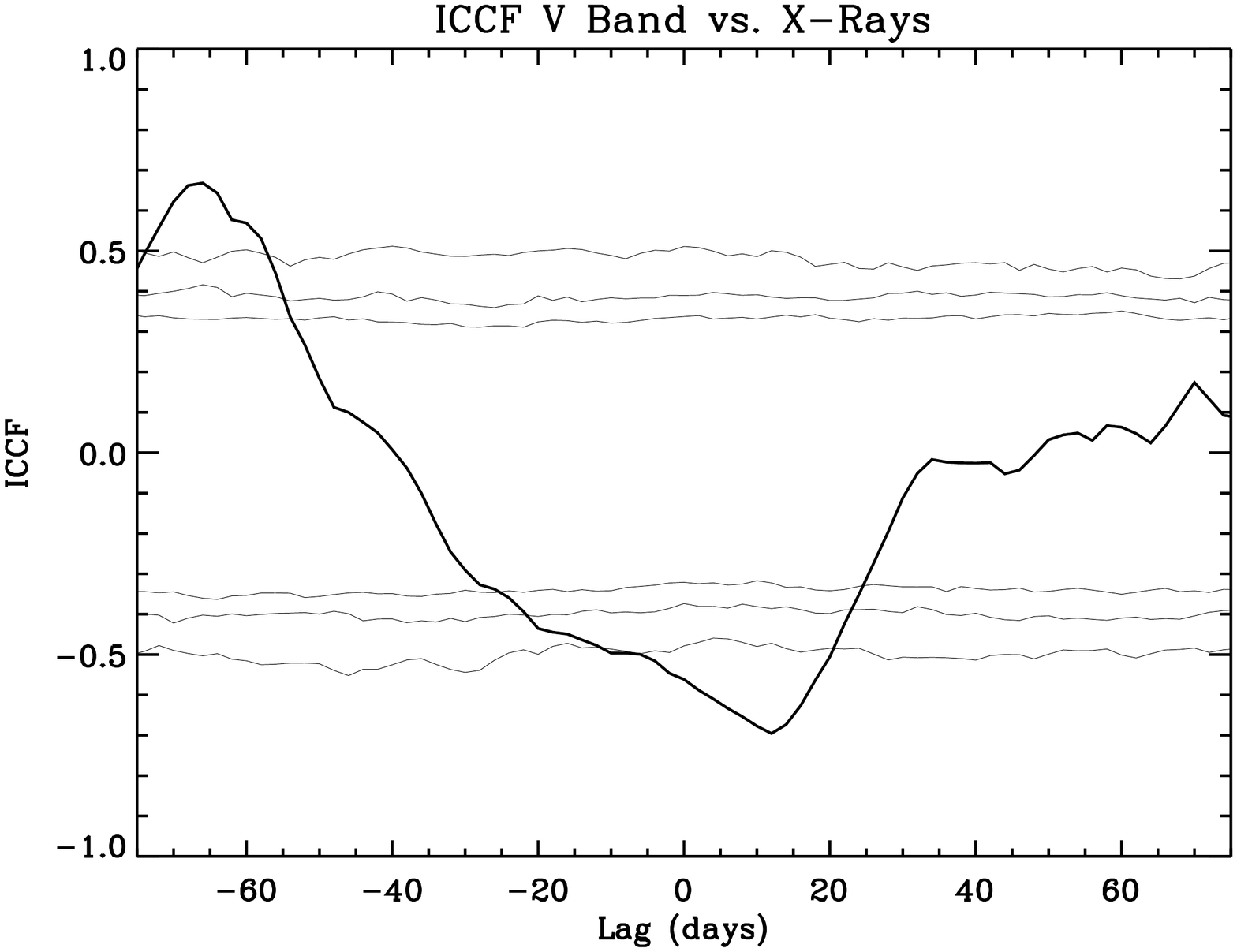}
\includegraphics[width=0.48\textwidth]{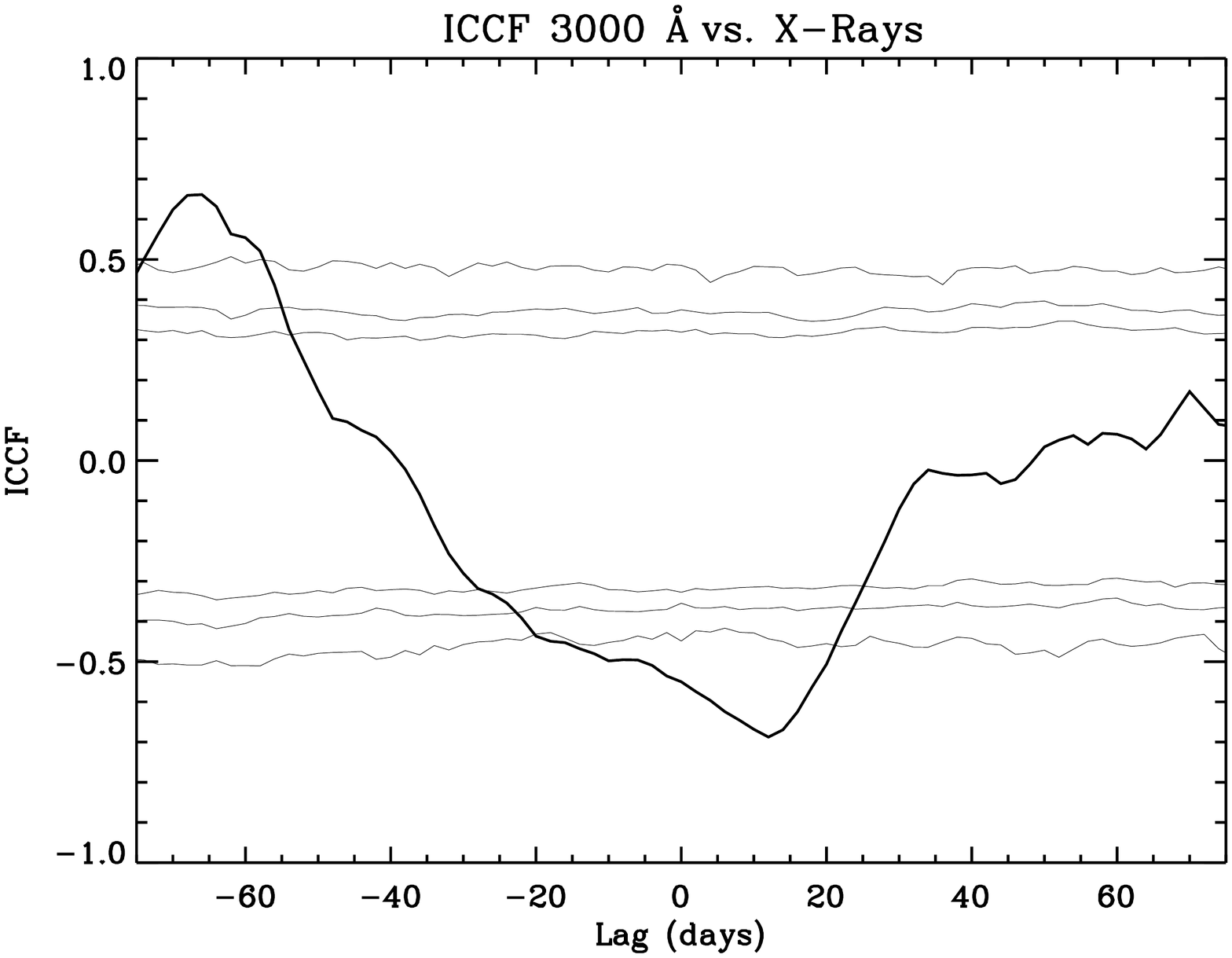}
\includegraphics[width=0.48\textwidth]{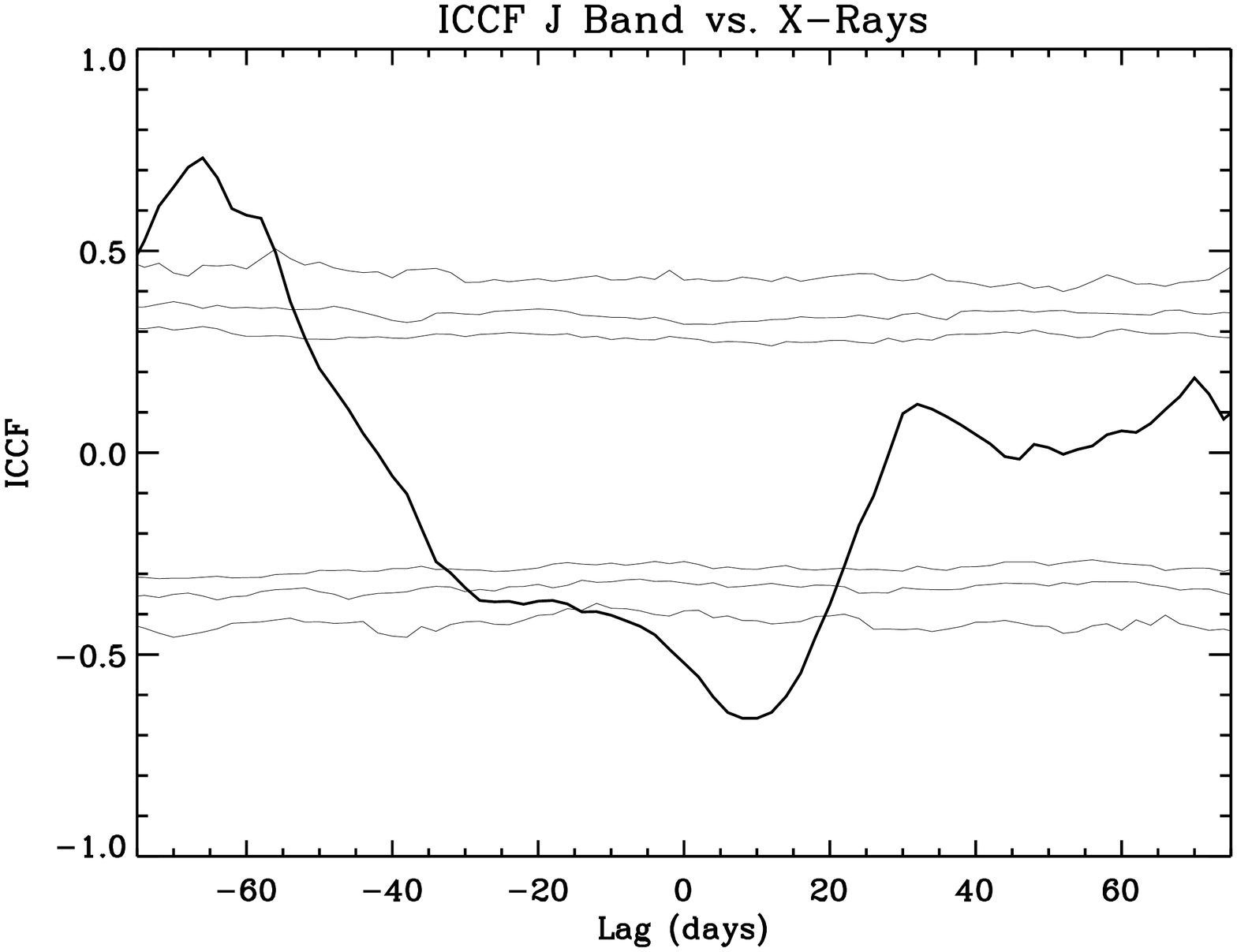}
\end{center}
\caption{Continuation of Fig.~\ref{CC_PC}.}
\label{CC_PC2}
\end{figure}

\clearpage

\section{Theoretical Study of the Gamma-Ray Opacity in the Jet of 3C 279}

\label{app:cross_sections}


\subsection{Cross-Section Calculations}

\subsubsection{Electron-Positron Pair Production}

The creation of a pair of positive and negative electrons must be interpreted as a transition of an ordinary electron from a state of negative energy to a state of positive energy. The energy necessary to create a pair of free electrons is larger than $2m_ec^2$ (where $m_e$ is the electron mass and $c$ is the speed of light in vacuum). It can be supplied through the absorption of a $\gamma$-quantum or by impact of a particle with kinetic energy greater than $2m_ec^2$. Energy and momentum conservation, however, are only possible if another particle is present (for instance, a nucleus). Thus pairs will be created by $\gamma$-rays or fast particles in passing through matter. We consider the most importance case; that is the creation of pairs by $\gamma$-rays in the presence of a nucleus with charge $Z$ (atomic number).  \\
\vspace{0.3cm}
	
Denoting the energy and momentum of the two electrons by $E_+$, $\vec{p}_+$, $E_-$, $\vec{p}_-$, the process in question is the following: a $\gamma$-quantum passing through the Coulomb field of the nucleus is absorbed by an electron in the negative state $E=-E_+$, $\vec{p}=\vec{p}_+$, then the electron going over into a state of positive energy $E_-$, $\vec{p}_-$.	\\
\vspace{0.3cm}

This process is closely related to the Bremsstrahlung process. The reverse process to the creation of a pair is obviously the transition of an ordinary electron in the presence of a nucleus from a state with energy $E_0=E_-$ to a state $E=-E_+$ emitting a light quantum,

\begin{equation}
k=E_0-E=E_++E_-
\end{equation}

Given that we can see the pair production as a modified version of a Bremsstrahlung process, from \cite{Heitler1954} we obtain the general expression for the Bremsstrahlung differential cross-section as

\begin{equation}
d\sigma=\frac{\alpha Z^2e^4}{\pi^2}d\Omega d\Omega_k kdk \abs{ \sum{\left( \frac{(u^*u')(u'^*\alpha'u'')}{E-E'} + \frac{(u^*\alpha'u'')(u''^*u_0)}{E_0-E''} \right)}  }^2
\label{eq:phi_bre}
\end{equation}

Where the $u$ terms refer to matrix elements for the Coulomb interaction matrix, $e$ is the charge of the electron, $	\alpha$ is the fine-structure constant and $d\Omega$ refers to a differential of solid angle.	\\
\vspace{0.3cm}

Pair production differs from ordinary Bremsstrahlung only in that the energy in the final state is negative. Now, the matrix elements for the reverse process are the conjugate complex expressions of those for the direct process. We can therefore take the cross-section for the creation of a pair directly from the Bremsstrahlung expression. In calculating this, however, we must insert a density function $\rho_F$ for the final state. Since in the case of pair production we have in the final state a positive and a negative electron, from \cite{Heitler1954} our density function is given by

\begin{equation}
\rho_F=\rho_{E_+}\rho_{E_-}dE_+
\end{equation}

Furthermore, we have now to divide by the velocity of the incident light quantum (i.e. by $c$). Thus, the differential cross-section has to be multiplied by

\begin{equation}
\frac{\rho_{E_+}\rho_{E_-}dE_+}{\rho_{E}\rho_{k}dk} \frac{p_0}{E_0}=\frac{p^2_-dE_+}{k^2dk}
\label{eq:density}
\end{equation}

Since $\vec{p}_0=\vec{p}_-$, $\vec{p}=-\vec{p}_+$, the angles $\theta$, $\theta_0$, $\phi$, denoting the direction of the electron in the initial and final state, are connected with the angles $\theta_+$, $\theta_-$, $\phi_+$, denoting the direction of the positive and negative electron, by

\begin{equation}
\theta_+=\pi-\theta , \quad \theta_-=\theta_0 , \quad \phi_+=\pi+\phi
\label{eq:angles}
\end{equation}

Then putting

\begin{equation}
E_0=E_- , \quad E=-E_+ , \quad \vec{p}_0=\vec{p}_- , \quad \vec{p}=\vec{p}_+
\label{eq:energies}
\end{equation}

and inserting Eqs.~\ref{eq:density},~\ref{eq:angles} and \ref{eq:energies} in the formula \ref{eq:phi_bre}, we obtain the differential cross-section for the creation of a pair $\vec{p}_+$, $\vec{p}_-$:

\begin{equation}
\begin{split}
& d\sigma = - \frac{\alpha Z^2 e^4}{2\pi} \frac{p_+p_-dE_+}{k^2} \frac{\sin{\theta_+} \sin{\theta_-} d\theta_+ d\theta_- d\phi_+}{q^4} \times       \\
& \times \bigg \{ \frac{p^2_+ \sin^2{\theta_+}}{(E_+ - p_+ \cos{\theta_+})^2} (4 E^2_- q^2) + \frac{p^2_- \sin^2{\theta_-}}{(E_- - p_- \cos{\theta_-})^2} (4 E^2_+ - q^2) +       \\
& + \frac{2 p_+ p_- \sin{\theta_+} \sin{\theta_-} \cos{\phi_+}}{( E_- - p_- \cos{\theta_-} )( E_+ - p_+ \cos{\theta_+} )} (4 E_+ E_- + q^2 - 2 k^2) -      \\  
& - 2 k^2 \frac{ p^2_+ \sin^2{\theta_+} + p^2_- \sin^2{\theta_-} }{ ( E_- - p_- \cos{\theta_-} ) ( E_+ - p_+ \cos{\theta_+} ) }			 \bigg \}
\end{split}
\label{eq:phi_pp1}
\end{equation}

where

\begin{equation}
q^2 = (\vec{k} - \vec{p}_+ - \vec{p}_-)^2
\end{equation}

Integrating over the angles, the cross-section for the creation of a positive electron with energy $E_+$ and a negative one with energy $E_-$ then becomes 

\begin{equation}
\begin{split}
& \sigma_{E_+} dE_+ = \overline{\sigma} \frac{p_+ p_-}{k^2} dE_+  \bigg \{ -\frac{4}{3} - 2E_+ E_- \frac{p^2_+ + p^2_-}{p^2_+ p^2_-} +    \\
& + \mu^2 \bigg( \frac{E_+ \epsilon_-}{p^3} + \frac{\epsilon_+ E_-}{p^3_+} - \frac{\epsilon_+ \epsilon_-}{p_+ p_-} \bigg) + L \bigg[ \frac{k^2}{p^3_+ p^3_-}	(E^2_+ E^2_- + p^2_+ p^2_-) -    \\
& - \frac{8}{3} \frac{E_+ E_-}{p_+ p_-} - \frac{\mu^2 k}{2 p_+ p_-} \bigg( \frac{E_+ E_- - p^2_-}{p^3_-}\epsilon_- + \frac{E_+ E_- - p^2_+}{p^3_+}\epsilon_+ + \frac{2k E_+ E_-}{p^2_+ p^2_-} \bigg) \bigg]		 \bigg \}
\end{split}
\label{eq:phi_pp1}
\end{equation}

where

\begin{equation}
\begin{split}
& \epsilon_+ = 2 \log{\bigg( \frac{E_+ + p_+}{\mu} \bigg)} , \quad L=2\ln{\bigg( \frac{E_+ E_- + p_+ p_- + \mu^2}{\mu k} \bigg)}  \\
& \quad \quad \quad \quad \quad \quad \quad \quad \quad \quad \overline{\sigma}=\frac{Z^2 r_0^2}{137} 
\end{split}s
\label{eq:phi_pp2}
\end{equation}

In the extreme relativistic case where all energies are large compared with the rest energy of the electron (which applies to our case) Eq.~\ref{eq:phi_pp1} becomes

\begin{equation}
\sigma_{E_+}dE_+=4\overline{\sigma} dE_+ \frac{E^2_+ + E^2_- + \frac{2}{3} E_+ E_-}{k^2} \bigg[ \ln{\bigg( \frac{2 E_+ E_-}{k\mu} \bigg)} - \frac{1}{2} \bigg]
\label{eq:phi_pp3}
\end{equation}

Applying equipartition, i.e. $E_+=E_-=E_0 /2$, this equation reduces to 

\begin{equation}
\sigma_{E_+}=Z^2 r_0^2 \alpha \bigg\{ \frac{28}{9} \bigg[ \ln{ \bigg( \frac{2E_0}{mc^2} \bigg) } \bigg] - \frac{109}{42} \bigg\}
\label{eq:phi_pp4}
\end{equation}


\subsubsection{Inverse Compton Effect}

The process we discuss here is the following: A primary light quantum $\mathbf{k_0}$ collides with a free electron which we can assume to be initially at rest:

\begin{equation}
\mathbf{p_0}=0, \quad E_0=\mu \quad (\mu=mc^2)
\label{eq:phi_ic1}
\end{equation}

The general case $\mathbf{p_0} \neq 0$ can be obtained from the special case of Eq.~\ref{eq:phi_ic1} by a Lorentz transformation. In the final state the light quantum has been scattered, so that we have a quantum $\mathbf{k}$ instead of $\mathbf{k_0}$. Since the momentum is conserved in the interaction of light with free electrons, in the final state the electron has a momentum $\mathbf{p}$ (energy $E$)

\begin{equation}
\mathbf{p}=\mathbf{k_0} - \mathbf{k}
\label{eq:phi_ic2}
\end{equation}

The conservation of energy states that

\begin{equation}
E+k=k_0 + \mu
\label{eq:phi_ic3}
\end{equation}

According to Eqs.~\ref{eq:phi_ic2} and \ref{eq:phi_ic3} the frequency of the scattered quantum cannot be the same as that of the primary quantum. Using the relativistic relation between momentum and energy $p^2=E^2 - \mu^2$ and denoting the angle between $\mathbf{k_0}$ and $\mathbf{k}$ by $\theta$, we obtain from Eqs.~\ref{eq:phi_ic2} and \ref{eq:phi_ic3}

\begin{equation}
k= \frac{k_0 \mu}{\mu + k_0(1-\cos{\theta})}
\label{eq:phi_ic4}
\end{equation}

which is the well-known formula for the frequency shift of the scattered radiation. In the relativistic case, the frequency shift increases with the angle of scattering $\theta$. 		\\
\vspace{0.3cm}

An electron moving with relativistic velocity with a given momentum $\mathbf{p}$ can exist in four states, corresponding to the fact that the electron may have either of two spin directions and also a positive or negative energy:

\begin{equation}
E = \pm \sqrt{(p^2 + \mu^2)}
\label{eq:phi_ic5}
\end{equation}

Now, we calculate the transition probability by constructing the transition matrix. The compound matrix element, which determines the transition probability is then given by

\begin{equation}
K_{FO}=\sum{\bigg( \frac{ H_{FI} H_{IO} }{ E_O - E_I } + \frac{ H_{FII} H_{IIO} }{ E_O - E_{II} } \bigg)}
\label{eq:phi_ic6}
\end{equation}

where $\sum$ denotes the summation over all four intermediate states, i.e. over both spin directions and both signs of the energy. $E_O$, $E_I$, ... represent the total energies in the initial and the intermediate states. The energy differences occurring in the denominator of Eq.~\ref{eq:phi_ic6} are given by

\begin{equation}
\begin{split}
& E_O - E_I = \mu + k_0 - E'		\\
& E_O - E_{II} = \mu + k_0 - (E'' + k_0 + k) = \mu - E'' - k
\end{split}
\label{eq:phi_ic7}
\end{equation}

where $E'$, $E''$ represent the energy of the electron in the states I, II,

\begin{equation}
E' = \pm \sqrt{(p'^2 + \mu^2)} , \quad E'' = \pm \sqrt{(p''^2 + \mu^2)}
\label{eq:phi_ic8}
\end{equation}

If we denote the Dirac amplitudes of the electron with the momenta $\mathbf{p_0}$, $\mathbf{p}$, $\mathbf{p'}$, $\mathbf{p''}$ by $u_0$, $u$, $u'$, $u''$ and the components of the matrix vector $\mathbf{\alpha}$ in the direction of the polarization of the two light quanta $\mathbf{k_0}$ and $\mathbf{k}$ simply by $\alpha_0$ and $\alpha$, respectively, from \cite{Heitler1954}, the matrix elements for the transitions $O \rightarrow I$, etc., are given by

\begin{equation}
\begin{split}
H_{FI}= -e \sqrt{ \bigg( \frac{2\pi \hslash^2 c^2}{k} \bigg) (u^* \alpha u') } , \quad
H_{IO}= -e \sqrt{ \bigg( \frac{2\pi \hslash^2 c^2}{k_0} \bigg) (u'^* \alpha_0 u_0) }			\\
H_{FII}= -e \sqrt{ \bigg( \frac{2\pi \hslash^2 c^2}{k_0} \bigg) (u^* \alpha_0 u'') } , \quad
H_{IIO}= -e \sqrt{ \bigg( \frac{2\pi \hslash^2 c^2}{k} \bigg) (u''^* \alpha u_0) }	
\end{split}
\label{eq:phi_ic9}
\end{equation}

The transition probability per unit time for our scattering process is,

\begin{equation}
w=\frac{2\pi}{\hslash} \abs{K_{FO}}^2 \rho_F
\label{eq:phi_ic10}
\end{equation}

where $\rho_F$ denotes the number of final states per energy interval $dE_F$. By conservation of momentum the final state is determined completely by the frequency of the scattered quantum $\mathbf{k}$ and the angle of scattering. Therefore we have

\begin{equation}
\rho_F dE_F = \rho_k dk
\label{eq:phi_ic11}
\end{equation}

where $\rho_k$ denotes the number of states for the scattered quantum per energy interval $dk$. It would be incorrect, however, to equate the energy intervals $dk$ and $dE_F$. Since the final energy is given as a function of $k$ and $\theta$ by

\begin{equation}
E_F = k+ \sqrt{( p^2 + \mu^2 )} = k+ ( k_0^2 + k^2 - 2k_0k\cos{\theta} + \mu^2 )^{\frac{1}{2}}
\label{eq:phi_ic12}
\end{equation}

we obtain

\begin{equation}
\bigg( \frac{\partial k}{\partial E_F} \bigg)_{\theta} = \frac{Ek}{\mu k_0}
\label{eq:phi_ic13}
\end{equation}

and hence

\begin{equation}
\rho_F = \rho_k \bigg( \frac{\partial k}{\partial E_F} \bigg)_{\theta} = \frac{d\Omega k^2}{ (2\pi \hslash c)^3 } \frac{Ek}{\mu k_0}
\label{eq:phi_ic14}
\end{equation}

$d\Omega$ is the element of solid angle for the scattered quantum. Collecting our formulae \ref{eq:phi_ic6}, \ref{eq:phi_ic7}, \ref{eq:phi_ic9}, \ref{eq:phi_ic10}, \ref{eq:phi_ic14} and dividing by the velocity of light, we obtain the differential cross-section for the scattering process,

\begin{equation}
d\sigma= e^4 \frac{E k^2}{\mu k_0^2}d\Omega \bigg\{ \sum{ \bigg[ \frac{ (u^* \alpha u')(u'^* \alpha_0 u_0) }{ \mu + k_0 - E' } + \frac{ (u^* \alpha_0 u'')(u''^* \alpha u_0) }{ \mu - k - E'' } \bigg] } \bigg\}^2
\label{eq:phi_ic15}
\end{equation}

By evaluating the matrix elements in Eq.~\ref{eq:phi_ic15}, we obtain for the summation

\begin{equation}
\Sigma = \frac{1}{2\mu} \bigg[ \frac{ (u^* \alpha K' \alpha_0 u_0) }{ k_0 } - \frac{ (u^* \alpha_0 K'' \alpha u_0 ) }{ k } \bigg]
\label{eq:phi_ic16}
\end{equation}

This can be further simplified by using the wave equation for $u_0$, $[(\mathbf{\alpha p_0})+\beta \mu] u_0 = E_0 \mu_0$. Since $\mathbf{p_0} = 0$, $E_0 = \mu$, $(1 - \beta) u_0 = 0$.	\\
\vspace{0.3cm}

However, this equation can only be used only when $\beta$ acts directly on $u_0$. Now $\beta$ anticommutes with $\alpha$ and $\alpha_0$ and hence the terms $1+\beta$ vanish and we obtain

 \begin{equation}
\Sigma = \frac{1}{2\mu} \bigg\{ u^* \bigg[ 2(\mathbf{e_0 e}) + \frac{1}{k_0} \alpha (\mathbf{\alpha k_0}) \alpha_0 + \frac{1}{k} \alpha_0 (\mathbf{\alpha k}) \alpha \bigg] u_0 \bigg\}
\label{eq:phi_ic17}
\end{equation}

The differential cross-section (\ref{eq:phi_ic15}) is proportional to the square of (\ref{eq:phi_ic17}). The value that Eq.~\ref{eq:phi_ic17} takes depends on the spin directions of the electron in the initial and final states. We are not, however, interested in the probability of finding the electron with a certain spin after the scattering process. We shall therefore sum $d\phi$ over all spin directions of the electron after the scattering process and shall average over the spin directions in the initial state. We must also take into account that the number of electrons available for the scatter is proportional to the atomic number ($Z$) of the most abundant elements, because those atoms will be the electron donators for the enviroment. Doing this, the differential cross-section now becomes

 \begin{equation}
d\sigma = \frac{1}{4} Z r_0^2 d\Omega \frac{k^2}{k_0^2}\bigg[ \frac{k_0}{k} + \frac{k}{k_0} - 2 + 4\cos^2{\Theta} \bigg]
\label{eq:phi_ic18}
\end{equation}

Eq.~\ref{eq:phi_ic18} represents the Klein-Nishina formula \citep{Klein1929} for our specific problem.	\\
\vspace{0.3cm}

It is convenient to consider the scattered radiation as composed of two linearly polarized components, one perpendicular and one parallel. Denoting the directions of polarization of $\mathbf{k_0}$ and $\mathbf{k}$ by $\mathbf{e_0}$ and $\mathbf{e}$ respectively, we can choose the following directions for $\mathbf{e}$:

\begin{itemize}
\item Perpendicular: $\mathbf{e}$ perpendicular to $\mathbf{e_0}$, $\cos{\Theta}=(\mathbf{e_0 e})=0$
\item Parallel: $\mathbf{e}$ and $\mathbf{e_0}$ in the same plane (i.e. in the $(\mathbf{k,e_0})$ plane),  \\
	$\cos^2{\Theta} =1 - \sin^2{\theta} \cos^2{\phi} $
\end{itemize}

Where $\phi$ is the angle between the $(\mathbf{k_0,k)}$ plane and the $(\mathbf{k_0,e})$ plane; and $\theta$ the angle of scattering $(\mathbf{k_0,k})$.	\\
\vspace{0.3cm}

To obtain finally the total intensity scattered into an angle $\theta$ we have to take the sum $d\phi = d\phi_{\bot} + d\phi_{\parallel}$. Also, substituting the fact that $\Gamma=\frac{k_0}{\mu}$, the differential cross-section becomes:

\begin{equation}
d\sigma = Z r_0^2 d\Omega \frac{1+\cos^2{\theta}}{2} \frac{1}{ [ 1 + \Gamma(1-\cos{\theta}) ]^2 } \bigg\{ 1 + \frac{ \Gamma^2 (1-\cos{\theta})^2 }{ (1+\cos^2{\theta})[1+\Gamma(1-\cos{\theta})] } \bigg\}
\label{eq:phi_ic19}
\end{equation}

To obtain the total scattering we have to integrate over all angles (solid angles and $\theta$); this yields to

\begin{equation}
\sigma_{IC}= 2\pi Z r_0^2 \bigg\{ \frac{1+\Gamma}{\Gamma^3} \bigg[ \frac{2\Gamma(1+\Gamma)}{1+2\Gamma} - \ln(1+2\Gamma) \bigg] + \frac{1}{2\Gamma} \ln(1+2\Gamma) - \frac{1+3\Gamma}{(1+2\Gamma)^2} \bigg\}
\label{eq:phi_ic20}
\end{equation}


\subsubsection{Cross-Sections Comparison}

\begin{figure}
\begin{center}
\includegraphics[width=0.49\textwidth]{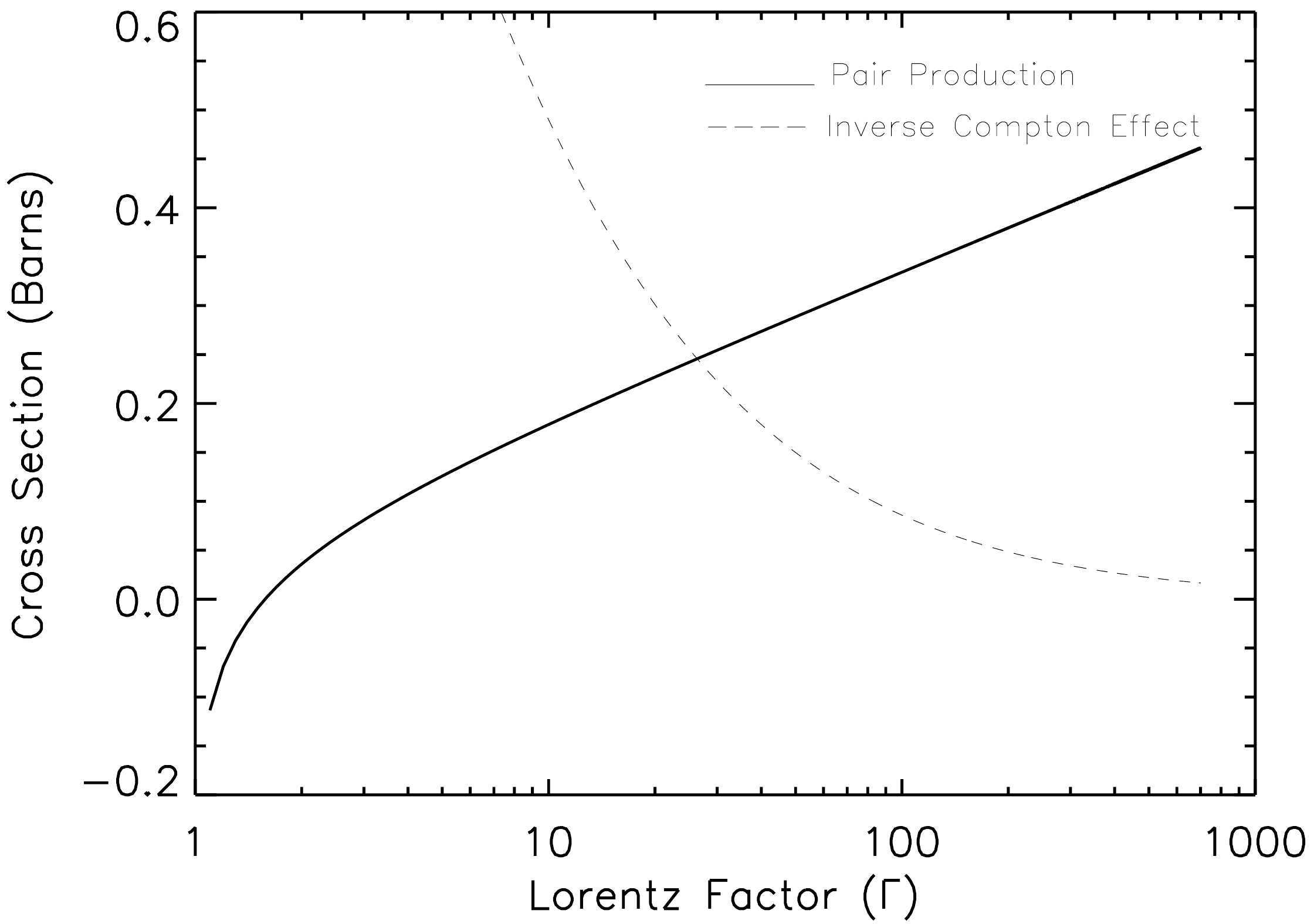}
\includegraphics[width=0.49\textwidth]{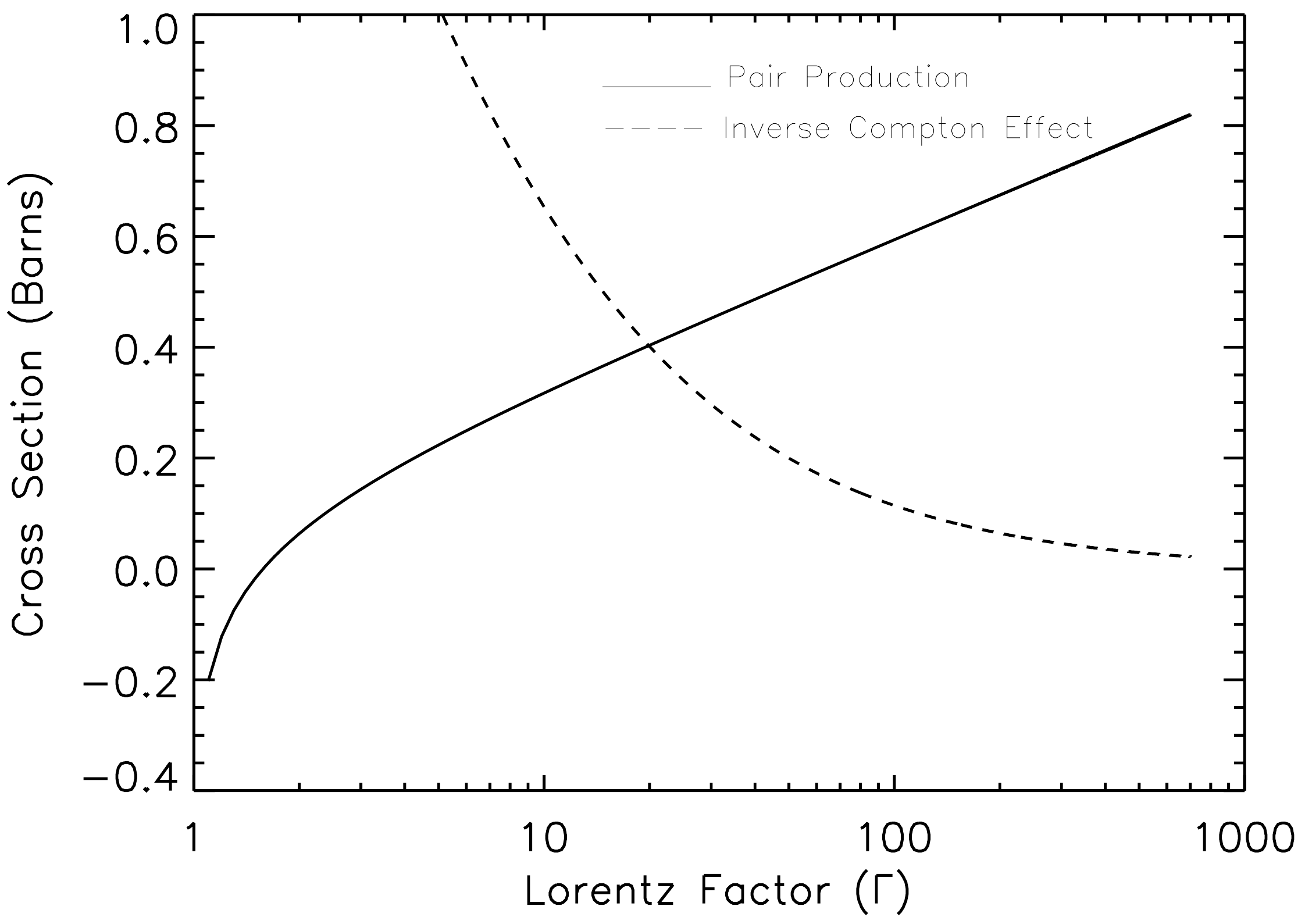}
\caption{Cross-Sections ($\sigma$) calculated as function of the Lorentz factor ($\Gamma$) for two cases. Left panel: The dominant element in the medium is Carbon. Right panel: The dominant element is Oxygen.}
\label{fig:CS}
\end{center}
\end{figure}

The cross-sections of both processes, electron-positron Pair Production (PP) and Inverse Compton Scattering (ICS) were calculated for a wide range of Lorentz factors. In the case of the pair production, the pair production referred is that from the produced electrons and positrons. We should take into account that both of these quantities depend on the atomic number of the most abundant element, in the case of the ICS as donators of electrons, and in the case of the PP, as the nucleus that gives place to the reaction. The calculated cross-sections for both processes can be seen in Fig.~\ref{fig:CS}.	\\
\vspace{0.3cm}

\begin{figure}
\begin{center}
\includegraphics[width=0.49\textwidth]{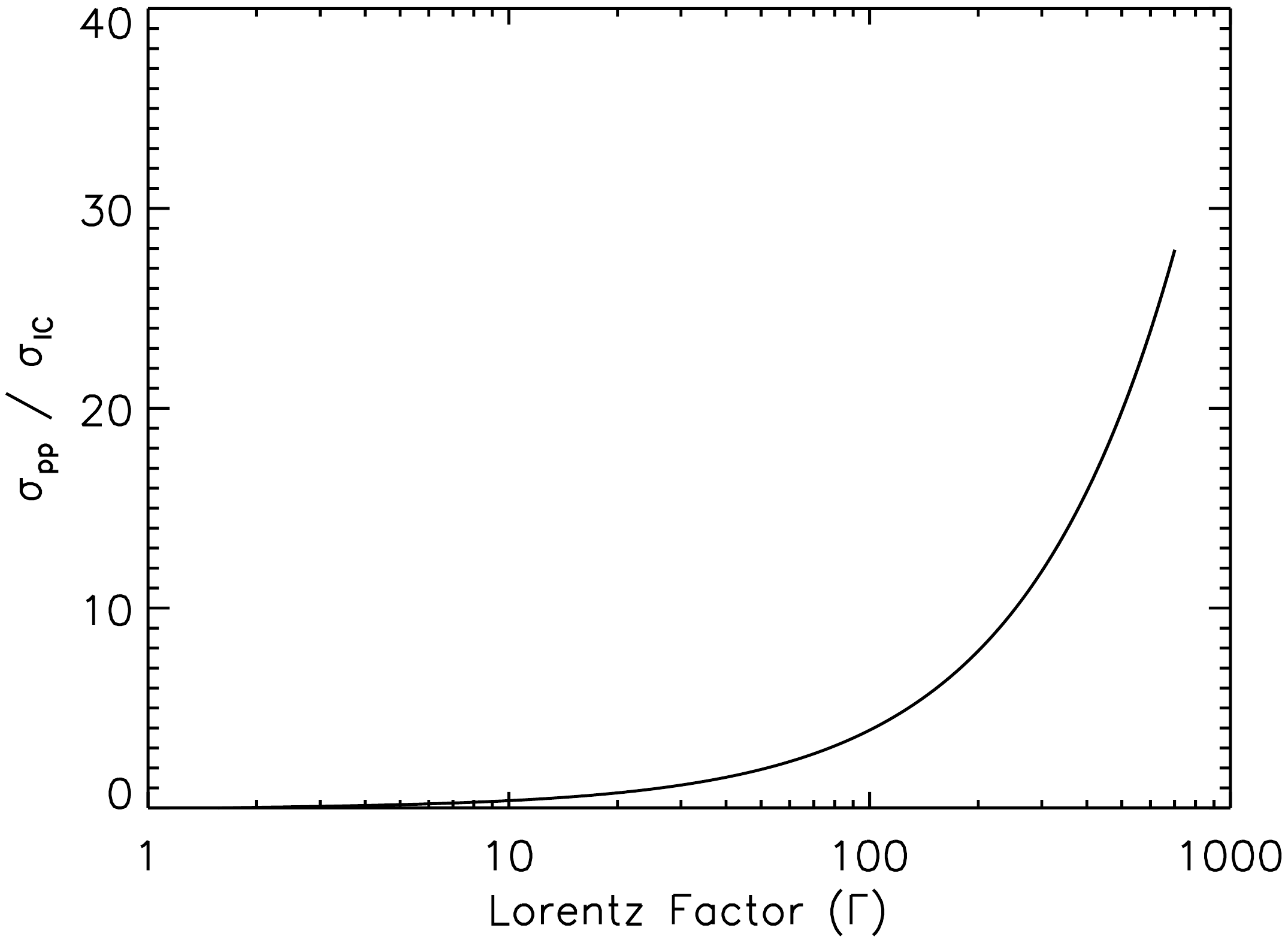}
\includegraphics[width=0.49\textwidth]{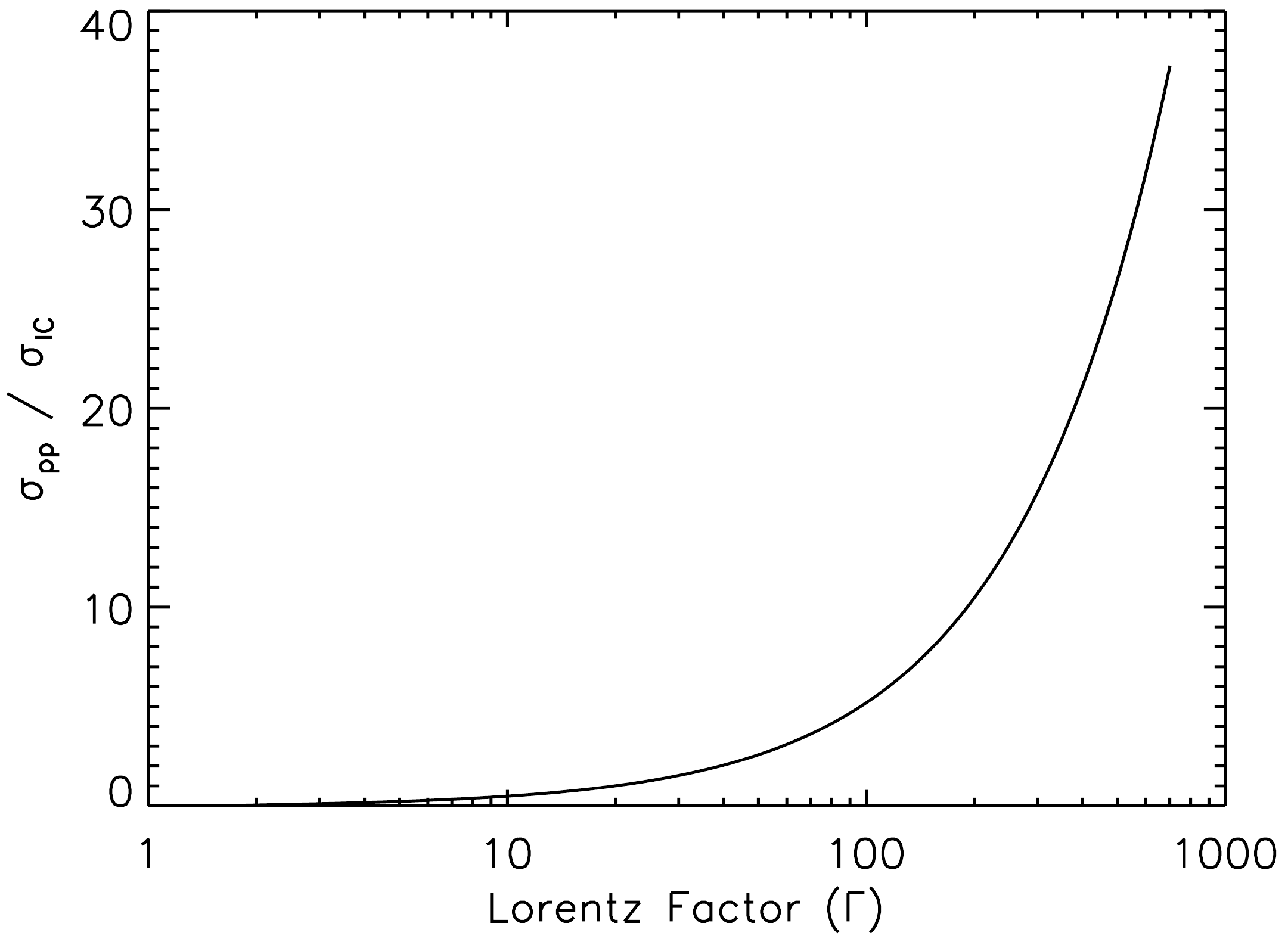}
\caption{Ratio of the PP to ICS cross-sections calculated as function of the Lorentz factor ($\Gamma$) for two cases. Left panel: The dominant element in the medium is Carbon. Right panel: The dominant element is Oxygen.}
\label{fig:CSR}
\end{center}
\end{figure}

For a better appreciation of how the cross-section vary with respect to each other, Fig.~\ref{fig:CSR} shows the ratio of the cross-sections as a function of the Lorentz factor.

\subsection{Lorentz Factor of the electrons radiating optical emission}

The frequency at which the maximum of the synchrotron emission is located, can be found as

\begin{equation}
\nu_{max}=0.29\nu_c , \quad \nu_c = \frac{c}{2\pi \rho} \Gamma^3
\label{eq:syn1}
\end{equation}

where $\nu_c$ is the critical frequency, $c$ is the speed of light in vacuum, $\Gamma$ is the Lorentz factor of the emitting electron, and $\rho$ is the curvature radius.	\\
\vspace{0.3cm}


The MOJAVE collaboration measured values of apparent velocity ($\beta_{app}$) for 3C 279 at different epochs \citep[see][]{Lister2009,Hovatta2014,Homan2015,Lister2016}, from which we can calculate Lorentz Factors. We first correct the apparent velocity by orientation, assuming a jet viewing angle of 2.4 degrees \citep{Hovatta2009}, and then calculate the Lorentz factors of the radio emitting electrons. The Lorentz factors obtained are in the range of $1.72 - 20.82$.

In order to obtain gamma-rays at the energies observed by Fermi, the seed photons for the ICS should be between the Near-Infrared and Near-UV.	\\
\vspace{0.3cm}

Under the assumption of a constant magnetic field, it should not come as a surprise that electrons whose maxima of synchrotron emission is in radio frequencies, have different Lorentz factor than those whose maxima is found in the optical regime; henceforth, we can estimate a Lorentz factor for the optical emitting electrons, by assuming that the maximum of emission from the radio emitting electrons is around 15 GHz (observation wavelength of MOJAVE).	\\
\vspace{0.3cm}

From Eq.~\ref{eq:syn1} we can calculate the curvature radius for the electrons in a magnetic field by using the calculated Lorentz factors, and a maximum frequency 15 GHz. Under the assumption of constant magnetic field throughout the emission region, the curvature radius should remain constant.	\\
\vspace{0.3cm}

By looking for a maximum frequency in the optical at 5000 \AA, and using the curvature radius obtained in the prior step, we obtain estimations of Lorentz factors in the range of $\sim58 - 711$ for the optical emitting electrons.		\\
\vspace{0.3cm}

As we can see from Fig.~\ref{fig:CSR}, the cross-section of the PP is similar to that of the ICS for the lower Lorentz factors obtained, while it can trump the ICS when looking at the higher Lorentz factors obtained. This supports our hypothesis that the electron-positron pair production can produce gamma-ray opacity in the source 3C 279. Unfortunately, all the observations used above were taken prior to the time-frame of this study.


\subsection{Comparison with the litetrature}
\label{comparison}

To the knowledge of the authors, there is no work that directly computes the dependence of the cross sections for the two aforementioned processes to the Lorentz factor. However, there are a number of works whose conclusions closely relate to the conclusion of this work regarding the absorption of gamma-rays via electron-positron pair production.
 \\

\cite{Protheroe1986} proved that the mean interaction length of electrons for Compton scattering increases with the electron energy (and therefore the Lorentz factor); which means that Compton interactions are less frequent the higher the energy (and Lorentz factor) of the electron. However, it is also worth mentioning that the calculations were made for interaction with the photon field of the CMB, and not that of an AGN.
 \\

\cite{Mastichiadis1991} talks about the triplet pair production ($\gamma e \rightarrow 3e$), and how its cross-section increases as $\sim ln \bigg( \frac{\gamma \epsilon_0}{m_e c^2} \bigg)$, while the cross-section for Compton scattering decreases as $\sim \frac{m_e c^2}{\gamma \epsilon_0}$. Even when the process is not exactly the same that we study in this work, this is in general agreement with our results, further supporting the idea of gamma-ray absorption at high Lorentz factors.
 \\

\cite{Zdziarski1993} study the gamma-ray emission of the blazar Mrk 421, in which they calculate the gamma-ray opacity due to pair production; and even when the opacity on this source should be lower than 1 in order for their interpretation of the spectrum to hold (at least at the observation time), they speculate that internal absorption by pair production may be the cause of the absence of TeV gamma-rays in other blazars. They also performed calculations showing that for a Lorentz factor of 700, and fast emission region expansion ($\beta_{exp}\gtrsim0.2$), pair production is an important factor in the output gamma-ray spectra.
 \\

\cite{Petry2000} point out that during a high state of the blazar Mrk 501 in 1997, the Klein-Nishina cross-section for Inverse Compton was greatly reduced, and therefore, there is no significant radiative output at (in the observer's frame) gamma-ray energies. This situation is similar to the activity period B of 3C 279 presented in this work, where the source is in a high state in all bands, with the gamma-rays being greatly reduced during this time-frame.

\label{lastpage}
\end{document}